\documentclass[]{jfm}
\input{amssym.def}
\input{amssym.tex}
\newcount\ndots
\def\drawline#1#2{\raise 2.5pt\vbox{\hrule width #1pt height #2pt}}

\def\trian{\raise 1.25pt\hbox{$\scriptscriptstyle\triangle$}\nobreak\ }

\def\square{${\vcenter{\hrule height .4pt
        \hbox{\vrule width .4pt height 3pt \kern 3pt
        \vrule width .4pt}
        \hrule height .4pt}}$\nobreak\ }
\def\plus{\raise 1.25pt \hbox{$\scriptscriptstyle +$}\nobreak\ }
\def\x{\raise 1.25pt \hbox{$\scriptscriptstyle \times$}\nobreak\ }

\def\solidtrian{\raise 1.25pt
   \hbox to 3bp{
\hfill}\nobreak\ }

\usepackage{natbib}
\usepackage{psfrag}
\usepackage{graphicx}
\usepackage{rotating}
\usepackage{color}

\def\der#1#2{ {{\partial #1 } \over {\partial #2}} }

\def\omr{\omega_r}
\def\ur{u_r}
\def\omz{\omega_z}
\def\uz{u_z}
\def\omt{\omega_\theta}
\def\ut{u_\theta}
\begin{document}           

\title[Three-dimensional Hills vortex]
{
The three-dimensional instabilities and
destruction of the Hill's vortex
}

\author[ P.~ Orlandi]
{
P.\ns  O\ls R\ls L\ls A\ls N\ls D\ls I ,
}
\affiliation
{
Dipartimento di Ingegneria Meccanica e Aerospaziale\\
Universit\`a La Sapienza, Via Eudossiana 16, I-00184, Roma
}
\date{\today}
\maketitle

\begin{abstract}

The Hill vortex is a three-dimensional vortex structure solution of the Euler equations.  For small amplitude axisymmetric disturbances on the external surface from the linear stability analysis by \citet{moff78} emerged the formation of a tail. Successive studies  confirmed this fact, and in addition observed  a different shape of the tail with azimuthal small amplitude disturbances.  In this paper the Navier-Stokes equations are solved at high values of the Reynolds number imposing large amplitude axisymmetric and three-dimensional disturbances on the surface of the vortex.  The conclusion is that the azimuthal disturbances produce a hierarchy of structures inside the vortex maintaining the shape of the vortex. On the other hand the axisymmetric disturbances are convected in the rear side, are dumped and form an axisymmetric wave longer higher the magnitude of the surface disturbance.  Simulations in a moving frame allow to extend the evolution in time leading to the conclusion that the Hill vortex is unstable and produces a  wide range of energy containing scales characteristic of three-dimensional flows.  
\end{abstract}

\section{Introduction}

The most studied mechanism by which a vortex may become unstable,
concern with an inviscid instability which
emerges when a vortex is strained, either by other
vortices, or as in the case of a vortex ring, 
by other portions of the curved vortex itself.
The long wave instability of a pair of counterotating 
longitudinal vortices, theoretically explained by \cite{CROW} 
and  reproduced in the laboratory by \cite{lew_wil98}
is an example of the former mechanism.
In presence of disturbances of  greater amplitude the strain produced 
by one of the vortices in the dipole affects the other 
and {\it vice versa} producing a sinusoidal modulation along the length of the dipole
that results in large-amplitude small-scale vortical structures that
strongly deform the primary vortices. This is 
the short-wave cooperative-instability that should be
generated to rapidly destroy the trailing vortices behind aircraft
(\citet{or_ca_le_sh01}).
A theoretical study of the formation of azimuthal instabilities
on vortex rings was given by
\cite{widnal74}. Their predictions were verified
by simulations of the Navier-Stokes equations by \cite{sh_ver_orl94}.
At high Reynolds number the vortex ring when a range of small
disturbances at wave numbers different from the most
unstable are added may reach a turbulent state.
In order to reach the turbulent state the simulation
should follow the ring for a long distance
requiring a large computational cost. To accelerate 
this process it may let the ring to impinge a solid
wall.  \cite{orl_verz93} described this  interaction that
is characterised by the  creation of vorticity patches on a wide range 
of scales leading to a rapid destruction of the initial vortex ring.
\cite{ren15} at higher Reynolds number reached a fully turbulent
condition that was corroborated by energy spectra 
decaying following a characteristic $k^{-5/3}$ power law.
Moreover, they  found that
the major contribution to the turbulent kinetic energy comes
from the azimuthal component that is mainly distributed
in the core region of vortex rings.

The vortex introduced by \cite{hill1894} differently than vortex
rings is a solution of the 3D Euler equations preserving
its shape while is moving with a constant velocity.
The cylindrical coordinate system $(\theta,r,z)$ is the most
appropriate coordinate system for its study.   
The azimuthal vorticity $\omega_\theta$ 
varies with $\sigma=\sqrt{r^2+(z-z_0)^2}$ 
within a spherical region of radius $a$ centered at $z=z_0$. 
Outside $\omega_\theta=0$
the consequent  sharp interface can not be reproduced
numerically in the solution of the Euler equations.
Several approaches to study the instability of inviscid vortices
are mentioned in the introduction of \cite{protas16},
therefore are not reported in this paper. Here
we are interested to solve the full set of Navier-Stokes equations at the highest
possible Reynolds number. This number is chosen in
such a way to reproduce and maintain for a sufficient
time the interface at the surface of the vortex as sharp as possible.
The previous studies were focused on the linearised response of the vortex
to perturbation applied on its external surface.
For instance \citet{moff78} derived a set of
ordinary differential equations, through
a linear perturbation theory, that allows to analyse
the time evolution of axisymmetric disturbances of the surface.
They found that, depending on the shape of
the disturbance, a sharp spike directed into or
out the vortex is localised at the rear stagnation
point. The same conclusions were reached by the
computations of \cite{pozrikidis86} of the non-linear regime.
\citet{fukuyu94} considered non-axisymmetric disturbances
leading to results consistent with those by \citet{moff78}.
They concluded that spikes form, and that their number is
equal to  the azimuthal wave number of the perturbation.
\citet{rozi99} by imposing small scale disturbances
reached the same conclusion on the deformation at the rear
stagnation point. They also found that the oscillatory motion
at the front observed in presence of polar  disturbances disappears
by adding azimuthal disturbances. They conclude that the Hill
vortex subjected to small disturbances
is unstable, and that the exponential growth
of the spike by varying the azimuthal wave 
number remains to be clarified, as  well as
to study the instability at the interior of the vortex.

The present study is devoted to understanding better the
obscure points before mentioned. Since any 
numerical method can not resolve a discontinuous
vorticity field a smoothing should be used
to let the initial $\omega_\theta$ to go to zero
outside the surface. 
\citet{stanaway88}  to reproduce the collision of two Hill's vortices
through pseudo-spectral methods applied the
smoothing to overcome the abrupt jump
between the vortical and the irrotational regions.
The smoothing function, was introduced 
by \citet{melander87} in two-dimensional simulations 
to connect the inner to the outer values of the 
vorticity within a layer of
thickness $\delta$. The amplitude of this layer 
can be varied in amplitude with radial and
azimuthal wave numbers, to mimic surface disturbances.

On solving the Navier-Stokes equations, a first consideration
should be done on which is the appropriate  value of the
Reynolds number in order to be as
close as possible to an inviscid simulation.
The first check of the accuracy and of the quality
of the numerical method implies that by
inserting large axisymmetric disturbances on the surface 
the vortex must remain axisymmetric in a fully three-dimensional
simulation. This occurrence can be corroborated by proving
that the other two vorticity components remain equal to zero. 
On the other hand, when a combination of axisymmetric
and azimuthal disturbances are assigned 
the direct numerical simulation allow to understand in which region 
of the vortex the secondary vorticity components grow
and become large enough to destroy the spherical
shape of the Hill vortex. In addition it
can be understood whether are generated flow structures 
characteristic of fully turbulent flows as those  in
the viscous collision of two orthogonal Lamb dipoles
(\citet{orlandi12}).
An efficient numerical method in presence of a single
wave number excited should maintain the initial
symmetries, and the comparison with the random wave number disturbances
allows to establish whether
the destruction of the Hill vortex is enhanced or not.
The response of the Hill vortex to large amplitude 
disturbances allow to investigate
whether similar conclusion of those obtained with
the assumption of a linear instability analysis are 
reached. The further advantage of the DNS is the
possibility  to look at the evolution of the flow inside the
vortex, to understand the mechanism transferring the
disturbances in space and to look at the preferential growth of secondary
vorticity in certain region. That is to study the entire process 
leading to the complete
destruction of the main vorticity distribution characteristic
of the Hill vortex.

\section{ Numerical scheme}

In  three-dimension  the  Navier-Stokes equations in
primitive variables reduces the number of operations to
advance the solution.
In cylindrical  coordinates, the definition of the variables
$q_{\theta}=rv_{\theta} , \ \ \ q_r= rv_r , \ \ \ q_z=v_z $,
as done by \cite{orlandi97} leads to the equation
in Chapt.10 of \citet{orlandi00}. The discretization of
the equation does not differ from that described in
\citet{verzicco96} where   the variable $v_{\theta}$.  was used. 
The quantities $q_i$ do not have the
same physical dimensions, and this could be a matter of confusion.
We wish to point out that these variables
were introduced for numerical reasons.  The introduction of $q_r$
simplifies the  treatment  of  the singularity at
$r=0$, and $q_{\theta}$ gives a better accuracy near the axis.
In the present simulations an external free-slip boundary
has been assumed at a distance equal to $\pi/a$, as
in \cite{orlandi97} periodicity holds in $z$, with  $-\pi < z < \pi$.
Even if finite differences allow to cluster the grid points
in regions of high gradients here  an uniform grid in $r$
has been used.

The numerical scheme is described in Chapt.10 of \citet{orlandi00}
here  the  main  features  of  the   method are summarised.
Viscous and advective terms are discretized
by  centred  second  order  schemes.
In the three-dimensional case,  in the limit of
$\nu  \rightarrow  0$,  energy is conserved
by the discretized equations.  The  system
of  equations  was  solved by a fractional step method.
In a first step a non-solenoidal velocity field $\hat q_i$ is computed,
and, if the pressure gradients at the previous time step
are retained, the boundary  conditions for $\hat q_i$
are simplified.  A scalar quantity $\Phi$ is
introduced to project the non-solenoidal field
onto a solenoidal one. The large band matrix
associated to the elliptic equation for $\Phi$
is reduced to a tridiagonal matrix by periodic $ FFTs $
in the azimuthal and in the
axial directions. This procedure is
very efficient for obtaining the solenoidal velocity.
The updated pressure is computed from the scalar $\Phi$.
A third order  Runge-Kutta  scheme, 
introduced by \citet{wray87},
was used to advance in time through three sub-steps.
The viscous terms are treated implicitly by the Crank-Nicolson
scheme. The equation are made dimensionless by taking
the radius of the vortex $a$ and a reference velocity
linked to the vorticity distribution given in the next
section.

\section{Initial conditions}

The Hill  vortex at $t=0$ is centered on the axis at $z=z_0=-0.5\pi$.
The  distribution of the unperturbed 
$\omega_\theta$, if $\sigma=\sqrt{r^2+(z-z_0)^2}$,  is

\begin{eqnarray}
\omega_\theta=    \lambda \sigma \left\{ 
\begin{array}{ll}
1 {\hskip1.5cm}  &  \sigma \leq a_i \\  
1 - f(\xi) {\hskip0.4cm}& a_i\leq \sigma < a_e  \\
0 {\hskip1.5cm} & a_e < \sigma  
\end{array} \right.
\label{IHILL}
\end{eqnarray}

\noindent where $f(\xi)$ is the smoothing function ($f(\xi )= 0$ 
for $\sigma=a_i$ and
$f(\xi)= 1$ for $\sigma=a_e$, with $a_i=a-\delta_0/2$ and $a_e=a+\delta_0/2$) 
defined by \citet{melander87}, that
connects the inner to the outer values of the vorticity.
$\lambda$ is a constant related to the uniform stream velocity 
at infinity ($\lambda=-15U/2a^2$) of a irrotational flow
past a sphere of radius $a$. The vortex therefore, if $\lambda=1$ 
moves with  a constant velocity  $U=2a^2/15$. The circulation
of the unperturbed vortex is equal to $\Gamma_0=2/3$, therefore
the reference circulation can be assumed equal to $\Gamma_*=3\Gamma_0/2$
and the Reynolds number is $Re=\Gamma_*/\nu$ with $\nu$ the 
kinematic viscosity.

The disturbance on the surface of the vortex is assigned by varying the
thickness $\delta_0=a_e-a_1$ of $f(\xi)$ through azimuthal $n_a$ and polar
$n_p$ wave numbers. By taking a spherical reference system
($\sigma,\phi,\theta$) located at center of the vortex the
smoothing function $f(\xi)$ is applied in the narrow
layer $\delta=\delta_0\cos(n_p \theta)\sin(n_a \phi)$.

\subsection{Reynolds number effect}

\begin{figure}
\centering
\vskip -0.0cm
\hskip -1.8cm
\psfrag{ylab} {$z_c-z_5  $}
\psfrag{xlab}{ $t-5 $}
\includegraphics[width=8.0cm]{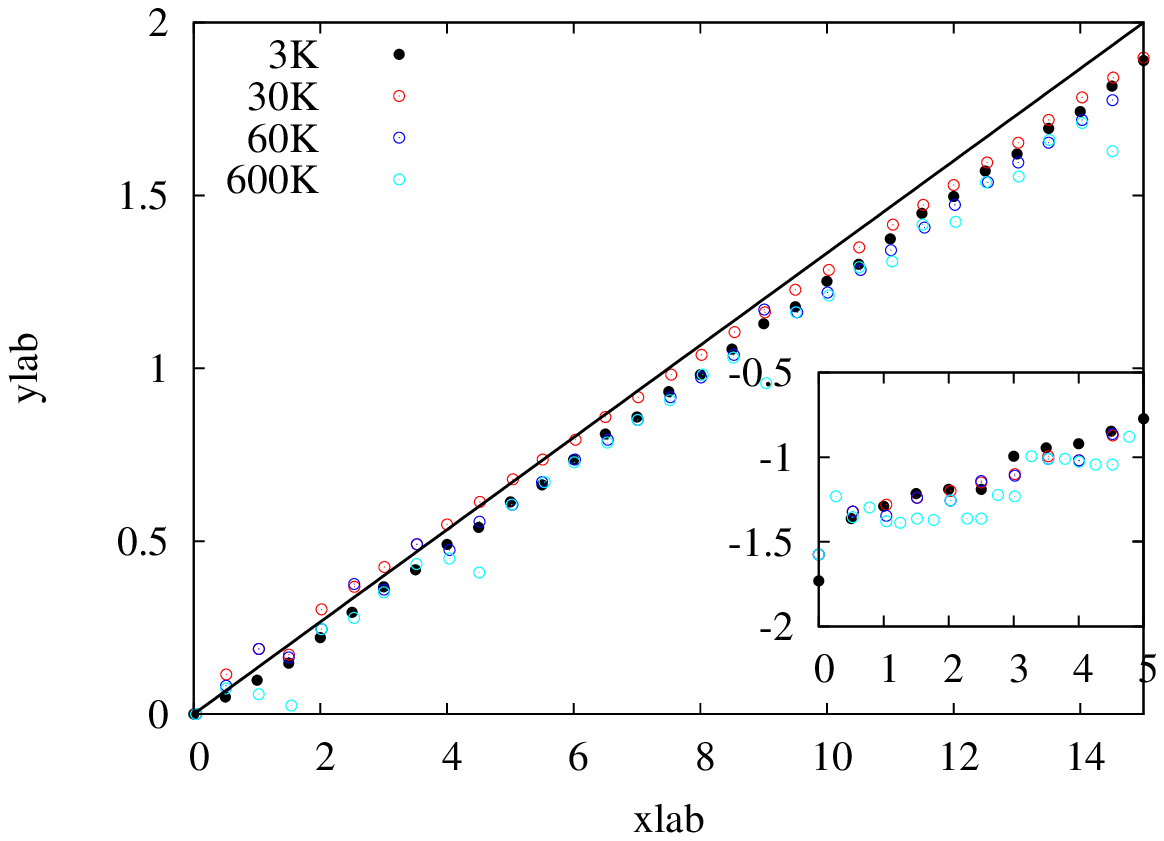}
\psfrag{ylab} {$ \omega_\theta  $}
\psfrag{xlab}{ $r $}
\includegraphics[width=8.0cm]{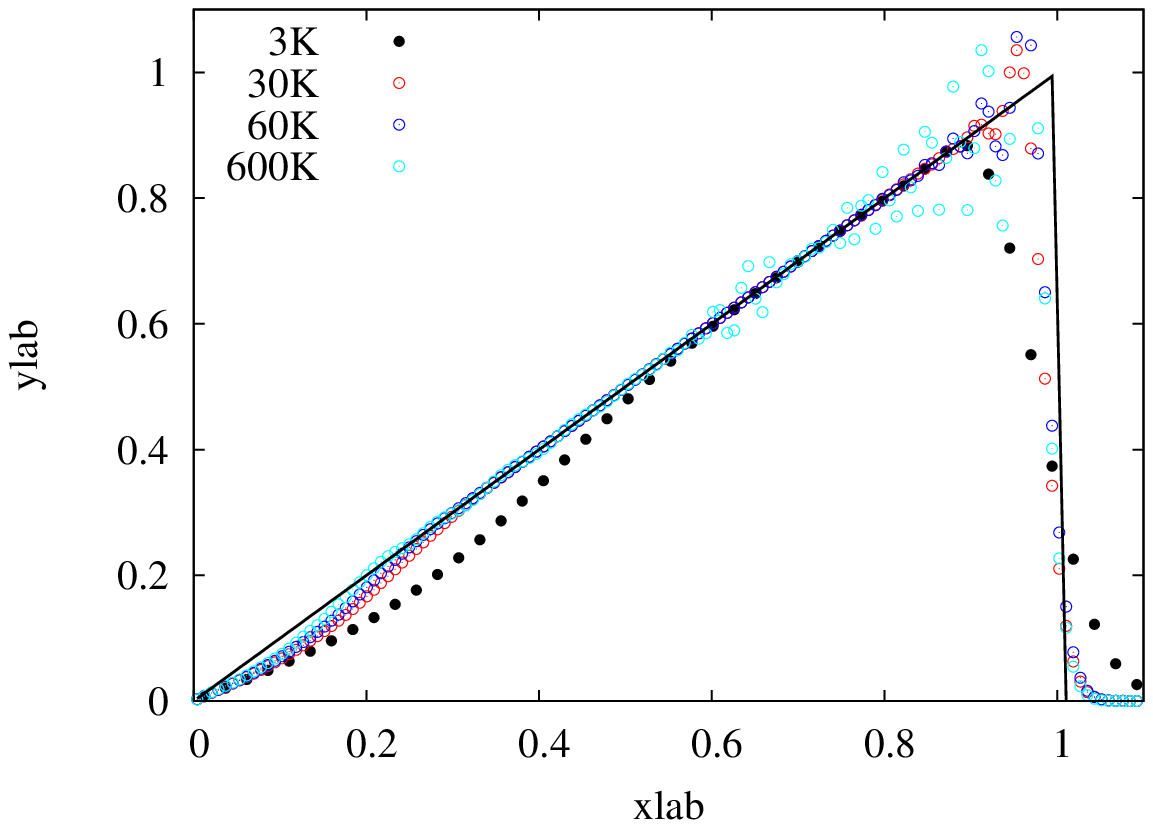}
\vskip -.5cm
\hskip 4.0cm  a)   \hskip 7.5cm  b)
\caption{Hill vortex with $\delta_0=0.01$ and $n_p=n_a=0$,
a) time evolution of the center of the vortex, b) profile of
$\omega_\theta$ versus $r$ at $z=z_c$ at $t=20$, the
solid symbols indicate the values obtained at the
$Re$ indicated in the legend.
}
\label{fig1}
\end{figure}

As above mentioned the three-dimensional viscous simulations should be performed
at a $Re$ number giving solutions not too different from
the theoretical ones in the inviscid case, and that without
any disturbance there is no any growth of the other
two vorticity components set equal to zero at $t=0$. 
A coarse grid $128\times 128 \times 256$ respectively
in the $\theta$ $r$ and $z$ directions was used at $Re=3000$, on the other
hand, at higher $Re$ numbers the grid was $192 \times 384 \times   768$
for the small $\delta_0$ and for others it was increased to
$256 \times 512 \times   1024$. With this mesh approximately
$330$ in $z$ and $160$ grid points in $r$ are used to
describe the vortex at $t=0$. The small amount of viscosity
helps to have an interface accurately described
by the mentioned grids.
A first validation of the numerical method and of the influence
of the viscosity can be performed by
comparing the distance travelled by the vortex with the
theoretical one determined by $U=2/15$.
The distance $z_c-z_0$ is evaluated by the discrete location of the maximum
vorticity. Figure \ref{fig1}a
shows that at any value of $Re$ there is a good agreement 
between the present one and that ($z_c-z_5=2/15 (t-5)$) in the inviscid case.
The inset of this figure explains why the first $5$ time
units are neglected. In fact, for any $Re$ number, 
the  oscillations on the $z_c$  are due to the inaccurate evaluation
of the center of the vortex, and also to the small amount of vorticity
shed during the motion. 
The small differences in figure \ref{fig1}a therefore are
due to the reduction of circulation 
due to the effect of the viscosity and to the small amount
of vorticity lost by the inaccurate representation of
the sharp interface at the boundary of the vortex.
This is emphasised in figure \ref{fig1}b by 
the  very large oscillations at $Re=600000$, that disappear at $Re=3000$
and become reasonable at $Re=30000$ and $Re=60000$. 
In addition, this figure shows that the viscosity affects the
vorticity profile not only at the edge but also in the central region. 
At $t=20$ the large smoothing at the edge, and the high deviation
from the linear profile for $0<r<5$ indicates that the results
at $Re=3000$ should be  quite far from those in the inviscid case. 
The deviation from the $\omega_\theta=r$ profile near the axis does not change
much going from $Re=30000$ to $Re=600000$. On the other
hand,  the oscillations near the edge of the vortex, those
we are mainly interested, at $Re=600000$ are too large. 
In both regions a good compromise is obtained at $Re=60000$
that has been used in all the perturbed simulations described in
the next sections.

\subsection{Polar perturbations}

\begin{figure}
\centering
\vskip -0.0cm
\hskip -1.8cm
\includegraphics[width=3.0cm,clip,angle=90]{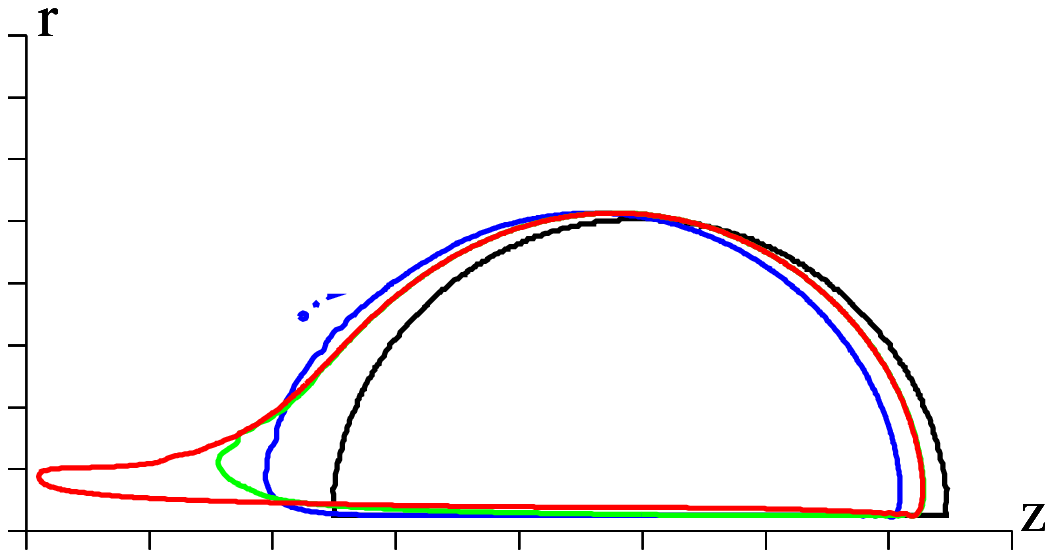}
\hskip 0.25cm
\includegraphics[width=3.0cm,clip,angle=90]{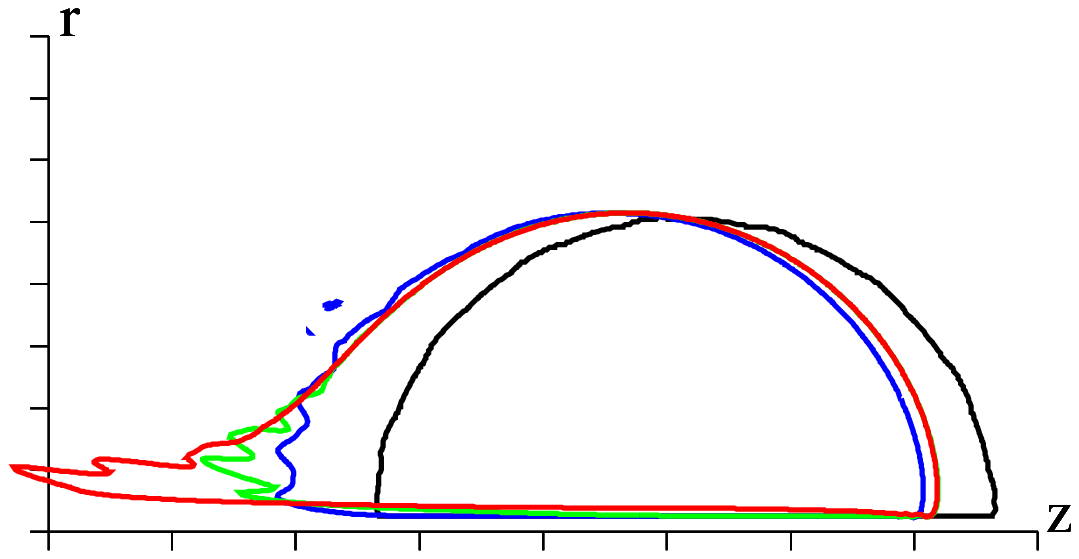}
\vskip 0.25cm
\hskip 3.3cm  a)   \hskip 5.3cm  b) 
\vskip -0.0cm
\hskip -1.8cm
\includegraphics[width=3.0cm,clip,angle=90]{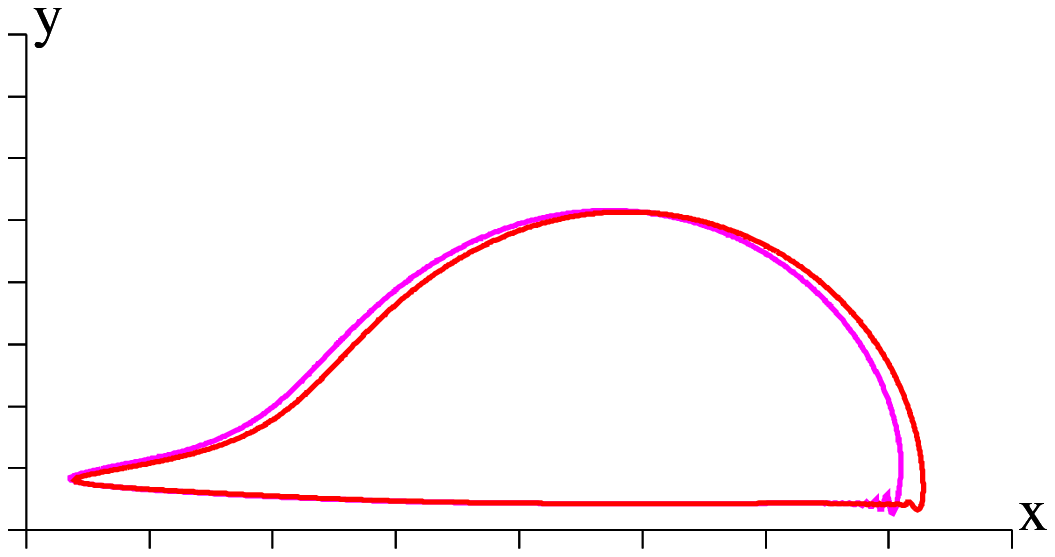}
\psfrag{ylab} {$z_F-z_R  $}
\psfrag{xlab}{ $t^* $}
\includegraphics[width=5.5cm]{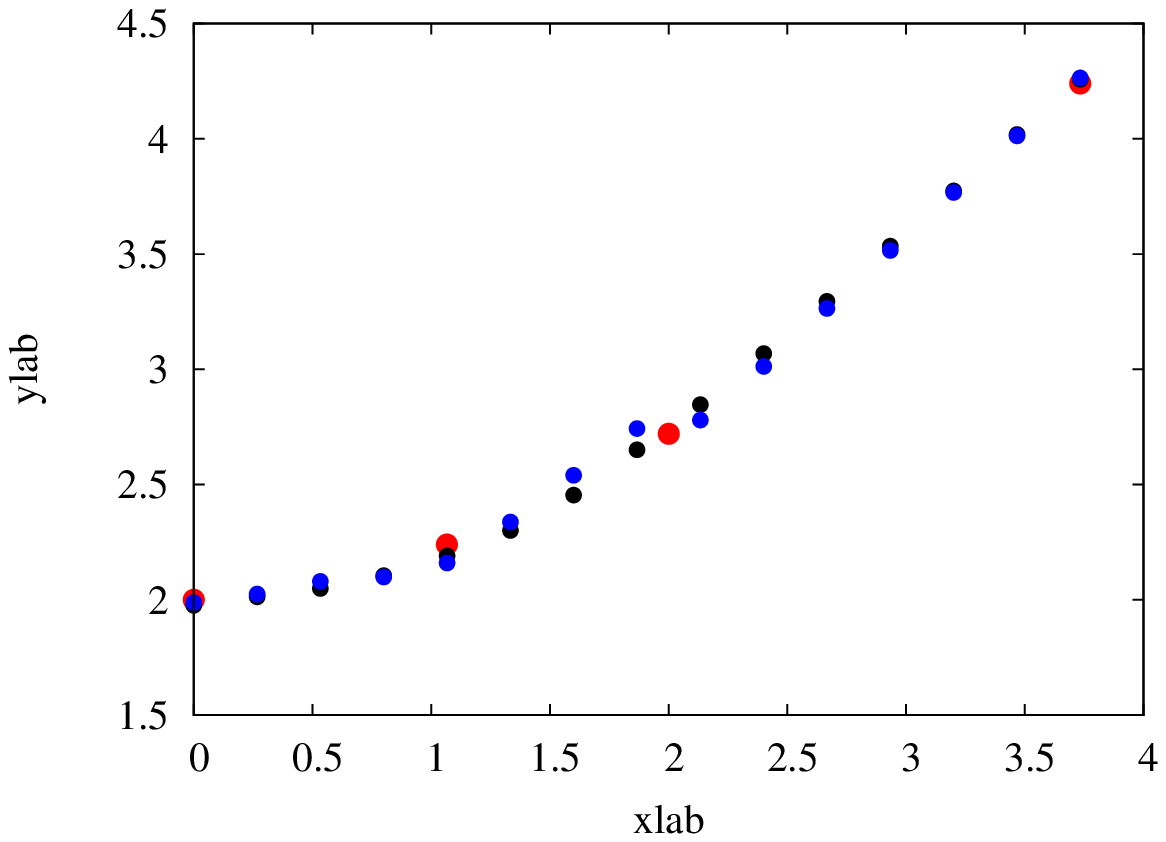}
\vskip 0.25cm
\hskip 3.3cm  c)   \hskip 5.3cm  d) 
\caption{ Contour of $\omega_\theta=0.05$ at $t=0$
black, $t=6$ blue $t=12$ green and $t=20$ red for a)
unperturbed $\delta_0=0.01$ ,b) perturbed $\delta_0=0.01$ and $n_p=16$,
c) $\omega_\theta=0.05$ at $t=30$ for unperturbed (red)
and perturbed (magenta), $r$ varies from $0$ to $1.5$ and
$-2 < z-z_c<1.0$;
d) time evolution of the distance between the front and
the rear back black $n_p=0$, blue $n_p=16$ and red
dots from figure 2 \cite{pozrikidis86}.
}
\label{fig2}
\end{figure}

\cite{moff78} assigned a linear disturbance combination
of Legendre polynomials demonstrating that for prolate disturbances
the vortex  detrained an amount of vorticity proportional
to the disturbance, that had a form of a spike growing in time
at the rear stagnation point. The same results were obtained by 
\cite{pozrikidis86} by large amplitude disturbances. Initially
the present simulations are devoted to considering  polar disturbances 
on the surface and to compare the evolution with 
that  of the unperturbed vortices.
The vortex perturbed with a $n_p=16$ wave number travels
for $30$ dimensionless time units corresponding to $t^*=Ut/a=2.66$.
In figure 2 of \cite{pozrikidis86} it can be 
estimated that the size of the spike is approximately equal
to the radius of the vortex. In figure \ref{fig2} the contours of 
$\omega_\theta=0.05$ at $t=0, 6, 12$ and $t=20$   show the growth
of the spike around the rear stagnation point. 
Even without polar disturbances the spike forms and grows in the rear
back of the vortex (figure \ref{fig2}a). This fact can be understood
by considering that the smoothing function $f(\xi)$ acts as a disturbance of
the inviscid solution.  The polar disturbance  with $n_p=16$ barely
visible on the surface at $t=0$ (black line), 
while the vortex travels, is advected
backward along the external boundary, and in figure \ref{fig2}b,
at $t=6$, the polar undulations are accumulated in
the rear. On the other hand, at the front the surface is  smooth  
and equal to  that in figure \ref{fig2}a at the same time. 
\noindent The amplitude of the undulation
grows in time during its translation, in fact it was small          
at $t=0$, and quite large at $t=6$ (blue line). Later on the perturbation
as well as the size of the spike continue to grow, and the undulations 
are accumulated on the surface of the
spike (green line).  At this high Reynolds number
these oscillations are not damped by the viscosity and
at $t=20$ produce a spike with shape different from that
without disturbances (red lines). However
at $t=30$ has been noticed that these disappear in the wake
and the vortices without and with polar disturbances have 
a similar form (figure \ref{fig2}c). The difference in
figure \ref{fig2}c between the perturbed and unperturbed
simulations are mainly visible at the front, in fact the
vortex with polar disturbances presents oscillations greater than
those in the unperturbed case. This oscillatory
motion at the front was also found by \citet{rozi99}.
In the inviscid simulations 
the form of the wake, 
after the vortex has moved for a long  distance,
at the rear, in the point where it is attached to the vortex,
shrinks considerably. 
Figure \ref{fig2}a and figure \ref{fig2}b show that 
in the present viscous simulation this sort of pinch-off
does not occur. However, the extension of the spike
measured as the distance from the front of the vortex up to
the end of the spike in \cite{pozrikidis86} agrees well
with the present unperturbed and perturbed $n_p=16$ results 
(figure \ref{fig2}d).                               

\subsection{Surface with azimuthal and polar perturbations}

\citet{rozi99} perturbed the surface of the vortex
with small amplitude azimuthal disturbances, they
presented the evolution of the surface at different time
in their figure 6 with $n_a=3$ and in figure 7 with $n_a=4$.
The inviscid results strengthened the modulation of the
spike by the azimuthal disturbance on the surface. In the
present simulations disturbances with $n_a=1,2,4$ and $n_a=8$ 
have been added to the previous discussed axisymmetric
disturbance with $n_p=16$. The combination of polar
and azimuthal disturbance implies that we are dealing
with a three-dimensional flow, therefore the simulations allow
to investigate whether and how the disturbances propagate
from the surface inside the vortex and in the attached spike. 
If the amplitude of the  disturbances do not grow it may be asserted
that the initial structure of the vortex is preserved even
if modifications of the surface, for instance the above
discussed spike, occur. On the other hand, if
the disturbances grow in magnitude there is a probability that
the vortex  breaks in a wide range of scales. This is the typical 
mechanism characterising the transition from laminar
to turbulent flows.

By imposing disturbances with a single
wave number the initial symmetry in the azimuthal direction should
be preserved. This a further satisfactory test required to validate
any numerical  method applied to the
Navier-Stokes equation written in polar coordinates. For the simulations
discussed in this section with azimuthal disturbances of
amplitude $\delta_0=0.01$ the resolution ($192 \times 384 \times   768$), i
in the previous section, is maintained. The DNS allow to look in
detail at the effects produced by the variation of $n_a$
on the distribution of the vorticity components
in planes normal to the streamwise motion. 
Before, however it is worth to looking at the global effects 
of the three-dimensional disturbances on the azimuthally averaged
vorticity components defined as 
$<\omega_j>(r,z)=\frac{\Sigma_i}{N_1} \omega_j$, having also
verified that  $<\omega_r>=<\omega_z>=0$.
\begin{figure}
\centering
\vskip -0.0cm
\hskip -1.8cm
\includegraphics[width=3.0cm,clip,angle=90]{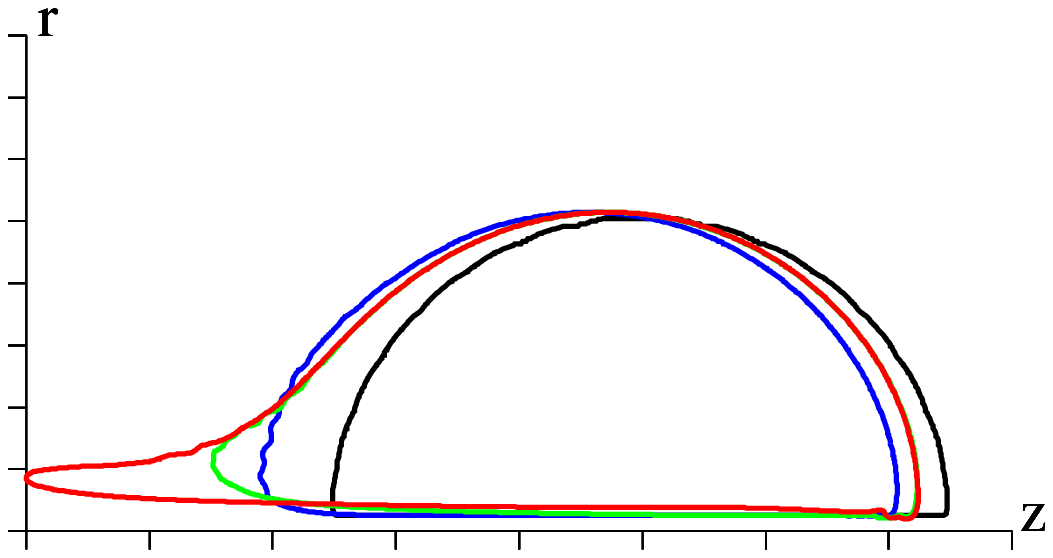}
\hskip 0.5cm
\includegraphics[width=3.0cm,clip,angle=90]{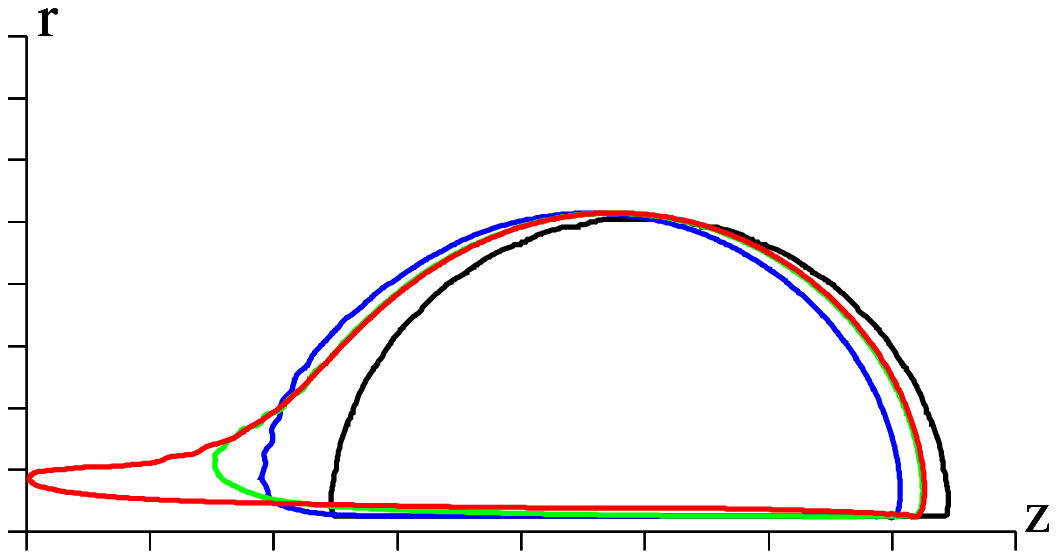}
\vskip 0.25cm
\hskip 3.3cm  a)   \hskip 5.3cm  b) 
\vskip 0.1cm
\hskip -1.8cm
\includegraphics[width=3.0cm,clip,angle=90]{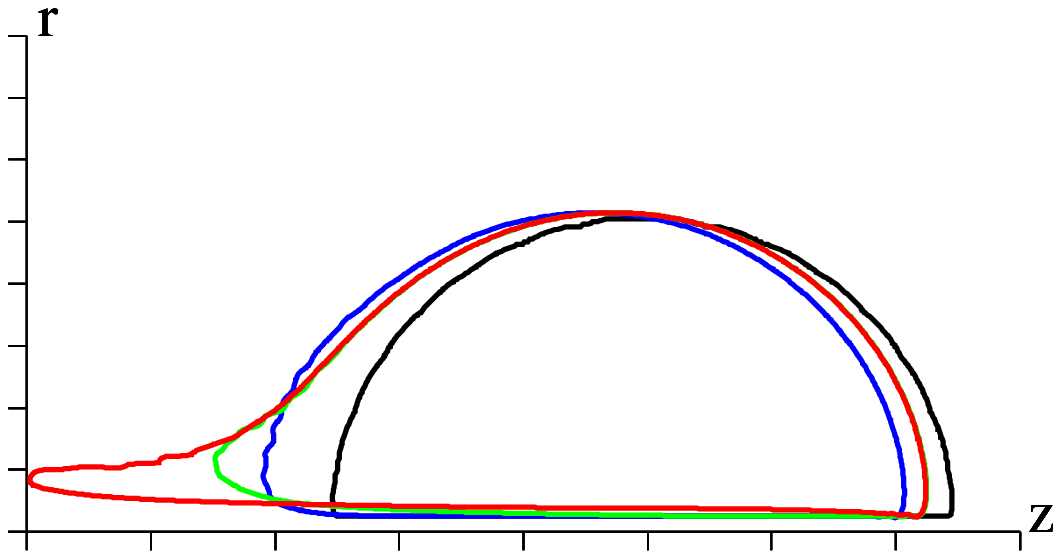}
\hskip 0.5cm
\includegraphics[width=3.0cm,clip,angle=90]{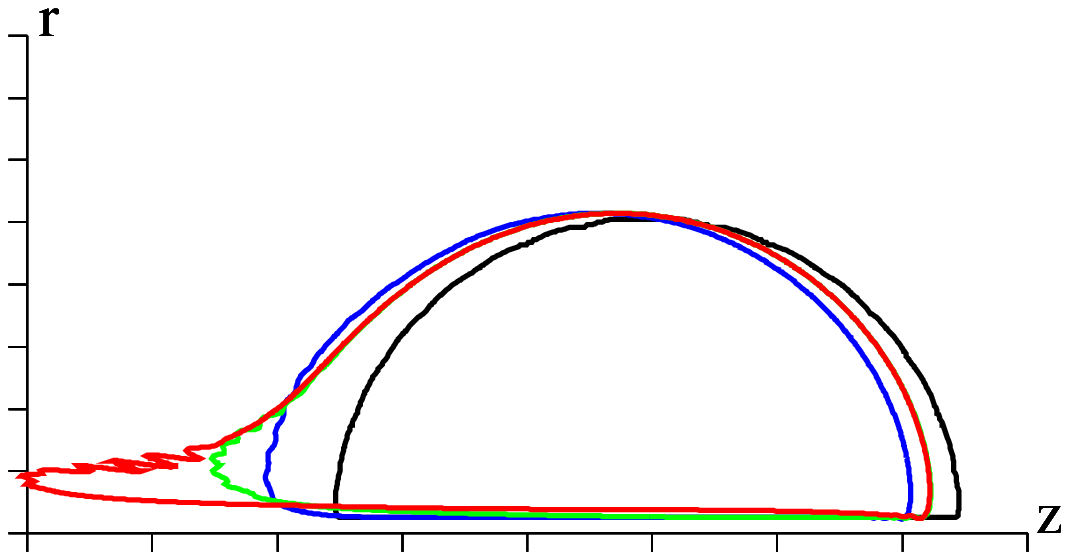}
\vskip 0.25cm
\hskip 3.3cm  c)   \hskip 5.3cm  d)
\caption{ Contour of $<\omega_\theta>=0.05$ at $t=0$
black, $t=6$ blue $t=12$ green and $t=20$ red for $\delta_0=0.01$
limits as in figure \ref{fig2}: 
a) $n_a=1$, b) $n_a=2$, c) $n_a=4$ and d) $n_a=8$.
}
\label{fig3}
\end{figure}
\noindent Figure \ref{fig3} similarly to  figure \ref{fig2} shows the
contours of $<\omega_\theta>$ at the  four $n_a$ considered.
The comparison between the contours in figure \ref{fig3}
an those in figure \ref{fig2}b , at a first glance,
shows that by adding azimuthal disturbances
the amplitude of the polar disturbances, while these
are transported towards the rear stagnation point, is reduced.
As a consequence the wake at $t=12$ (green line) does not have
the large amplitude disturbances depicted in figure \ref{fig2}b, 
and the  spike has a form similar to that of the unperturbed 
vortex in figure \ref{fig2}a. At the end of the simulation for 
$t=20$,  at any value of $n_a$, the red lines in 
in figure \ref{fig3} show that the amplitude of the
disturbances in the spike has been largely reduced and
becomes smaller lower $n_a$ is.

\begin{figure}
\centering
\vskip -0.0cm
\hskip -1.8cm
\includegraphics[width=2.3cm,clip,angle=90]{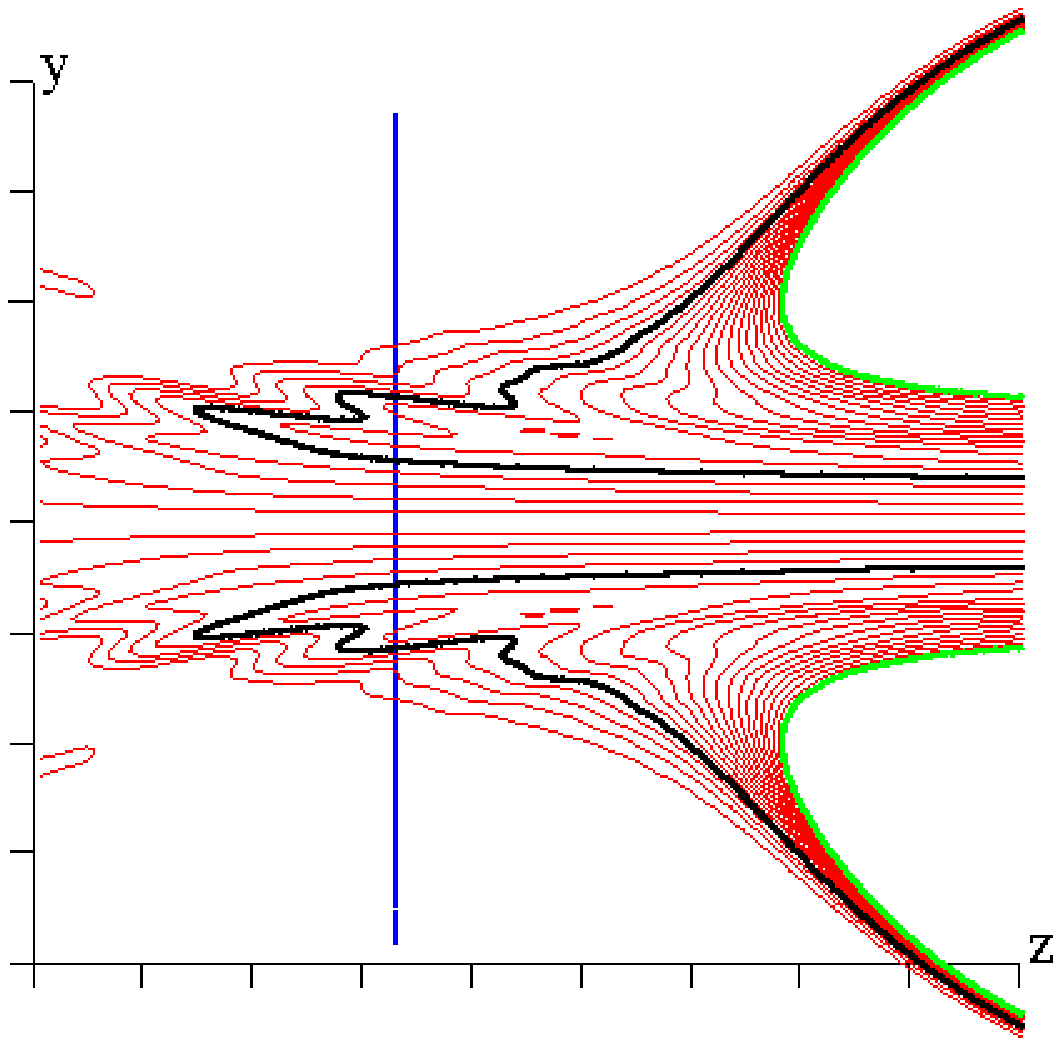}
\hskip 0.2cm
\includegraphics[width=2.3cm,clip,angle=90]{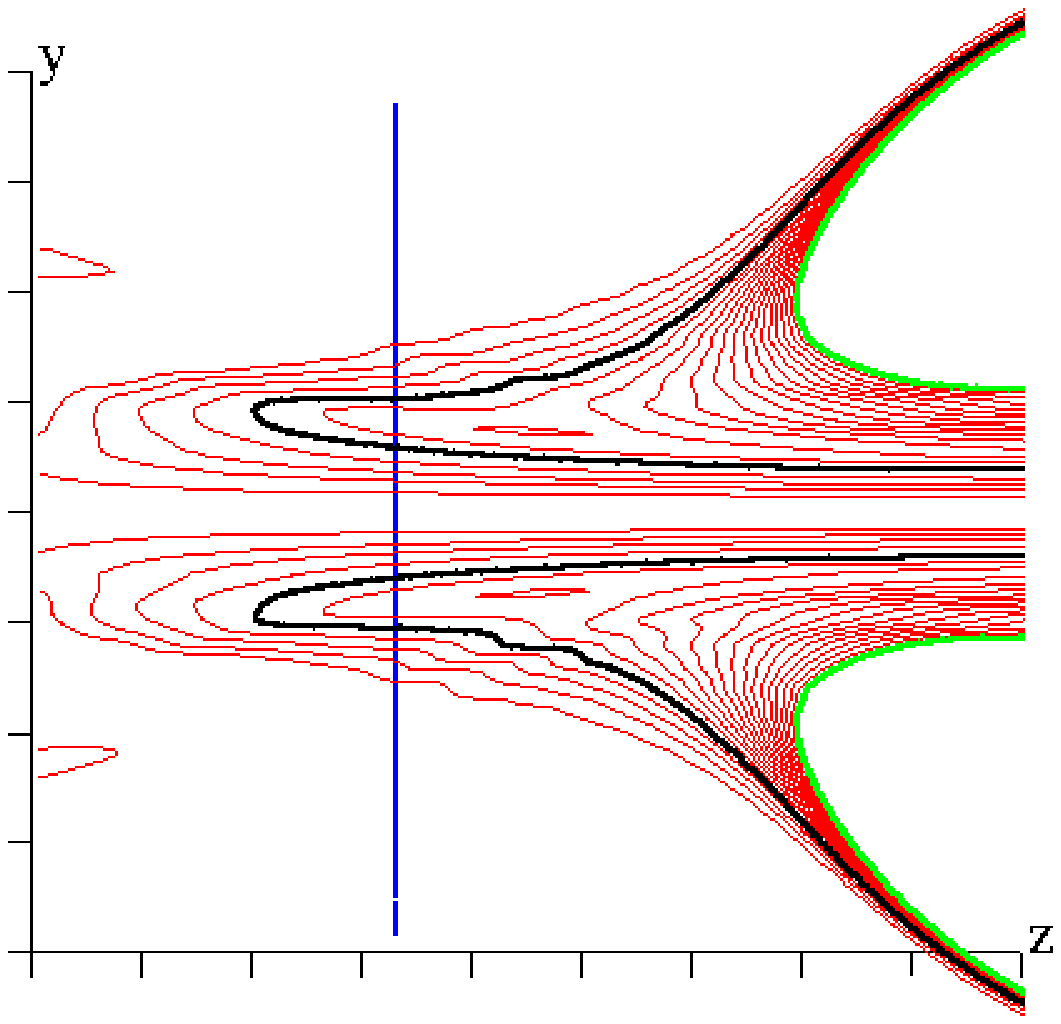}
\hskip 0.2cm
\includegraphics[width=2.3cm,clip,angle=90]{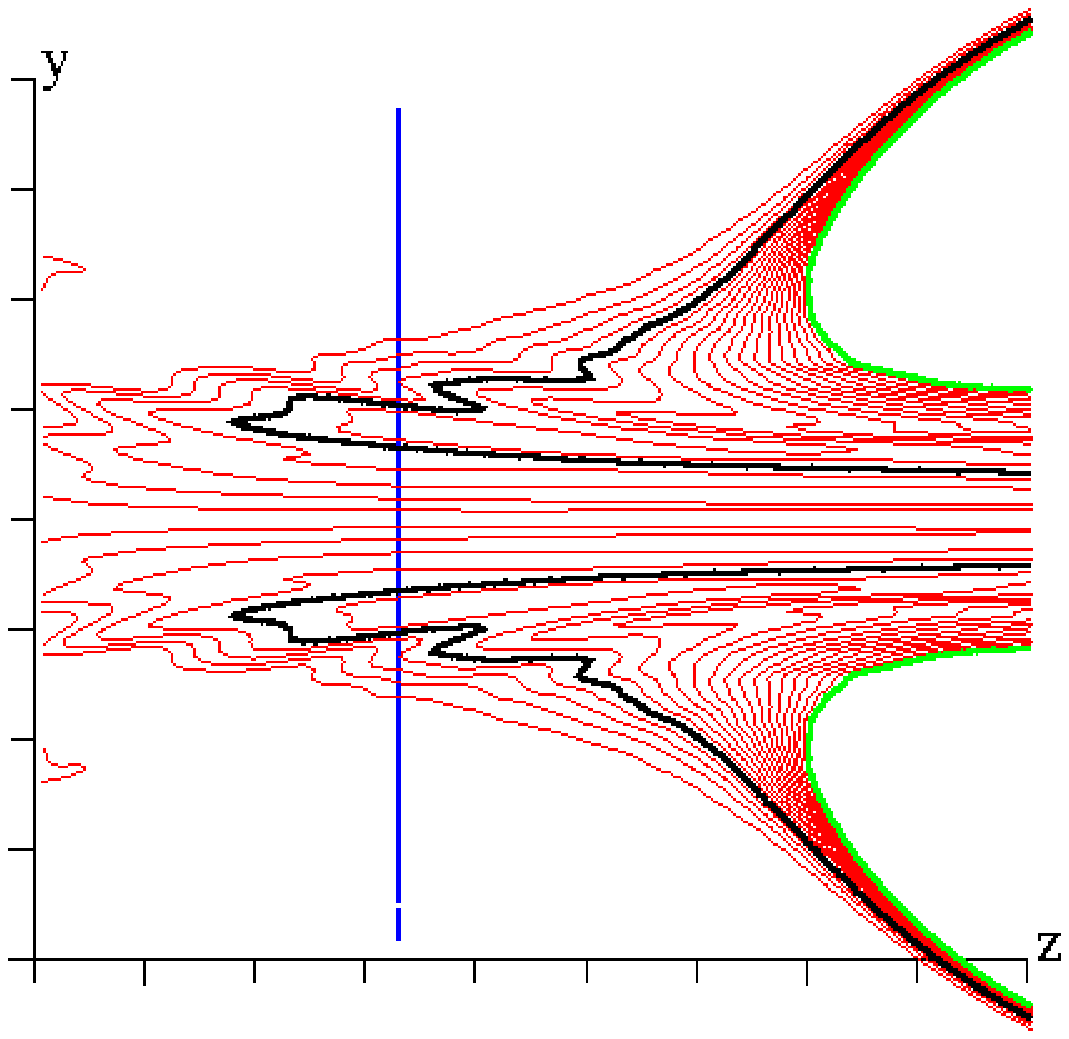}
\hskip 0.2cm
\includegraphics[width=2.3cm,clip,angle=90]{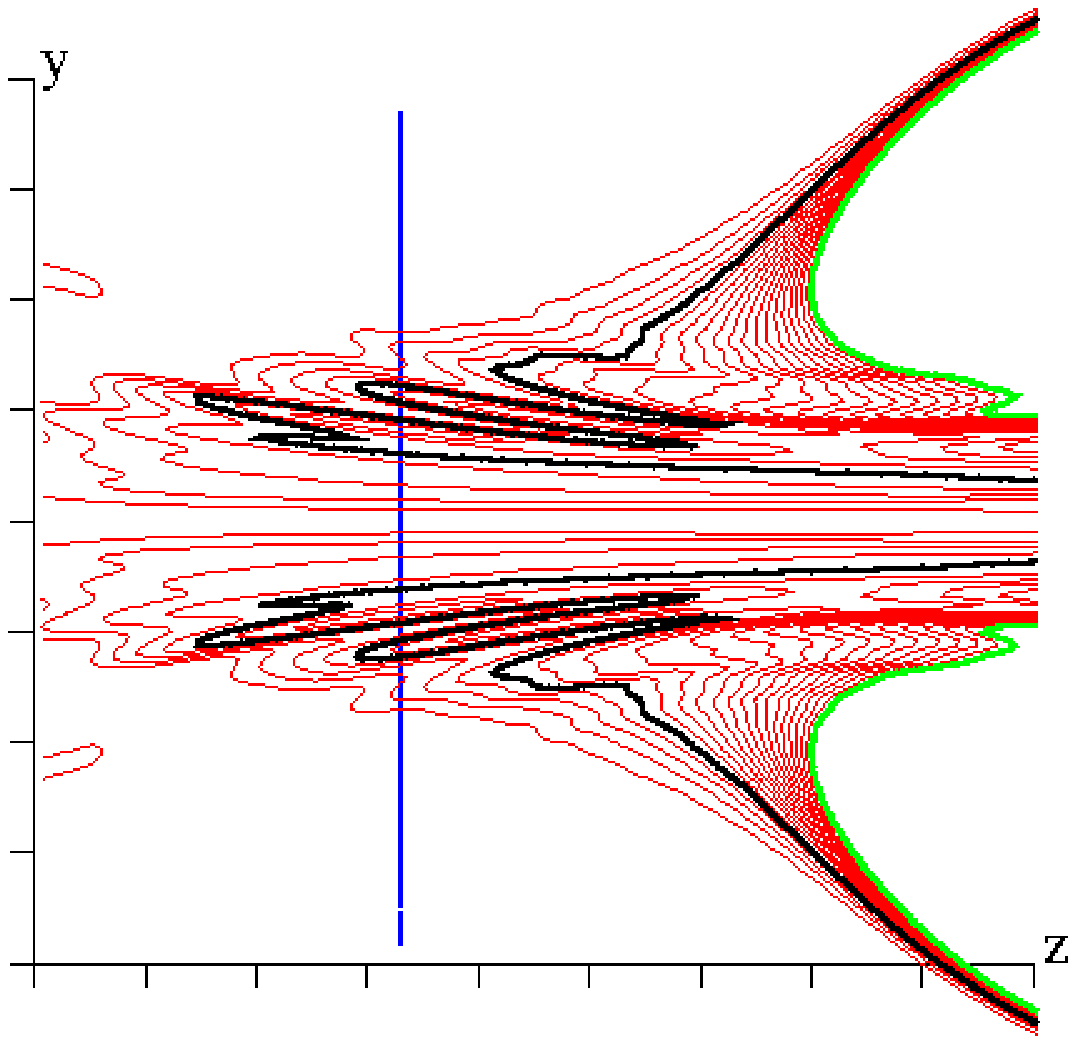}
\hskip 0.2cm
\includegraphics[width=2.3cm,clip,angle=90]{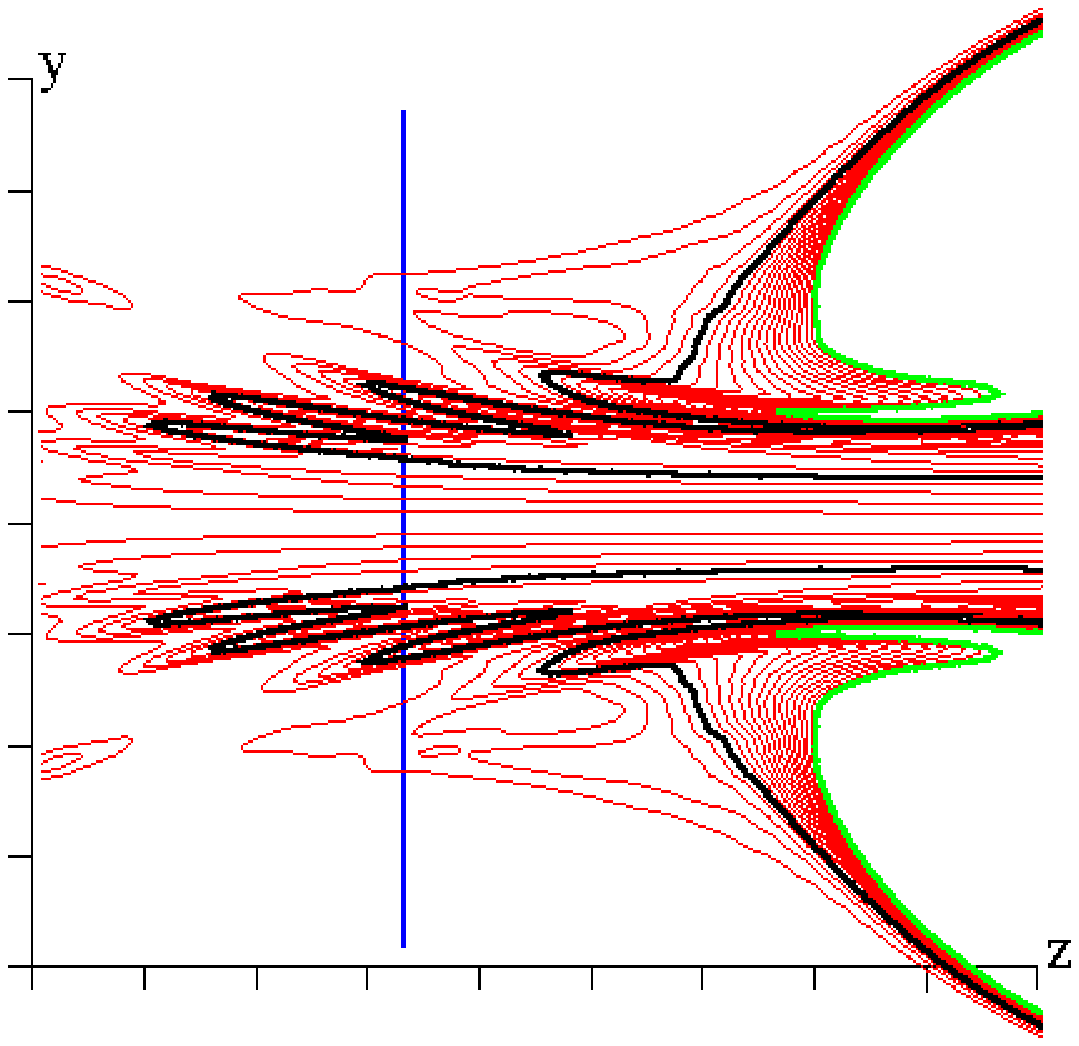}
\vskip 0.25cm
\hskip 0.7cm  a)   \hskip 2.3cm  b) \hskip 2.3cm  c) \hskip 2.3cm  d)
\hskip 2.3cm  e)
\vskip 0.1cm
\hskip -1.8cm
\includegraphics[width=2.3cm,clip,angle=90]{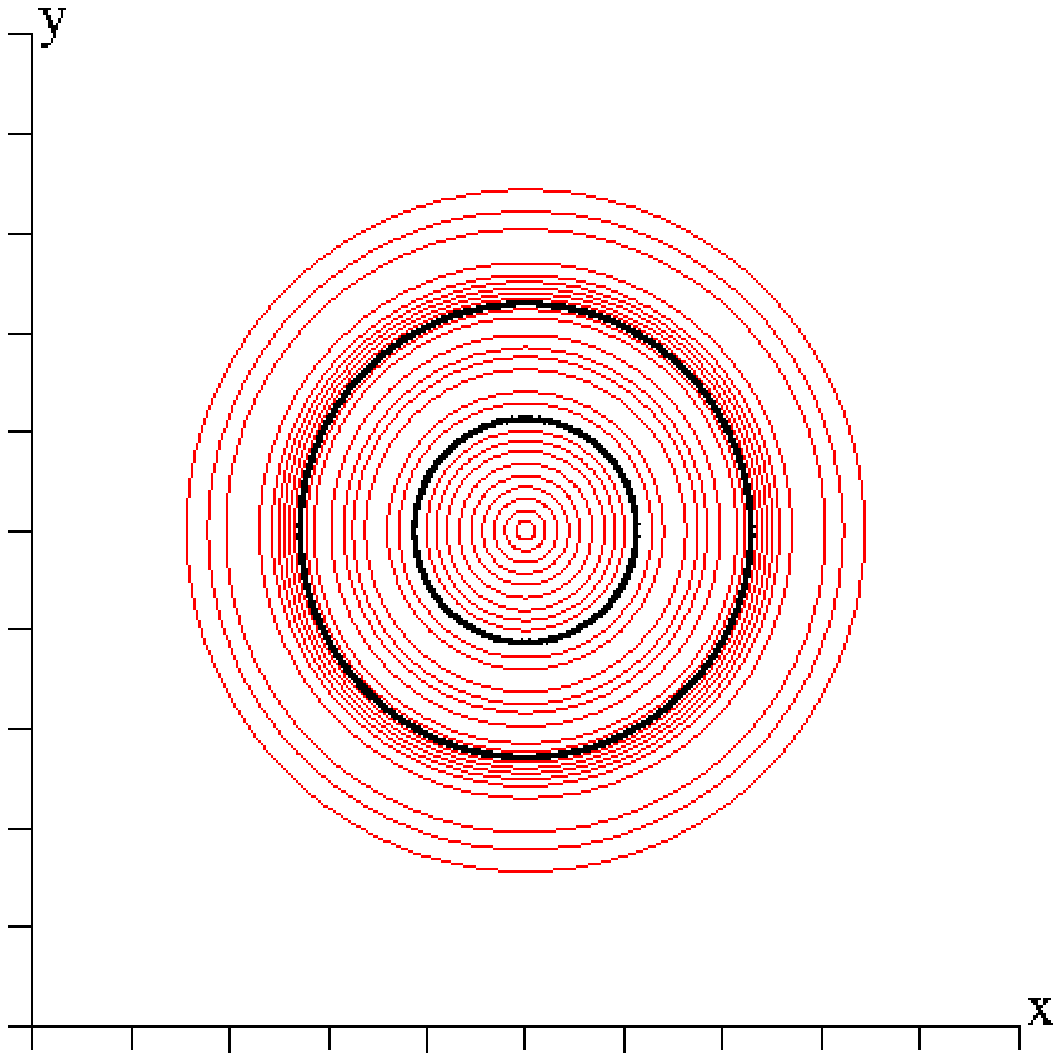}
\hskip 0.2cm
\includegraphics[width=2.3cm,clip,angle=90]{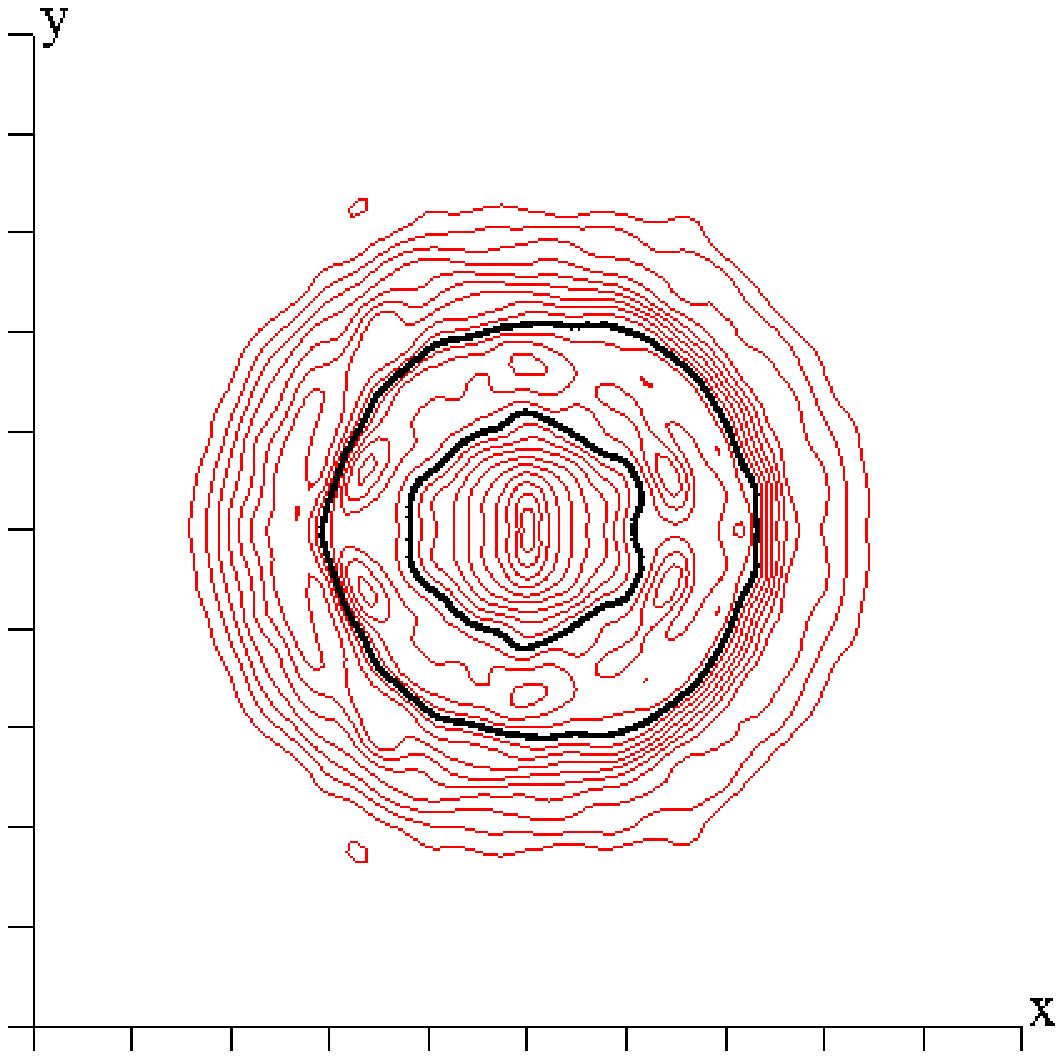}
\hskip 0.2cm
\includegraphics[width=2.3cm,clip,angle=90]{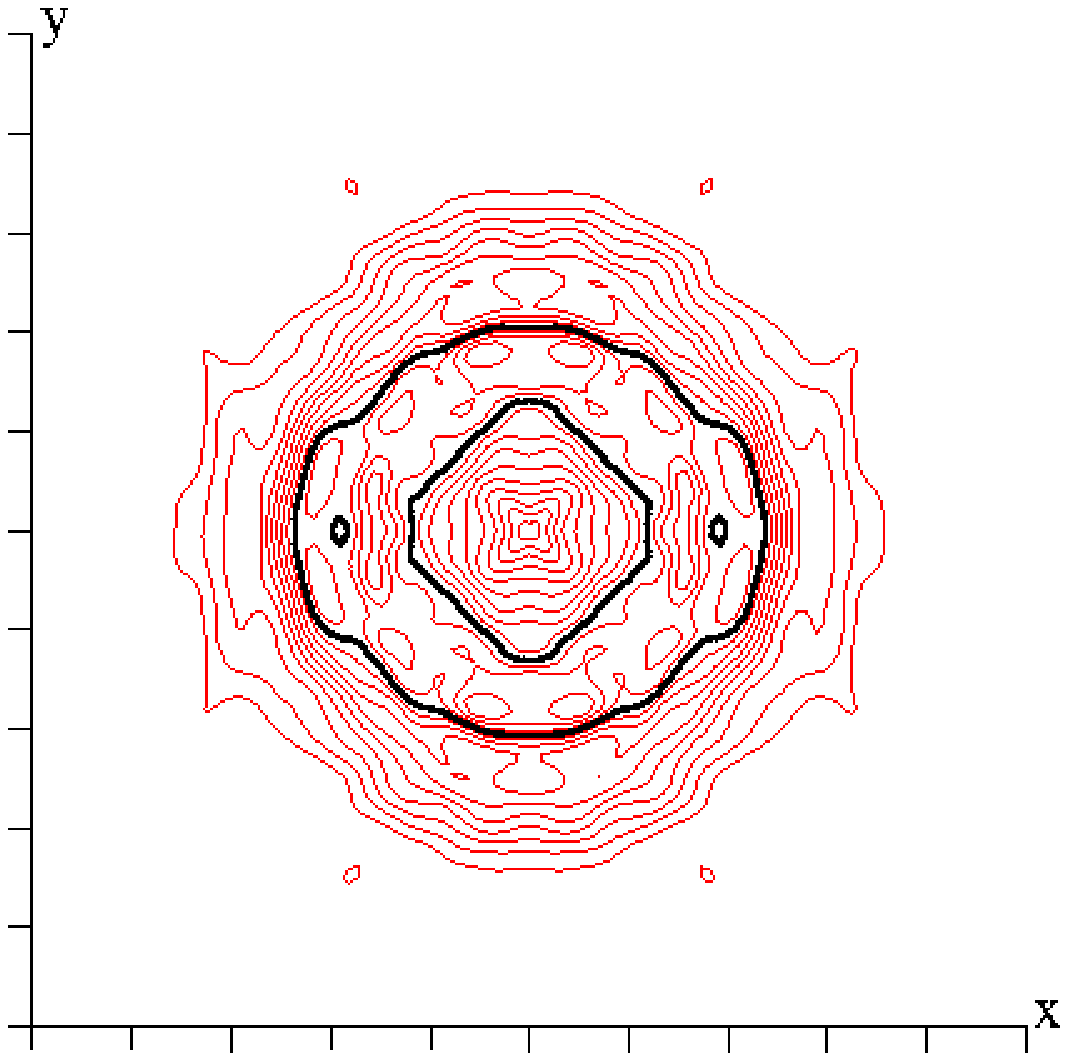}
\hskip 0.2cm
\includegraphics[width=2.3cm,clip,angle=90]{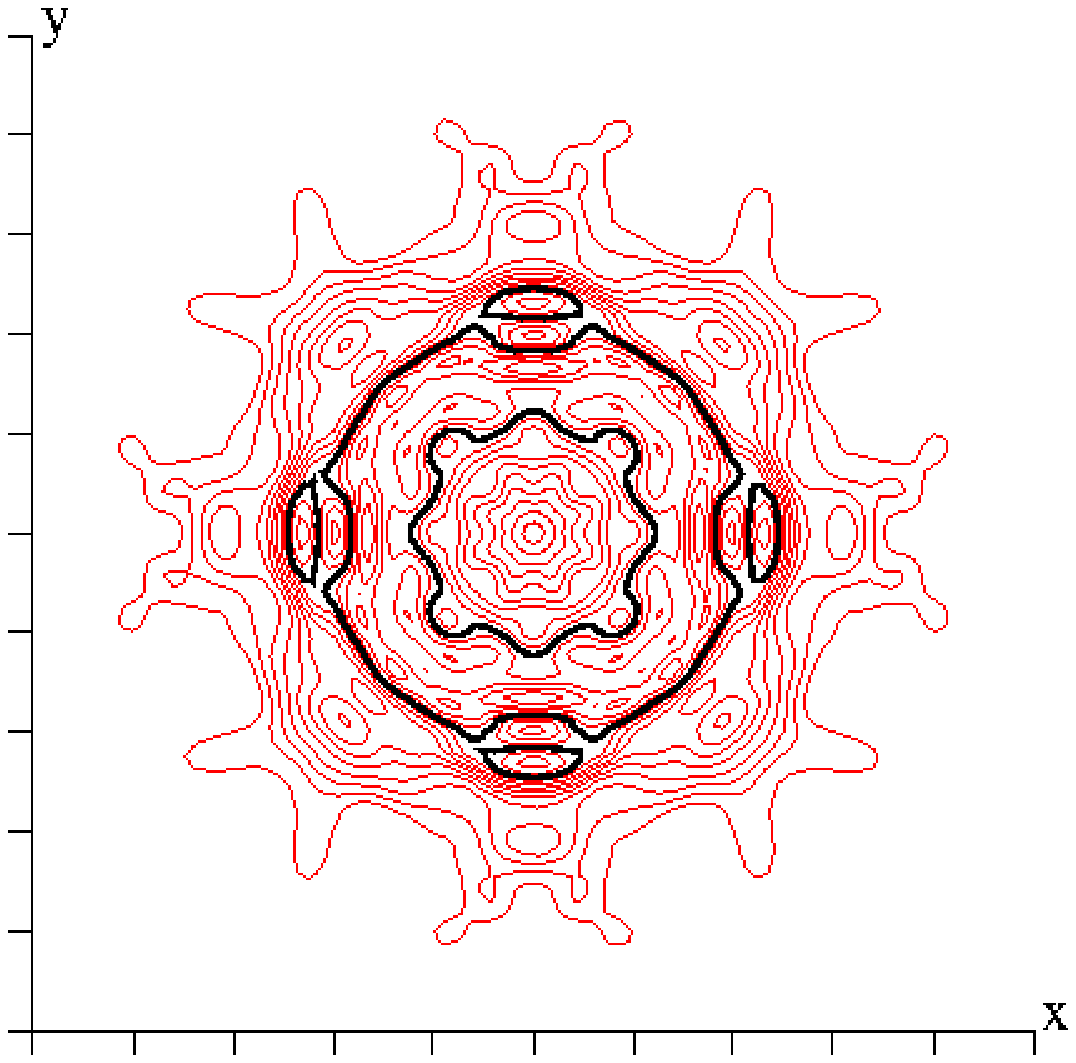}
\hskip 0.2cm
\includegraphics[width=2.3cm,clip,angle=90]{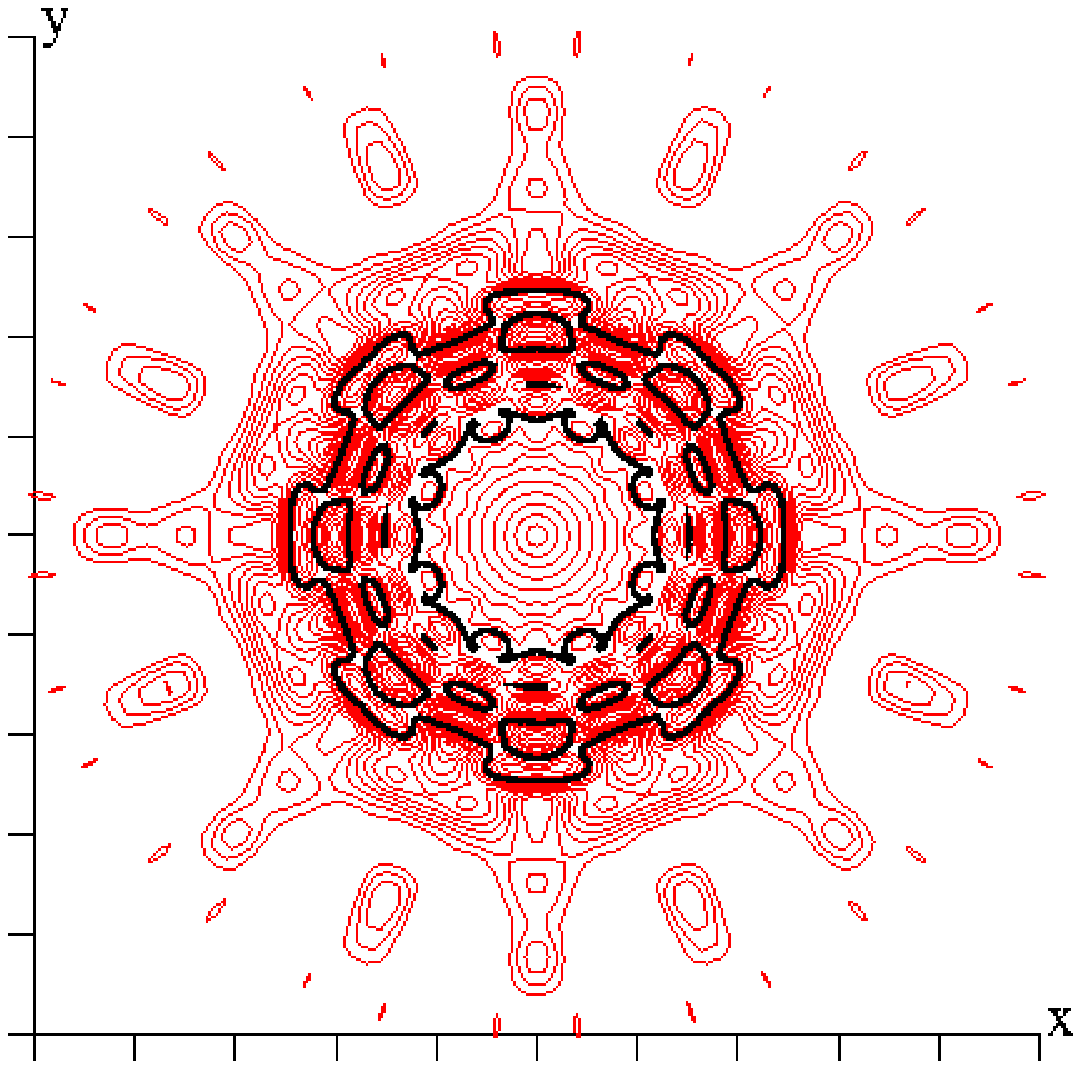}
\vskip 0.25cm
\hskip 0.7cm  f)   \hskip 2.3cm  g) \hskip 2.3cm  h) \hskip 2.3cm  i)
\hskip 2.3cm  j)
\caption{ Contour of $\omega_\theta$ in   $z-y$ with $-2.25<z-z_c<-0.75$
and $-0.75<y<0.75$ in a), b), c), d), e),
and $x_y$ with $-0.5 < x <0.5$ and $-0.5 < y <0.5$ in
f), g), h), i), j), planes at $t=20$ to enlight shape of the
spike red lines with $\Delta=0.005$, thick black $\omega_\theta=0.05$,
thick green $\omega_\theta=0.2$, with $n_p=16$ and;
a,f) $n_a=0$, b,g) $n_a=1$, c,h) $n_a=2$, d,i) $n_a=4$ and e,j) $n_a=8$.
}
\label{fig4}
\end{figure}

The linear and non-linear instability analysis demonstrated
that spikes of different shape form at the rear stagnation point, 
but can not describe the distribution of the vorticity field
at their interior and also, as before mentioned, inside
the vortex. The DNS does not have these limitations and
allows to understand better, during the evolution,
the transfer of the effects of the disturbances from the
surface to any region of the Hill vortex. Since a lot
of emphasis in the previous studies was devoted to
the spike formation, in figure \ref{fig4} a zoom of the rear
region of the vortex is presented at $t=20$.  
This is the time at which in figure \ref{fig3} 
the red lines $<\omega_\theta>=0.05$ are plotted. 
The complex shape of the true spike can not be described 
by contours of $<\omega_\theta>$, but should be reproduced
by distributions of $\omega_\theta$  in  $y-z$ planes crossing the axis
or in $x-y$ sections of interest.
To distinguish the spike from the vortex in figure \ref{fig4}
the solid green line corresponding to $\omega_\theta=0.2$ has been plotted.
In all cases, with the exception of the $n_a=8$ disturbance (figure \ref{fig4}e)
the green lines indicate that the interior of vortex is
not largely affected by the disturbances. 
If there is an effect, figure \ref{fig4}e shows that
it is localised near the axis and not in the external region. 
The complexity of the spike can be,
more deeply, investigated by contour plots of $\omega_\theta$
in $x-y$ planes corresponding to the blue lines in
figure \ref{fig4}a-e. The images in figure \ref{fig4}f-j depict
the formation of a large variety of structure directly
linked to the wave number $n_a$. In the axisymmetric simulation
with $n_a=0$ circular contours appear in figure \ref{fig4}f, 
that, together with figure \ref{fig4}a enlighten     
the complex shape of the spike due to the persistence,
in this region, of the polar disturbances. 

A substantial smoothing of the contour $\omega_\theta=0.05$ 
appears in figure \ref{fig4}b, that  is corroborated 
by the two quite circular black lines in figure \ref{fig4}g.
However, both figures explain that the increased smoothness
is also due to the choice to locate the $y-z$ plane at $\theta=\pi/2$
and $\theta=3\pi/2$.
Indeed, structures are visible in figure \ref{fig4}g,    
but these are weak with respect to those forming
for $n_a>1$. At $n_a=2$ four structures form at the center of the 
spike, that deform the inner $\omega_\theta=0.05$ contour
pushing it towards the second one. The intensification
of the azimuthal vorticity in certain regions and the
decrease in others can not be ascribed to the viscous
terms in the $\omega_\theta$ transport equation. Therefore
it should be due to the balance between advection and
vortex tilting and stretching. This budget is presented for
the $n_a=4$ later on, in the spike it has been observed
that the convective transport prevails on the other
non-linear terms.
At $n_a=4$ the structures are more 
complex and figure \ref{fig4}i depicts the 
oscillations of the black line in figure \ref{fig4}d.
The tendency to have in the central region  
a more circular spike in the azimuthal direction,
by increasing $n_a$ can be drawn by figure \ref{fig4}j
for $n_a=8$. On the other hand this figure shows
also that large oscillations appear at the edge of the
spike. These are attached at the vortex in a complex
way  as figure \ref{fig4}e demonstrates.
From these images it has been decided that the $n_a=4$
could be the wave number  more appropriate  
to investigate the response of the Hill vortex
to disturbances of different amplitude, 
described in the next section.

\begin{figure}
\centering
\vskip -0.0cm
\hskip -1.8cm
\includegraphics[width=5.0cm,clip,angle=90]{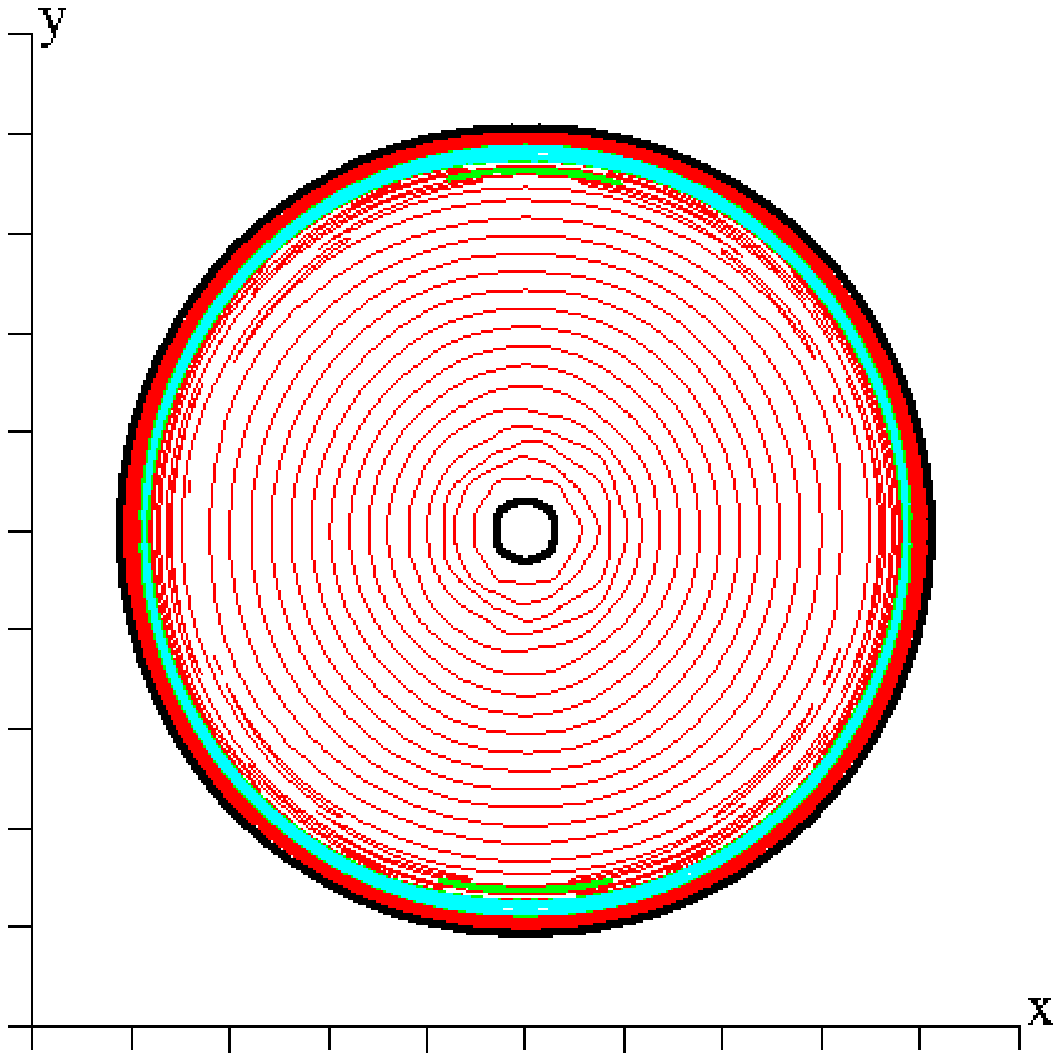}
\hskip 0.5cm
\includegraphics[width=5.0cm,clip,angle=90]{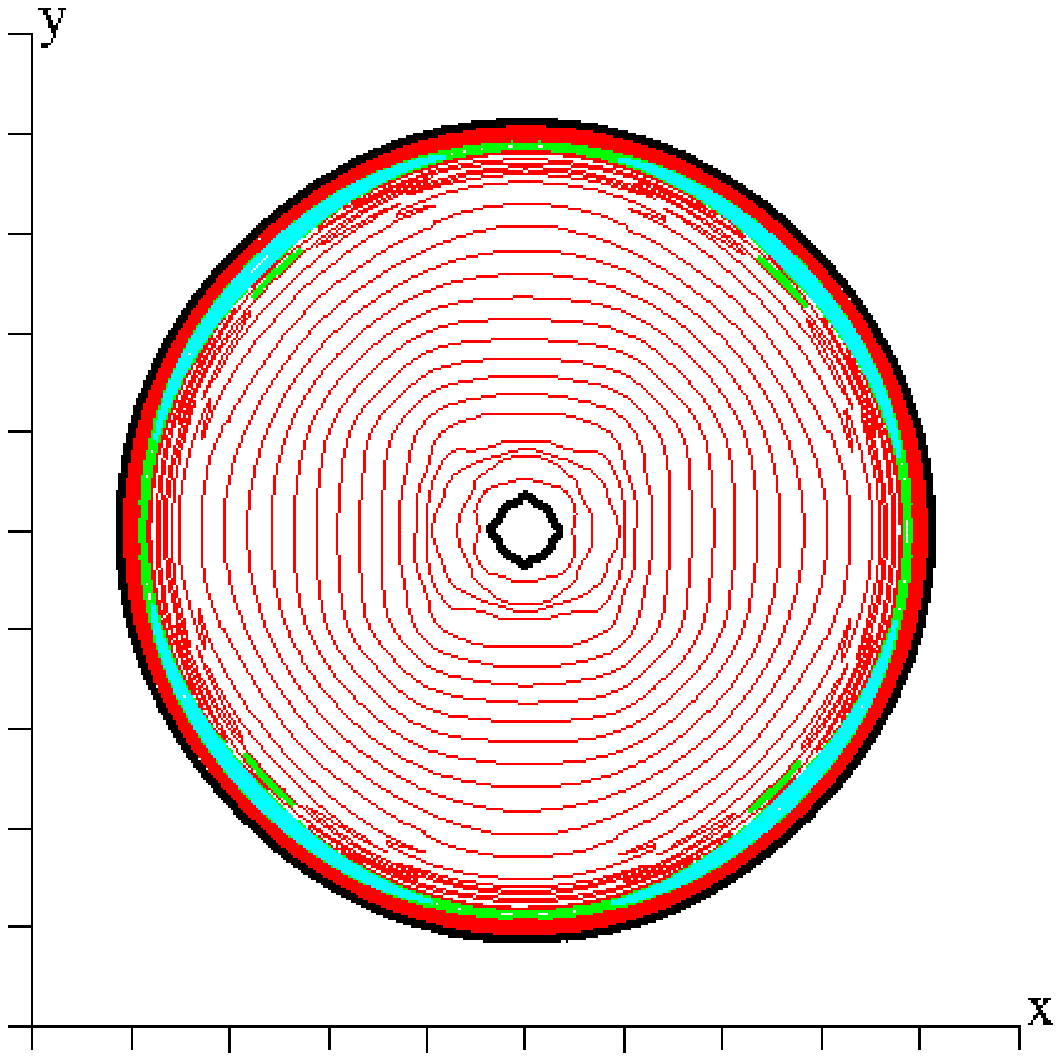}
\vskip 0.25cm
\hskip 3.5cm  a)   \hskip 5.5cm  b)
\vskip 0.1cm
\hskip -1.8cm
\includegraphics[width=5.0cm,clip,angle=90]{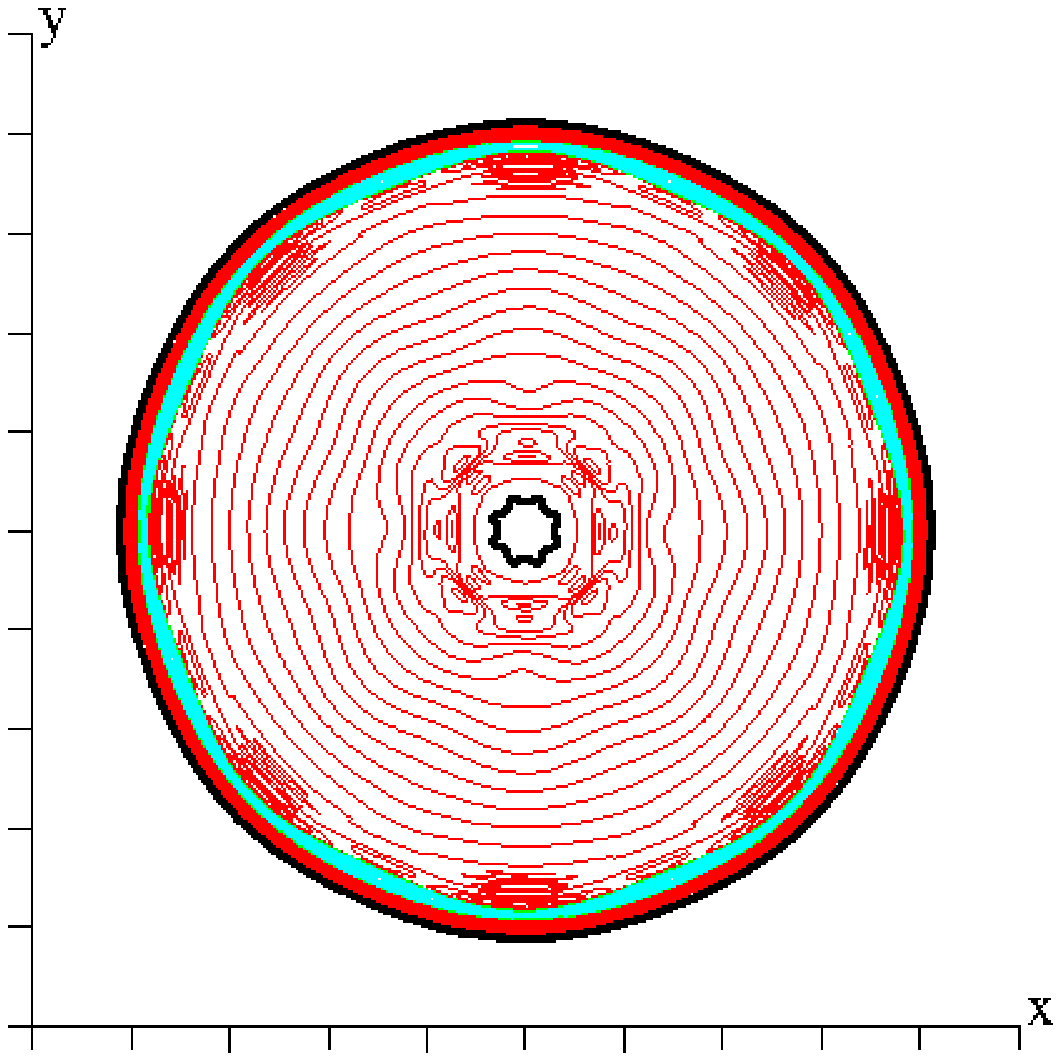}
\hskip 0.5cm
\includegraphics[width=5.0cm,clip,angle=90]{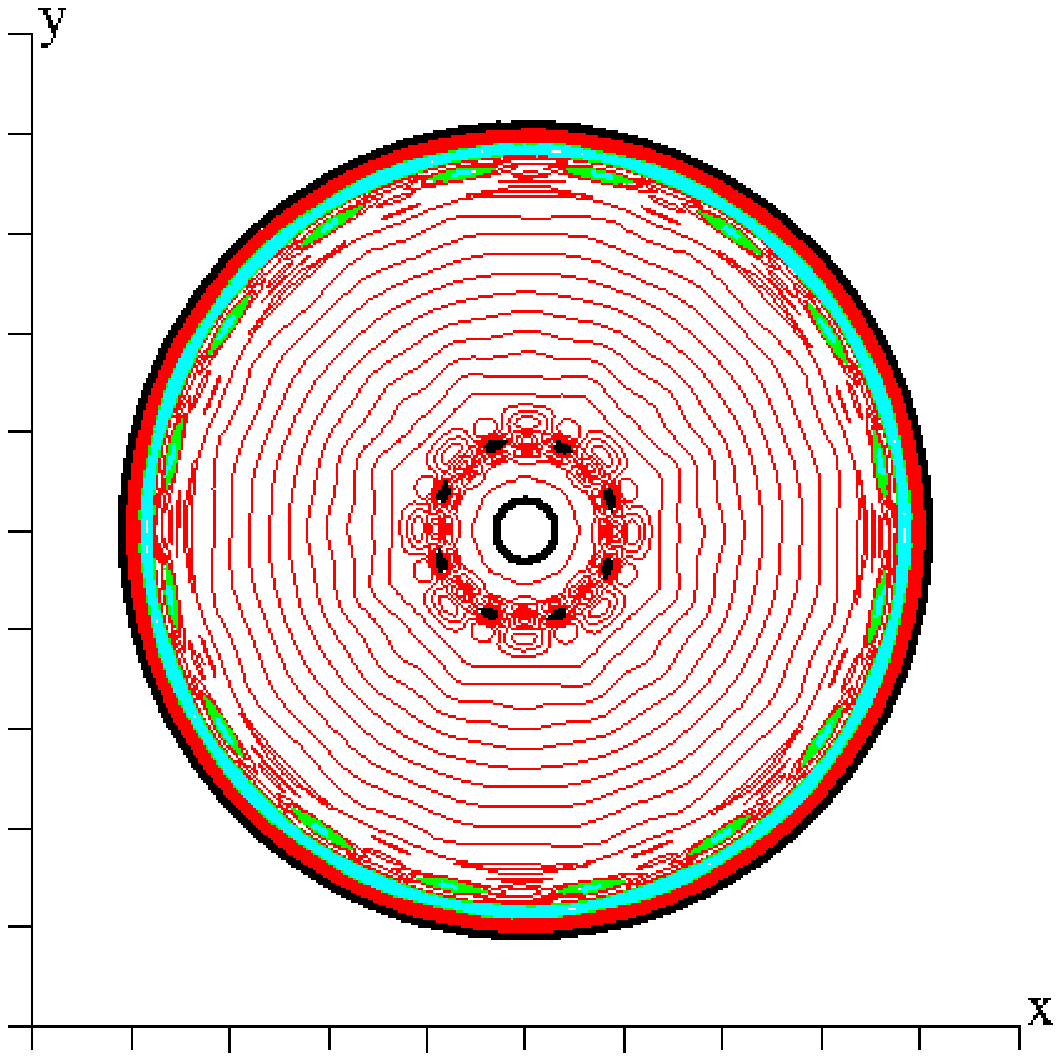}
\vskip 0.25cm
\hskip 3.5cm  c)   \hskip 5.5cm  d)
\caption{ Contour of $\omega_\theta$ in a $x-y$ plane and $x_y$ 
with $-1.2 < x <1.2$ and $-1.2 < y <1.2$ at $t=20$
at $z$ corresponding to the maximum of $<\omt>$ for $\delta_0=0.01$: 
a) $n_a=1$, b) $n_a=2$, c) $n_a=4$ and d) $n_a=8$,
red contours have an increment $\Delta=.05$ up to $\omega_\theta=0.95$,
the black thick lines is at $\omega_\theta=0.05$, the green at
$\omega_\theta=1.$ and the cyan have increments $\Delta=.05$
for $\omega_\theta>1.$.
}
\label{fig5}
\end{figure}

The complex physics in the interior of
the vortex may reveal  important issues during
the instability, that my lead to the destruction
of the Hill vortex.
From figure \ref{fig3} it may be inferred that the 
disturbance during the motion is convected in the rear,
and is not any more visible in a large part of the surface.
The disturbances, at later times, are visible
in the complex structures
of the spike above described, whose number is strictly connected
to the wave  number $n_a$. We wish to investigate,
through visualizations of $\omega_\theta$ in $x-y$ planes,
whether structures similar to those in the spike are
encountered in the central region of the vortex.
Figure \ref{fig5}a-d indeed show these structures
both near the surface and near the axis. As
well as in the spike the number is strictly linked
to the azimuthal wave number of the initial perturbation.
The common behavior for any wave number $n_a$ can be
drawn by looking at the time evolution of $\omega_r$ and
$\omega_z$ (not shown for sake of brevity). 
Up to $t=8$ $\omega_z$ grows near the external
surface, instead $\omega_r$ is concentrated in very thin regions 
near the axis. These regions grows in time while      
they  move far from the axis. At $t=10$ the size of the $\omega_r$
patches are almost equal to those depicted in 
the central region of figure \ref{fig5} and to those
of $\omega_z$ near the surface. At $t=20$, from the saved 
field,   any term in the transport equation of
$\omega_\theta$ have been evaluated. The equation is

\begin{equation}
\der{\omt}{t}=
-\underbrace{
(\frac{\ut}{r}\der{\omt}{\theta} 
+\ur\der{\omt}{r}-\frac{\ur\omt}{r}
+\uz\der{\omt}{z})
}_{C_\theta}
+
\underbrace{
\frac{\omt}{r}\der{\ut}{\theta} 
+\omr\der{\ut}{r}-\frac{\ut\omt}{r}
+\omz\der{\ut}{z}
}_{S_\theta}
+\underbrace{
\frac{1}{Re} \nabla^2 \omt 
}_{V_\theta}
\label{eqot}
\end{equation}

\begin{figure}
\centering
\vskip -0.0cm
\hskip -1.8cm
\includegraphics[width=4.5cm,clip,angle=90]{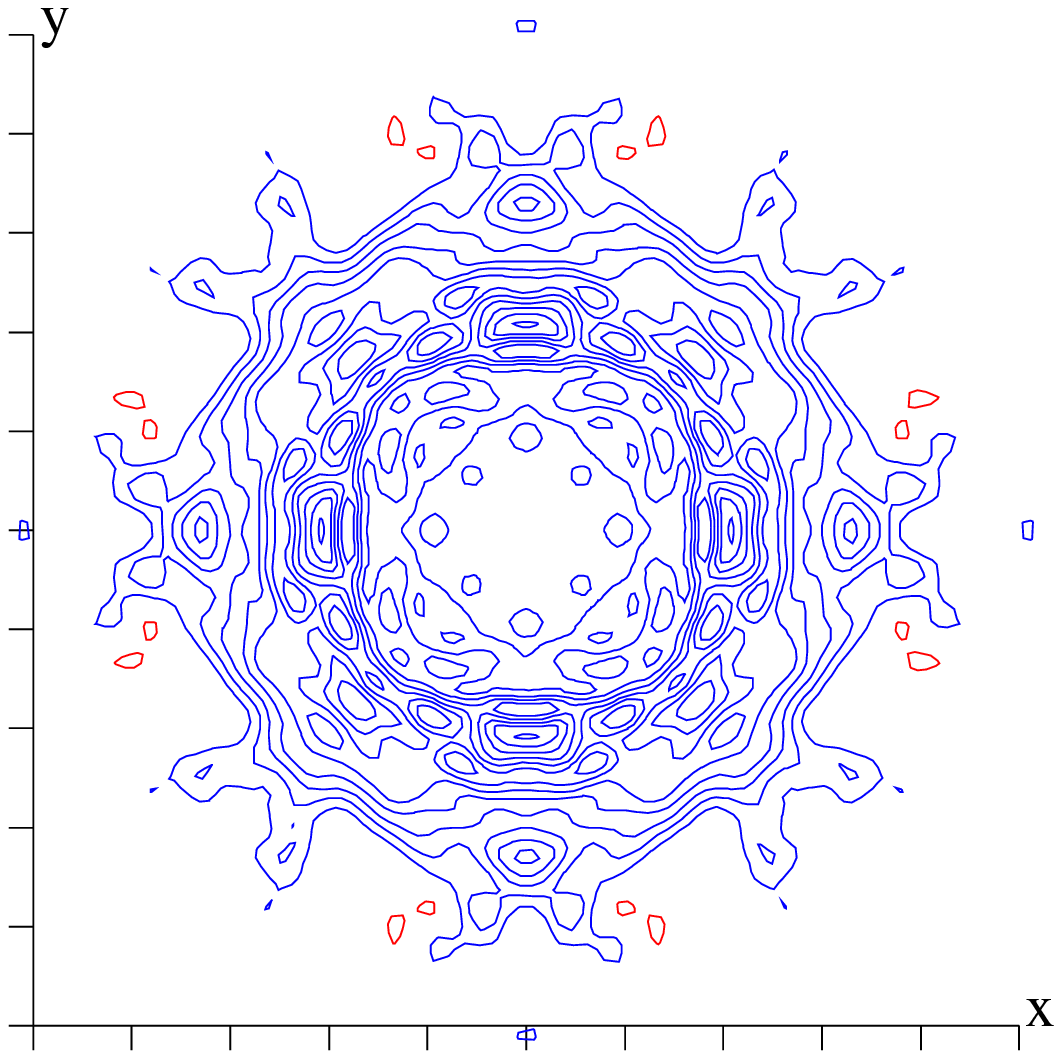}
\hskip 0.2cm
\includegraphics[width=4.5cm,clip,angle=90]{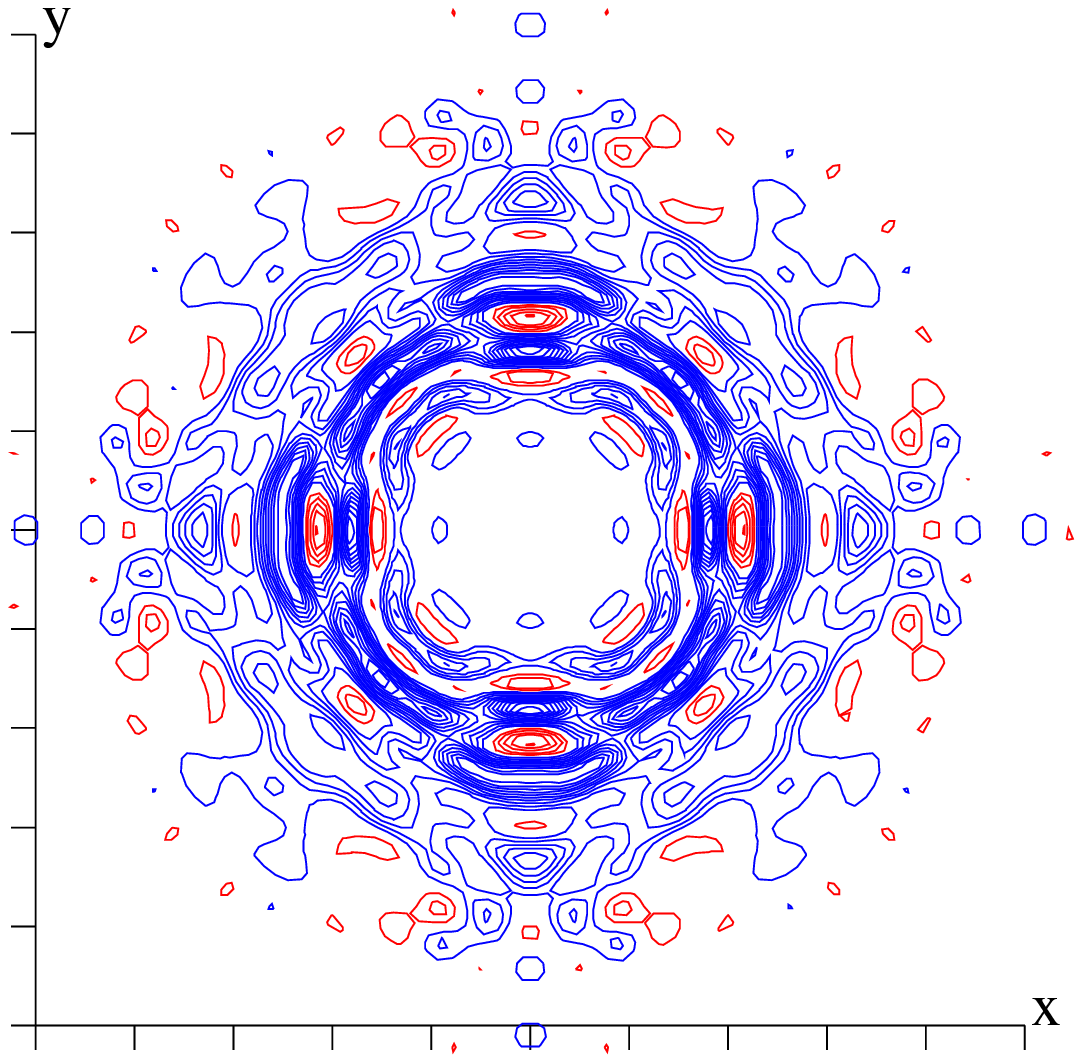}
\hskip 0.2cm
\includegraphics[width=4.5cm,clip,angle=90]{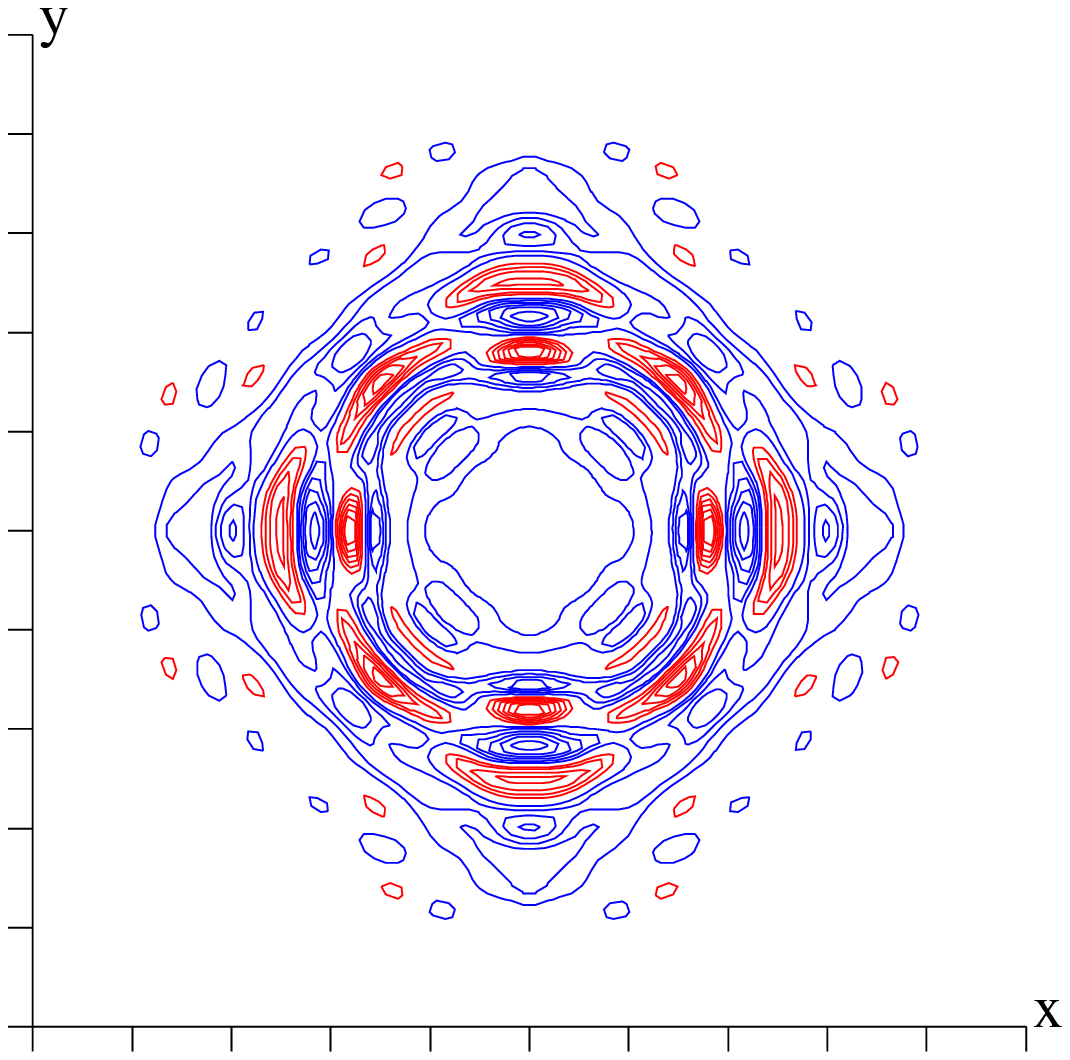}
\vskip 0.25cm
\hskip 1.8cm  a)   \hskip 3.9cm  b)  \hskip 3.9cm c)
\vskip 0.1cm
\hskip -1.8cm
\includegraphics[width=4.5cm,clip,angle=90]{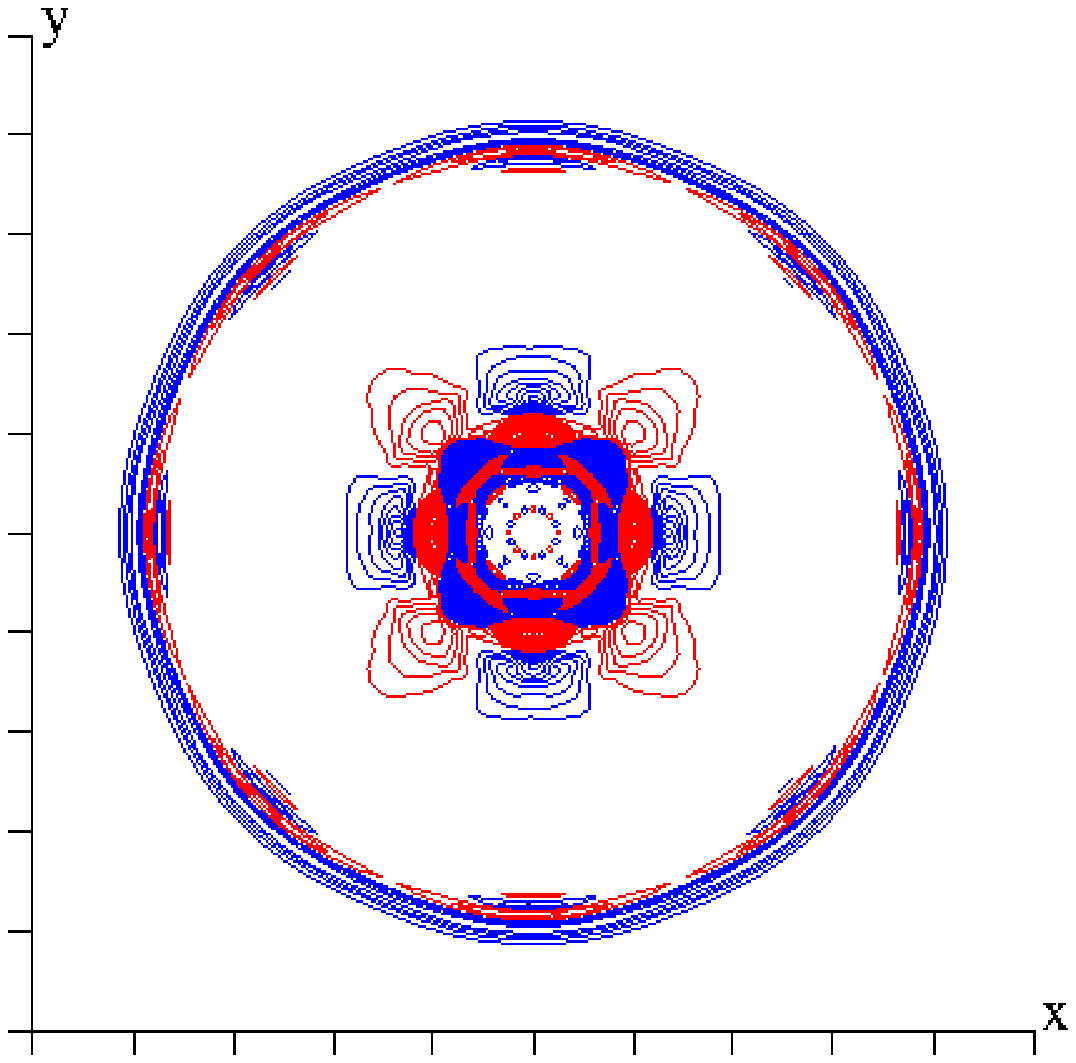}
\hskip 0.2cm
\includegraphics[width=4.5cm,clip,angle=90]{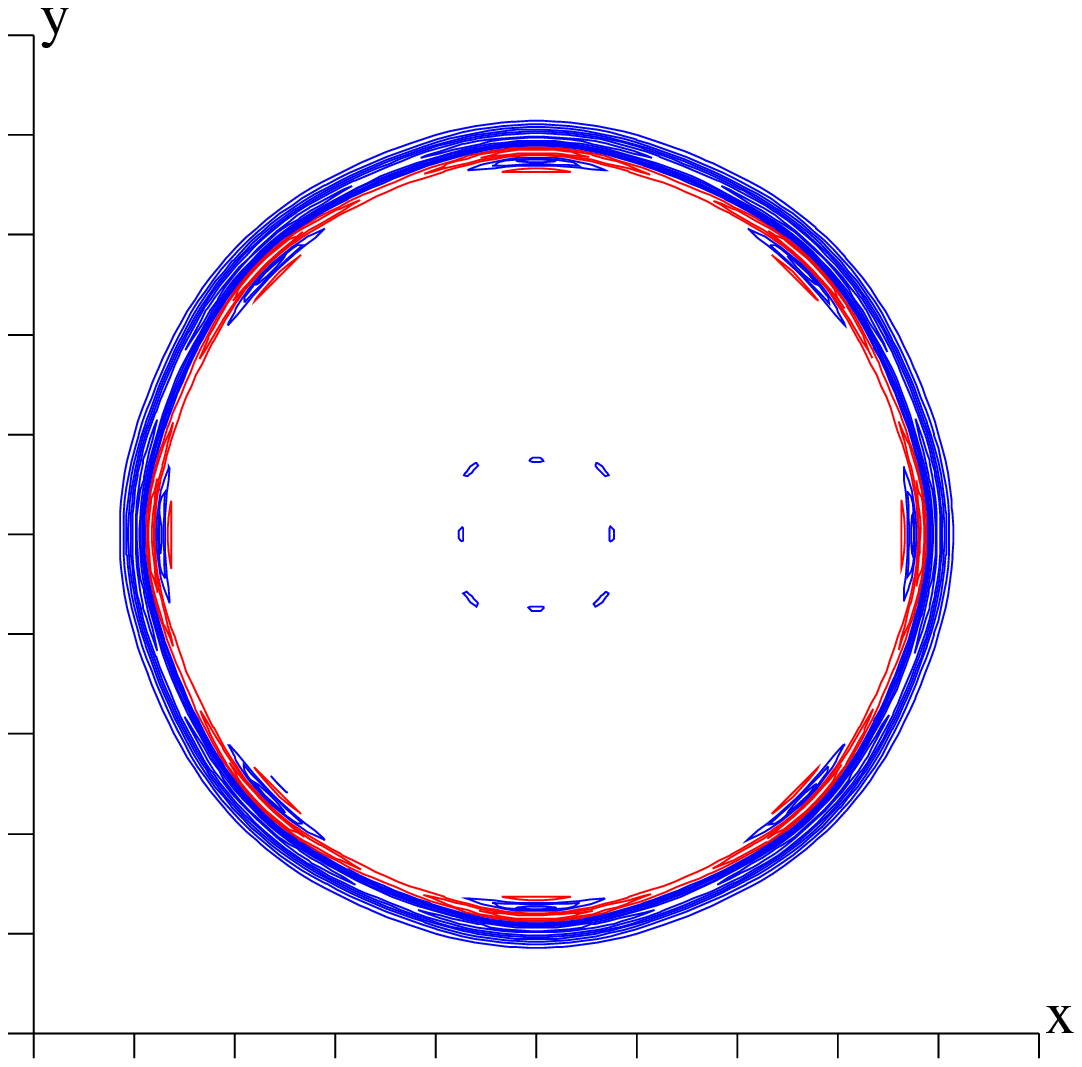}
\hskip 0.2cm
\includegraphics[width=4.5cm,clip,angle=90]{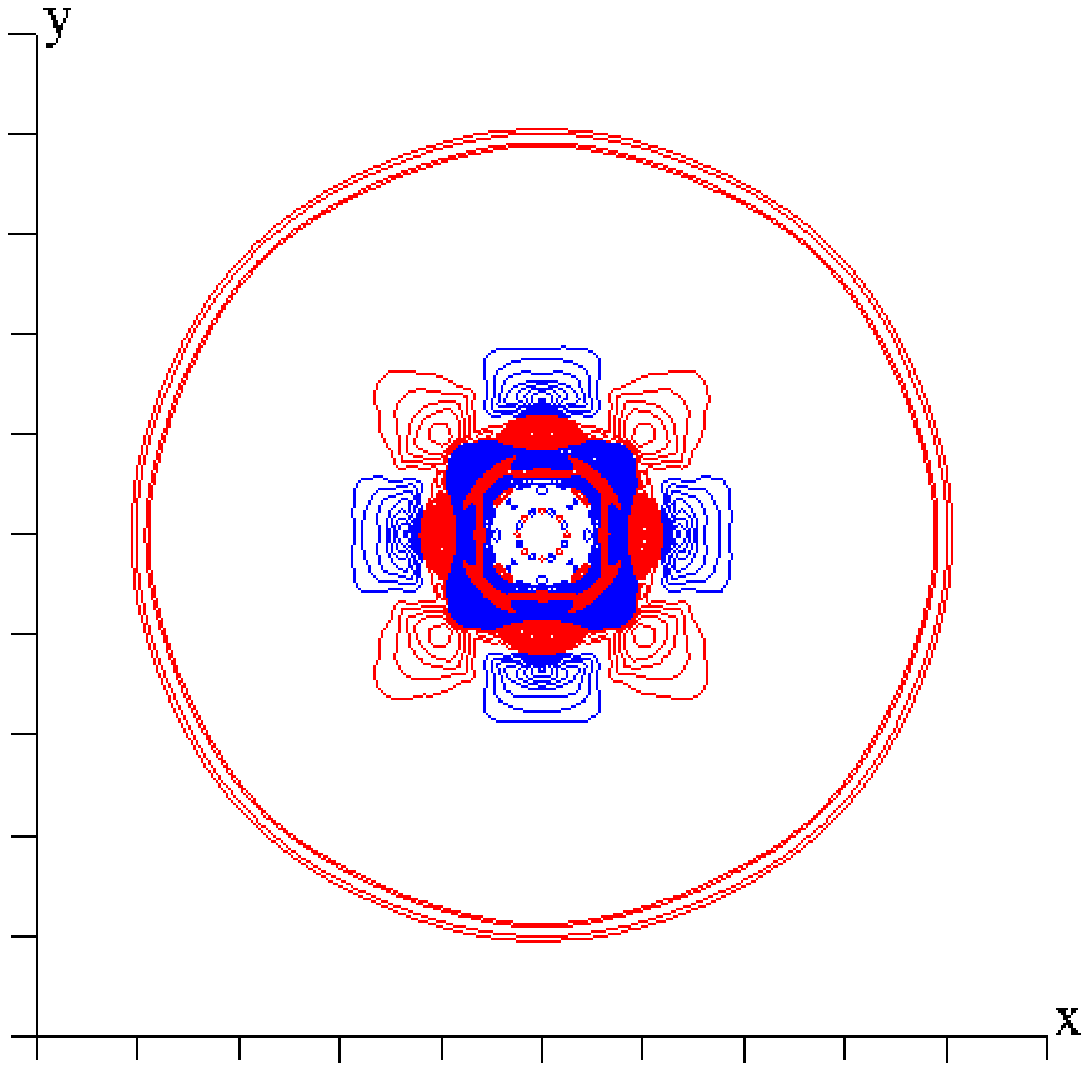}
\vskip 0.25cm
\hskip 1.8cm  d)   \hskip 3.9cm  e)  \hskip 3.9cm f)
\caption{For $n_a=4$ disturbance contour plots 
of the terms  in the  Eq.(\ref{eqot}): top
figures at the same $z$ location and $x,y$  limits
as in  figure \ref{fig4}
with increment $\Delta=.001$ , bottom at the same $z$ 
and $x,y$  limits as in  figure \ref{fig5}
with increment $\Delta=.01$ blue negative, red positive  
a) and d) $C_\theta$, b) and e) $\ur\der{\omt}{r}-\frac{\ur\omt}{r}$
c) and f)$\uz\der{\omt}{z}$.
}
\label{fig6}
\end{figure}

The large Reynolds number imply
a small effect of the viscous term, then 
the convective $C_\theta$ and the stretching and tilting
terms ($S_\theta$) should be relevant.  Since,
the total contribution of $S_\theta$ in each location
is smaller than that of $C_\theta$, the $S_\theta$ distribution
is not shown. Each term of $C_\theta$
contribute at a different rate depending on the region, and
in particular, $\frac{\ut}{r}\der{\omt}{\theta}$ 
is negligible everywhere, therefore this is not presented.
The entire $C_\theta$ contribution in figure \ref{fig6}a 
reproduces with a good
approximation the $\omt$ contours in figure \ref{fig4}a,
in any location is negative and this implies that
the strength of the spike is reduced by convection; it 
becomes longer as it was found by the present results in 
figure \ref{fig2}c. Looking at each contribution to $C_\theta$,
$\ur\der{\omt}{r}-\frac{\ur\omt}{r}$ in figure \ref{fig6}b
prevails on   $\uz\der{\omt}{z}$ in figure \ref{fig6}c.
The latter in certain points is positive, but the opposite action
of the former prevails resulting in $C_\theta<0$ everywhere.
At the center of the vortex the figure \ref{fig5} shows the
accumulation and depletion of $\omt$ near the axis and near the
external surface. These correspond to the red and blue regions of 
$C_\theta$ in figure \ref{fig6}d. Figure \ref{fig6}e shows that
$\ur\der{\omt}{r}-\frac{\ur\omt}{r}$ is mainly concentrated near
the external surface, instead  $\uz\der{\omt}{z}$ acts 
near the axis, as is seen in figure \ref{fig6}f.
\begin{figure}
\centering
\vskip -0.0cm
\hskip -1.8cm
\psfrag{ylab} {$\Delta \omt  $}
\psfrag{xlab}{ $ x   $}
\includegraphics[width=6.5cm,clip,angle=0]{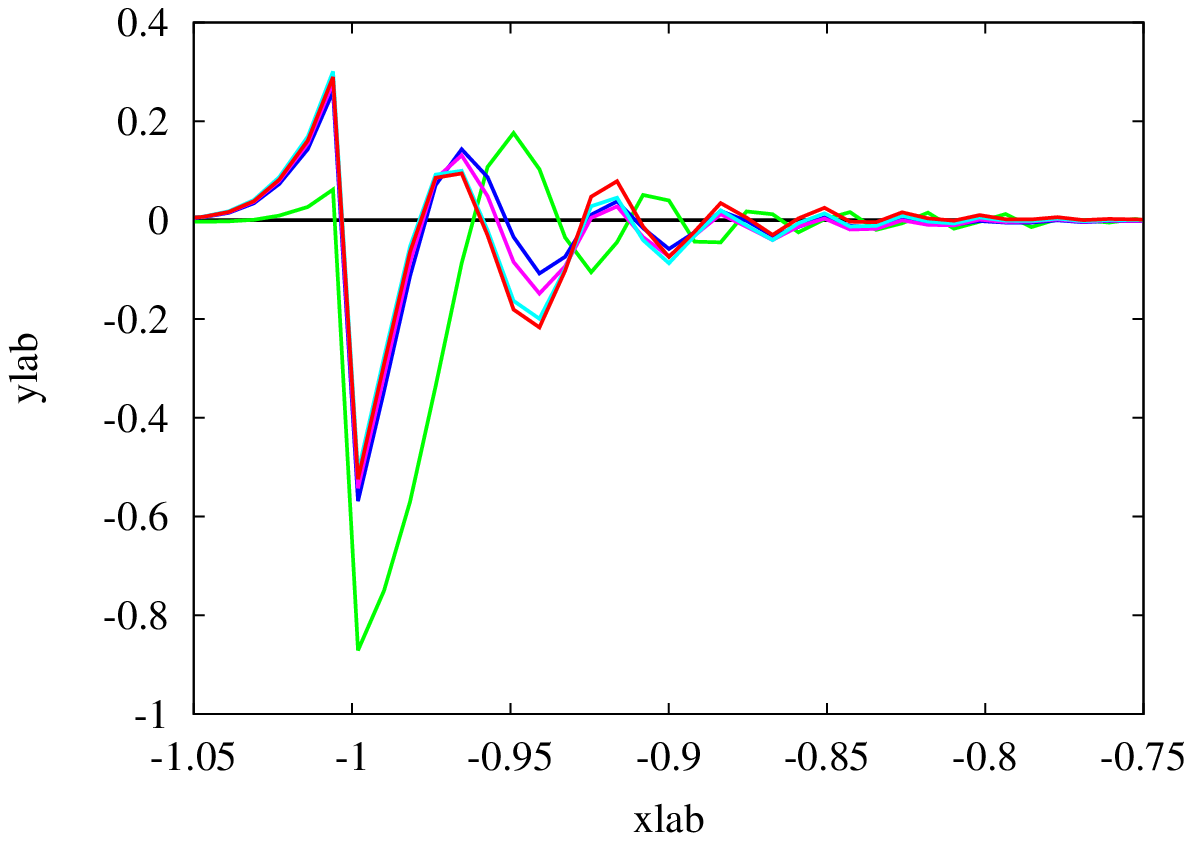}
\hskip 0.25cm
\psfrag{ylab} {$ $}
\psfrag{xlab}{ $ x   $}
\includegraphics[width=6.5cm,clip,angle=0]{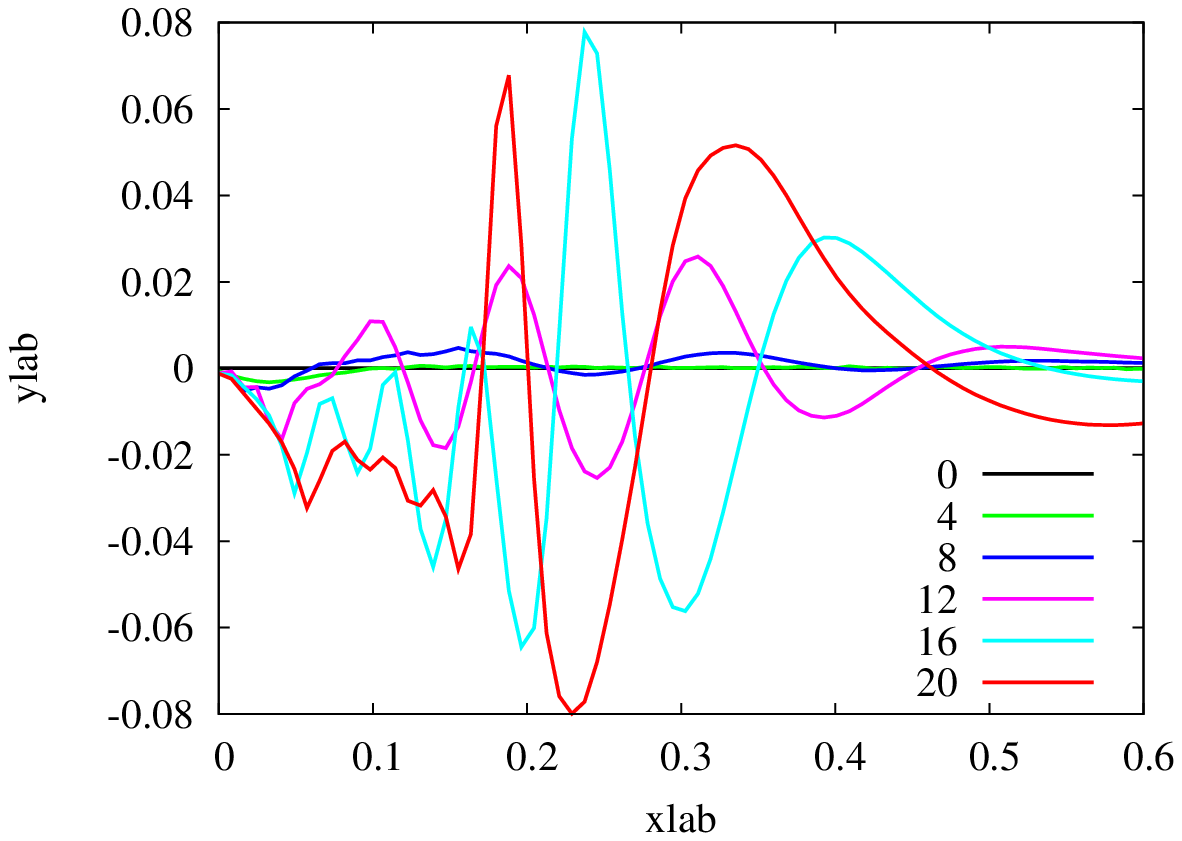}
\vskip -0.05cm
\hskip 3.3cm  a)   \hskip 6.3cm  b) 
\caption{ Profiles of $\Delta \omt$ along   $x$ at $y=0$ 
every $4$ time units indicated in the legend, in a) the layer
in the left, in b) the layer in the center of figure \ref{fig5}.
}
\label{fig7}
\end{figure}
\noindent From this figure it can be inferred that inside the vortex
the $C_\theta$ generates waves transferring disturbances from the interior
towards the surface. In order to demonstrate that these
are true travelling waves the red and blue regions should 
be concentrate at a certain time in a small region, 
and later on should move from this to other regions. 
Instead if these remain fix at a certain location these
could be related to insufficient resolution. Indeed it was mentioned
that at $Re=60000$ small oscillation remains
near the edge of the vortex due to the impossibility to
describe the sharp interface. Instead of presenting at each time
the distribution of $\omt$ in a $x-y$ plane a simple and
more efficient vision can be given by the
profiles, at $\theta=0$, of $\Delta \omt=\omt(x,0,t)-\omt(x,0,0)$.  
Figure \ref{fig7}a
confirms that the numerical oscillations near the interface do not grow and
do not move far from the region where are produced.  On the
other hand, the same plot in the central region ($0<x<0.6$) shows that at
$t=4$ (green) $\omt$ did not change much and that at $t=8$ (blue)
a small undulation starts to grow. At $t=12$ (magenta) the oscillation
of reasonable amplitude is depicted and the prof that it is a true wave
is given by the peaks shifted on the right with respect to those at
$t=8$. The other two profiles at $t>12$ emphasise that the disturbance
grow at the center and moves towards the external surface of
the vortex. The amplitude of the disturbances are rather
small thus it can be concluded that for small disturbances applied
on the surface  and, after a short translation, the
vortex has  a perturbed shape not too different from that at $t=0$.

\subsection{Effect of the amplitude of the perturbations}

In the previous section it has been observed that the azimuthal disturbances
are triggering the basic state of the Hill vortex producing in certain
regions an increase and in other a decrease
of $\omega_\theta$. This should be similar to what take place
in vortex rings. In this vortex the azimuthal disturbances grow in time,
the circular shape is not preserved, and at a certain time the complete
destruction of the vortex ring occur. This evolution is clearly described by
the pictures  at Page 67 of \cite{vandyke_82}. The difference
between a ring and the Hill vortex is that the former
is not and the latter is a form-preserving solution of the Euler equations.
A perfect Hill vortex can not be created in a laboratory, therefore numerical
experiments are the only way to investigate whether the
Hill behaves as a toroidal ring. From the, above discussed,
results in presence of small disturbances it seems that the instability 
of Hill vortices give rise to a sort of wake but 
large part of the structure retain
its initial $<\omt>$. In two dimensions it has been observed that
the Lamb dipole (\cite{lamb32}), a form preserving
solution of the Euler
equations, subjected to strong disturbances leaves
behind a wake, proportional to the disturbance, and
a new structure with the same characteristic of
the original one forms. This has been investigated by \cite{cavazza92}
by assigning different kind of disturbances, and finally getting
a dipole characterised by a linear relationship between vorticity
and stream function. 
\begin{figure}
\centering
\vskip -0.0cm
\hskip -1.8cm
\includegraphics[width=4.0cm,clip,angle=0]{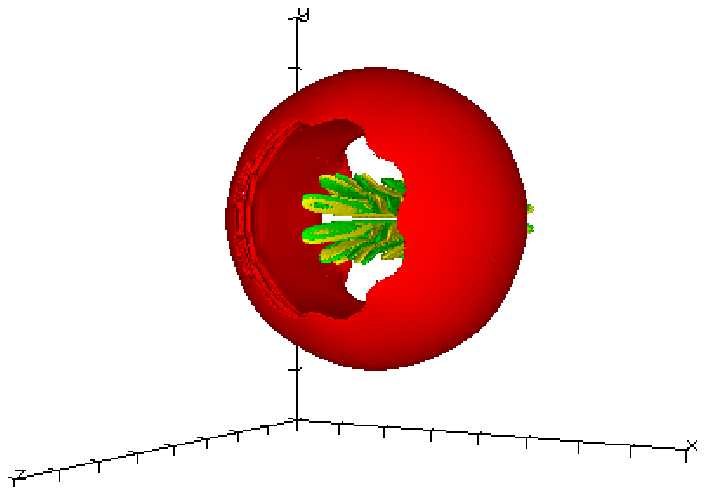}
\hskip 0.5cm
\includegraphics[width=4.0cm,clip,angle=0]{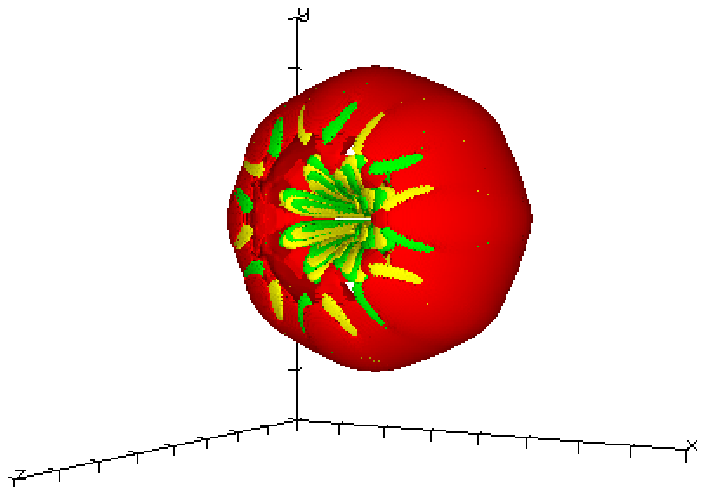}
\hskip 0.5cm
\includegraphics[width=4.0cm,clip,angle=0]{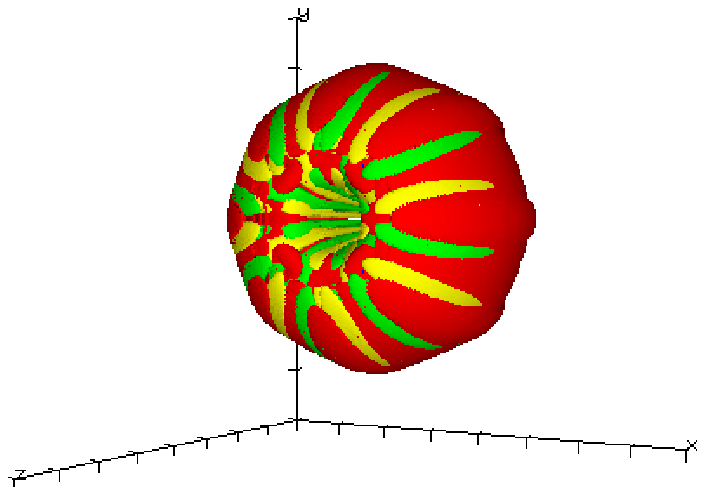}
\vskip 0.25cm
\hskip 1.7cm  a)   \hskip 3.9cm  b)   \hskip 3.9cm c)
\vskip 0.1cm
\hskip -1.8cm
\includegraphics[width=4.0cm,clip,angle=0]{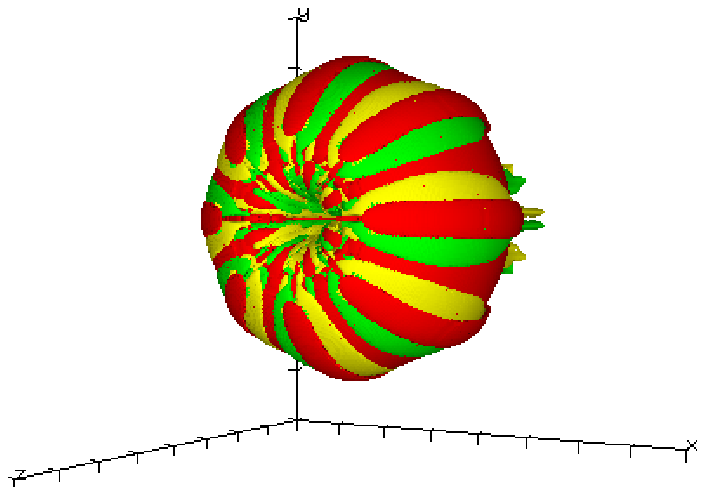}
\hskip 0.5cm
\includegraphics[width=4.0cm,clip,angle=0]{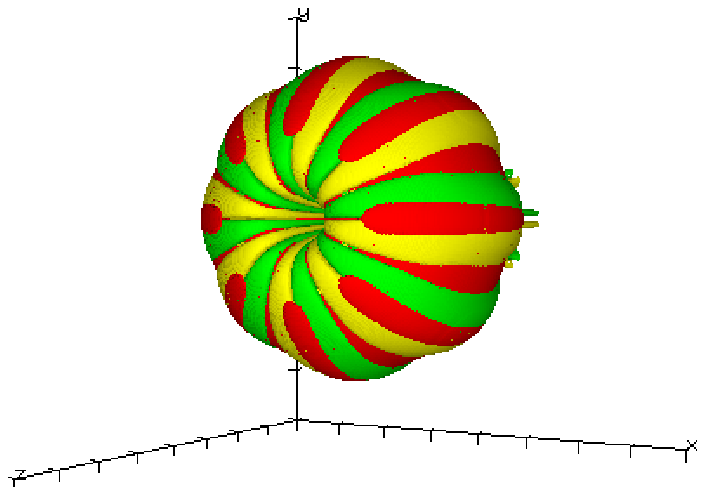}
\hskip 0.5cm
\includegraphics[width=4.0cm,clip,angle=0]{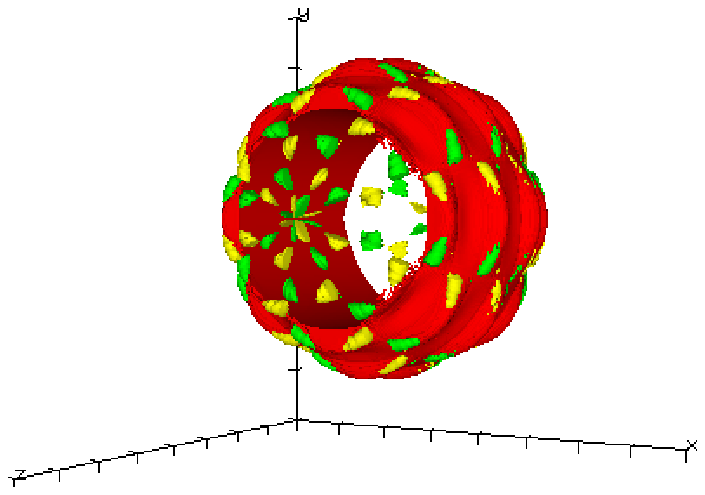}
\vskip 0.25cm
\hskip 1.7cm  d)   \hskip 3.9cm  e)  \hskip 3.9cm  f)
\caption{ Surface contour of $\omega_\theta=0.7$ (red) 
and $\omega_z=\pm 0.1$ positive yellow negative red at $t=20$
for:
a) $\delta_0=0.01$, b)$\delta_0=0.03$, c) $\delta_0=0.06$, d) $\delta_0=0.09$,
e) $\omega_\theta=0.7$ (red) and $\omega_r=\pm0.1$ at $t=20$,
f) $\omega_\theta=0.7$ (red) and $\omega_z=\pm0.1$ at $t=0$
}
\label{fig8}
\end{figure}
\noindent In 2D there is no vortex stretching in the vorticity
transport equations. In 3D the vortex stretching and tilting
terms are fundamental to go from the laminar through
an instability to the turbulent regime. For small disturbances
and a short evolution it has been found that the $S_\theta$
terms are smaller than $C_\theta$.  By increasing the amplitude of the 
disturbance the transition process may be faster, this explains why 
large disturbances on the surface of the
Hill vortex are applied and discussed in this section. 
At $t=0$ the contour $<\omega_\theta>=0.1$ for $\delta_0=0.09$ 
compared to the black
line  in figure \ref{fig2}b for $\delta_0=0.01$  emphasise that
a very corrugated surface has been generated. This plot
is not presented because the average in $\theta$ reduces the vision
of the complexity of the surface, that instead is appreciated
by a $\omega_\theta=0.7$ contour in figure \ref{fig10}f. It is important to recall
that this value is located at a distance from the centre
where the $\omt$ sharply goes to zero. 
To understand the large modifications of the  vorticity components 
while the vortex  evolves from $t=0$ up
to $t=20$ the surfaces of $\omega_z=\pm 0.1$
are superimposed to that of $\omt$ in figure \ref{fig8}f.
The effects of the polar and azimuthal 
disturbances are depicted by small patches
of $\omega_z$ in the regions of high curvature of $\omega_\theta$.
The visualizations at $t=20$ allow to understand the dependence
of the secondary vorticity on the amplitude of the disturbance,
namely are shown for $\delta_0=0.01$
in figure \ref{fig8}a, for $\delta_0=0.03$ in figure \ref{fig8}b,
for $\delta_0=0.06$ in figure \ref{fig8}c and
for $\delta_0=0.09$ in figure \ref{fig8}d.
At $t=20$ the polar disturbances are barely visible
on the red surface of $\omega_\theta=0.7$, 
these have been convected in the rear side forming
the long spike region with a corrugation more complex higher the amplitude of
the disturbance.
The azimuthal disturbances cause the formation of secondary 
vorticity components near the axis and near the surface of the vortex. 
Those near the axis propagate towards the surface, and as can
be inferred in figure \ref{fig8}, depending on their magnitude
are convected around the front stagnation point and remain close
to the external surface of the vortex. In this region 
the secondary vorticity contributes to the deformation 
of the surface of the vortex
with an amplitude of the azimuthal waviness greater than that
at $t=0$.  By comparing 
figure \ref{fig8}d and figure \ref{fig8}e it is rather difficult
to establish which component of the two secondary vorticity
components prevail to deform $\omega_\theta$.

\begin{figure}
\centering
\vskip -0.0cm
\hskip -1.8cm
\includegraphics[width=4.0cm,clip,angle=90]{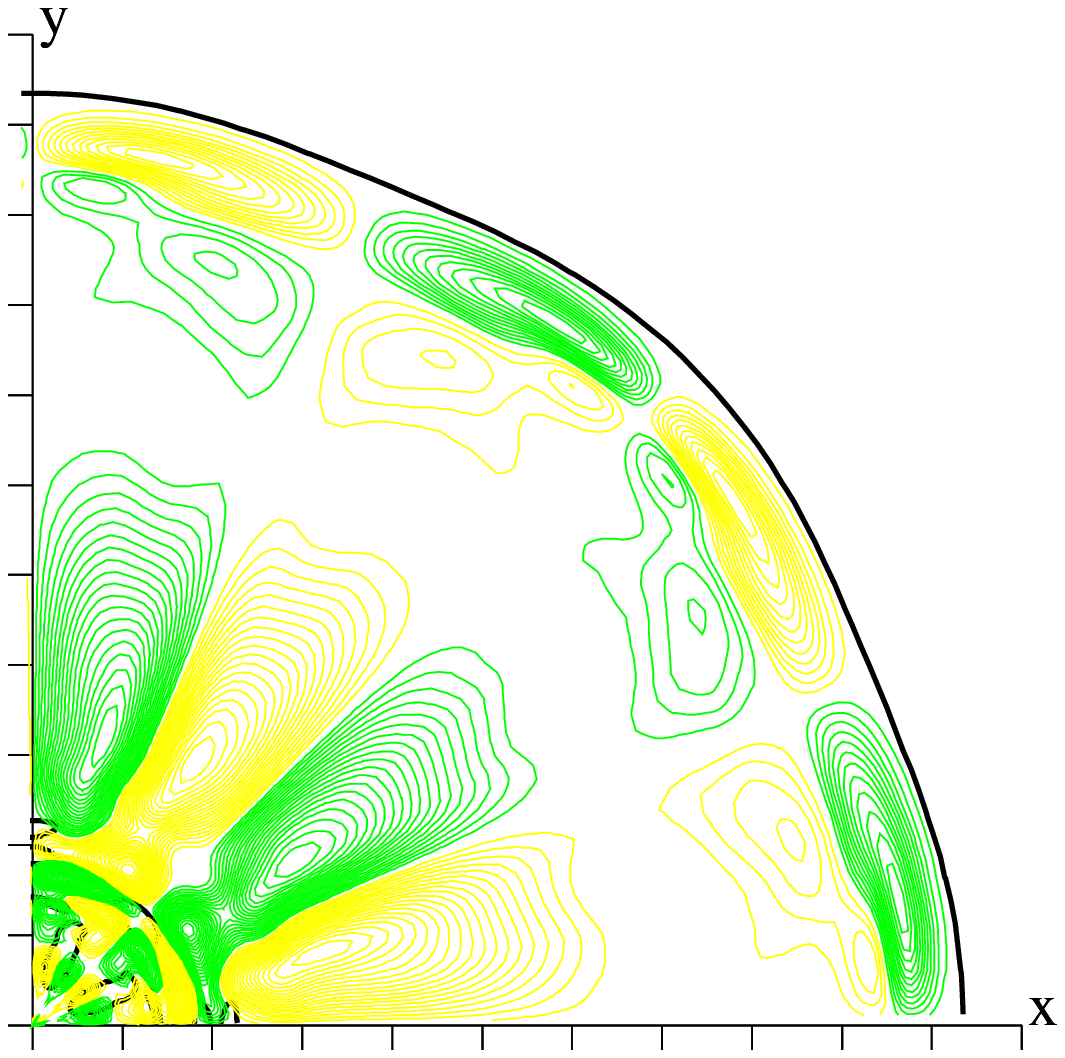}
\hskip 0.5cm
\includegraphics[width=4.0cm,clip,angle=90]{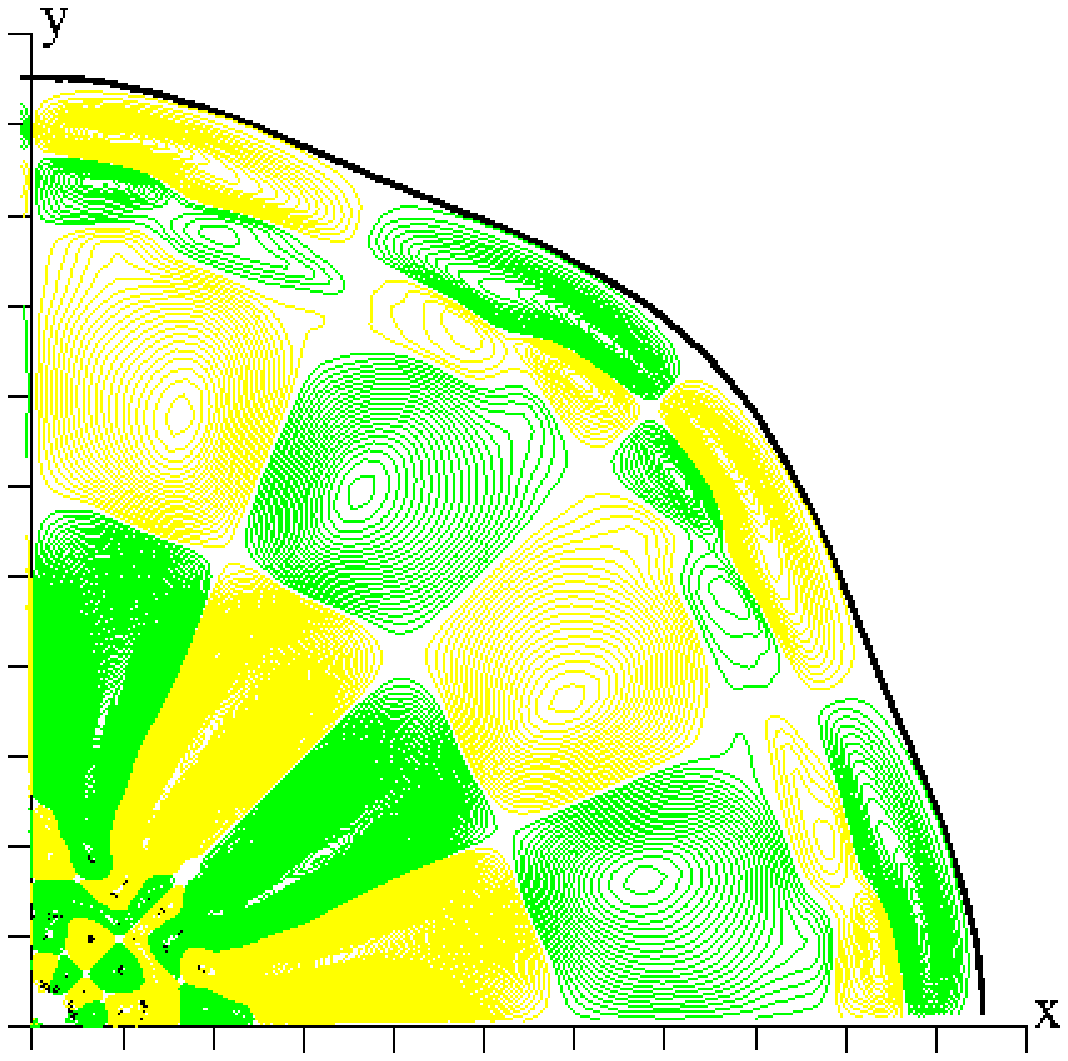}
\hskip 0.5cm
\includegraphics[width=4.0cm,clip,angle=90]{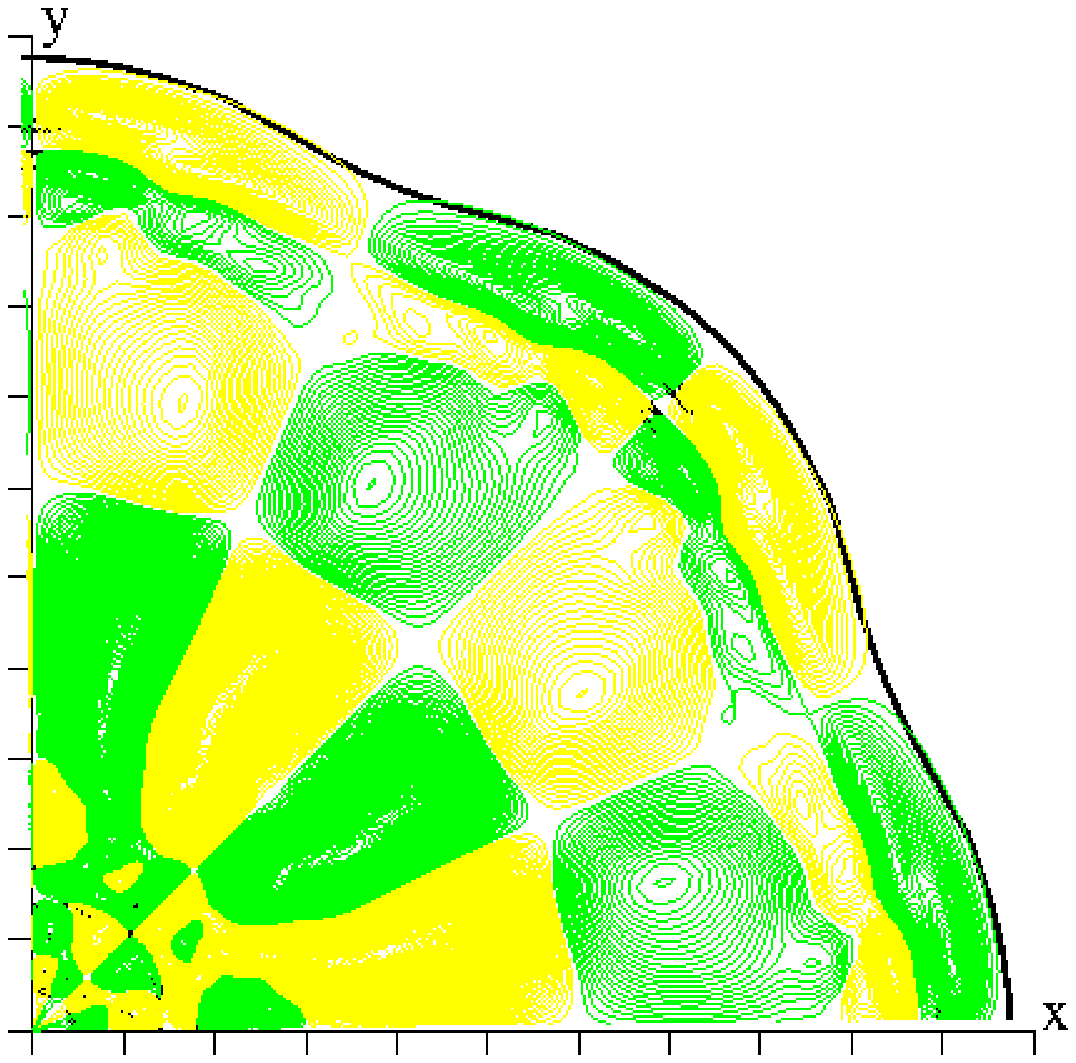}
\vskip 0.25cm
\hskip 3.0cm  a)   \hskip 4.5cm  b)   \hskip 4.5cm c)
\vskip 0.1cm
\hskip -1.8cm
\includegraphics[width=4.0cm,clip,angle=90]{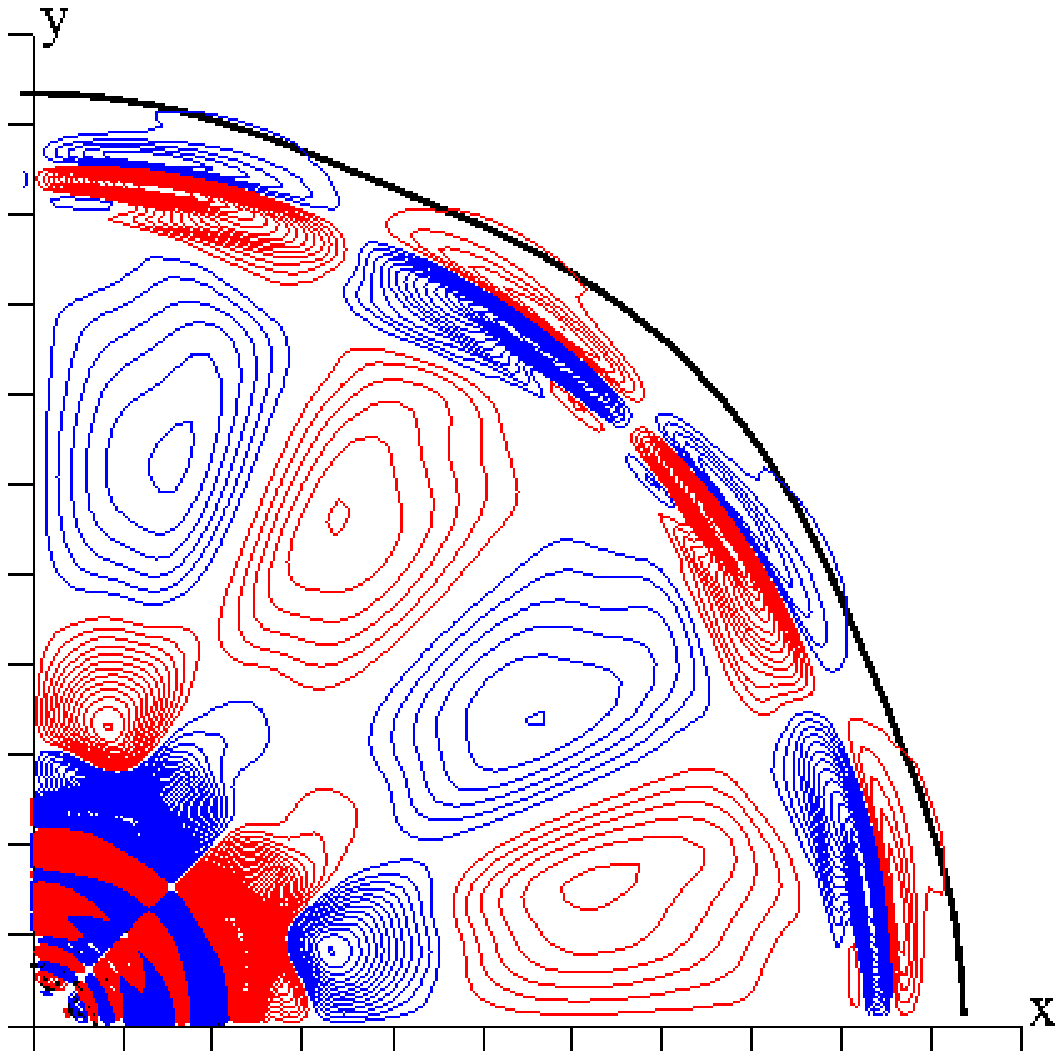}
\hskip 0.5cm
\includegraphics[width=4.0cm,clip,angle=90]{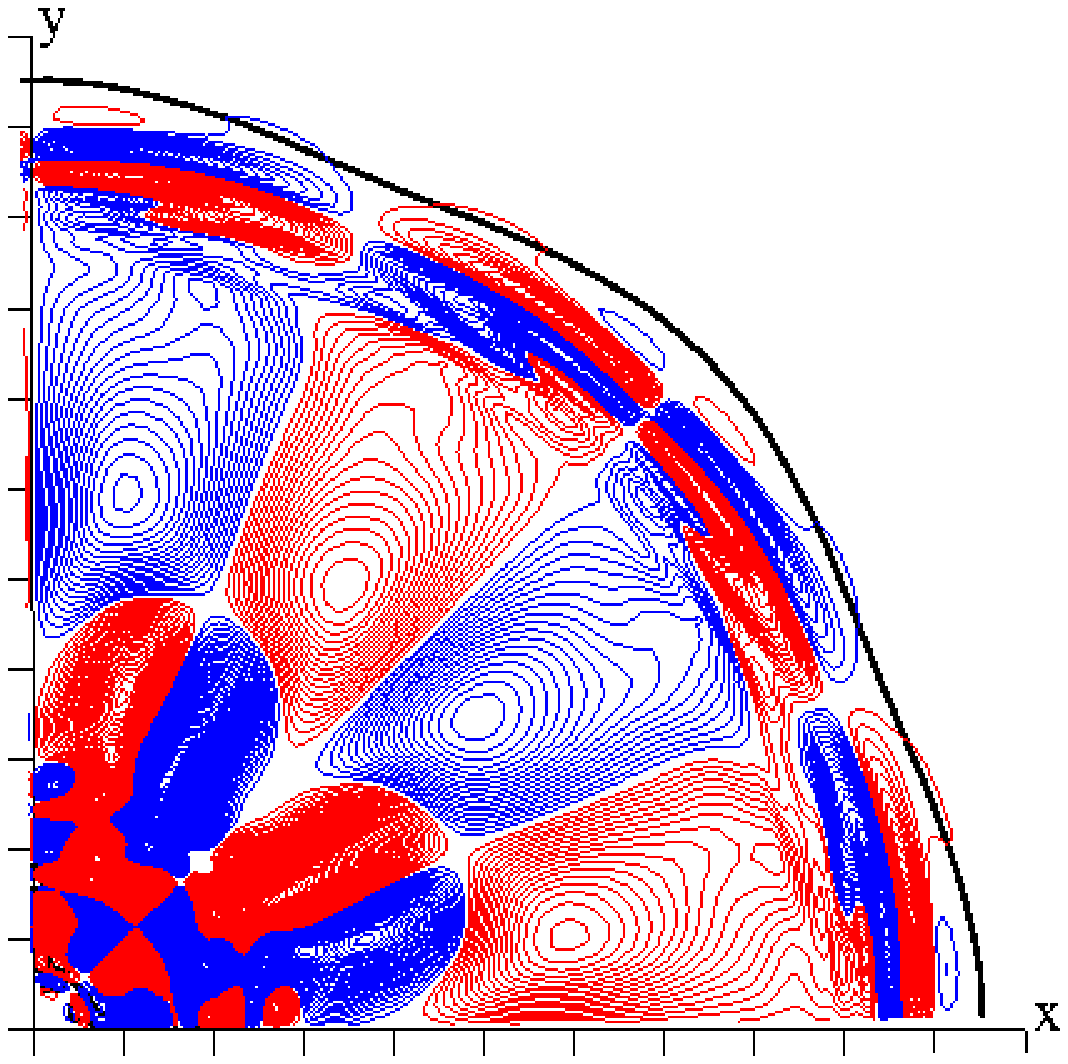}
\hskip 0.5cm
\includegraphics[width=4.0cm,clip,angle=90]{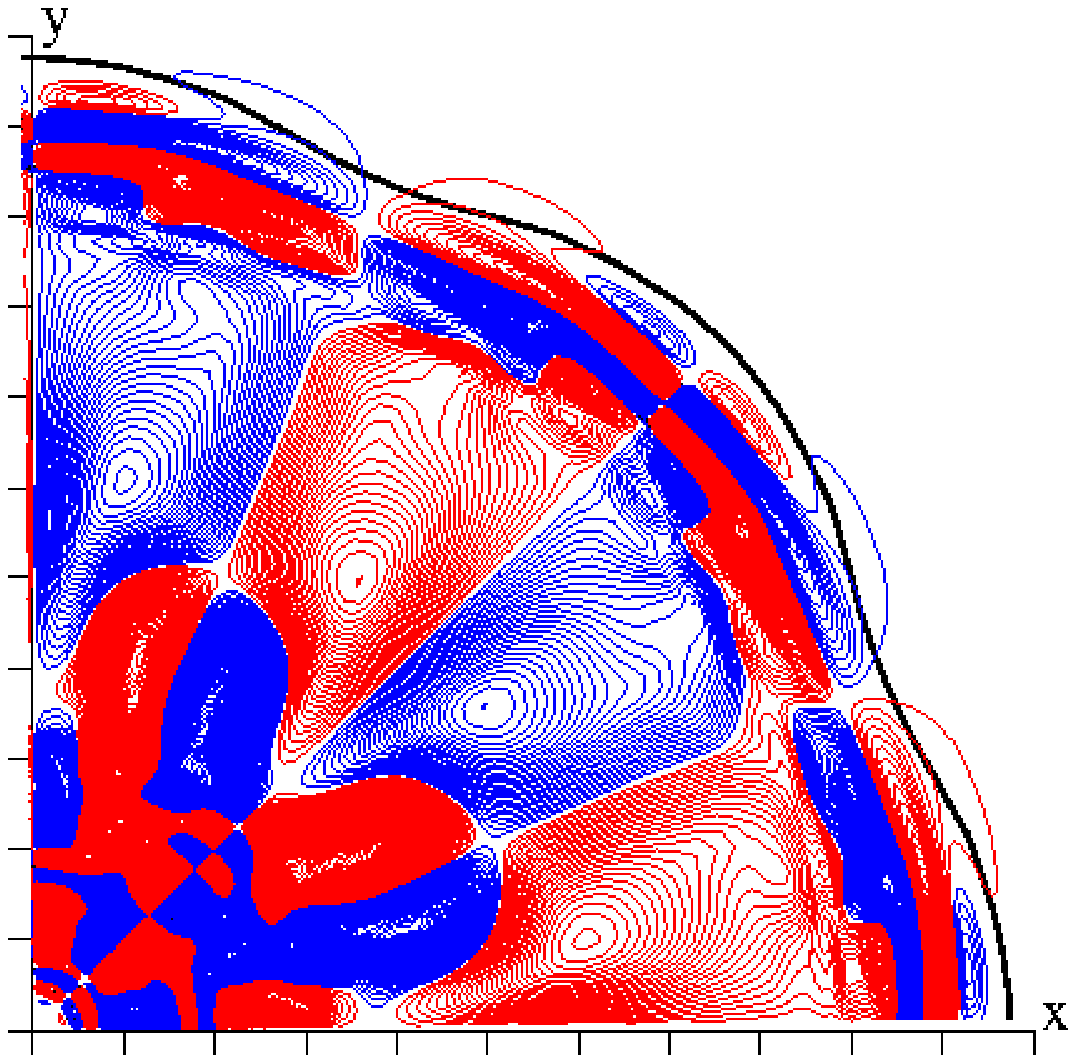}
\vskip 0.25cm
\hskip 3.0cm  d)   \hskip 4.5cm  e)  \hskip 4.5cm  f)
\caption{ Surface contour, in one quarter of the domain,
of $\omega_\theta=0.05$ (black) 
together with:
top $\omega_z$ positive red negative blue,
bottom $\omega_r$ positive green negative yellow
at $t=20$ at $z=z_c$, with increments $\Delta =.01$
a), d) $\delta_0=0.03$, b), e) $\delta_0=0.06$, c), f) $\delta_0=0.09$.
}
\label{fig9}
\end{figure}

To see in more detail the connection between the secondary vorticity
fields near the axis and near the surface of the Hill vortex,
one quarter of the box in a $x-y$
plane at the $z$ location of maximum $<\omt>$ has been considered. 
Indeed the figure \ref{fig9} shows the strong connection
between the small cylindrical patches at the axis, and the
ribbon like structure near the surfaces. Differently than in
figure \ref{fig8} the black line is at a value 
$\omega_\theta=0.05$,
that depicts better the deformation of the surface of
the vortex for any kind of disturbance assigned.
The shape of the secondary vorticity patches near the surface 
does not depend much on the amplitude of the disturbances.
However,  at high $\delta_0$ small patches  of $\omz$ 
are responsible of the increase of the deformation
of the surface.  The patches at an  intermediate distance
are less intense than those near the axis and therefore the
undulations of $\omt$ are smaller than those near
the axis , and near the surface.
The common features drawn
by figure \ref{fig9} is that the two vorticity components,
at this time, are almost equal in the region
around the axis, and that by increasing $\delta_0$    
$\omega_z$ affects in a large measure
the deformation of external vortex surface. 

These simulations, performed in a fixed reference frame,
are evolving for $20$ time units, to prevent that the
vortex, exiting from the domain, reenters encountering
part of the spike. In this lapse of time
it has been noticed that the deformation of the surface
increase up to $t=6$ and later on does not change
up to $t=20$. From this observation it can be concluded
that strong disturbances applied to the Hill vortex 
produce secondary vorticity components 
($\omega_r$ and $\omega_z$) 
at its interior that grows in amplitude and move towards
the external surface. However, the main vortex
continues to travel remaining compact    
without breaking in small scales typical of  turbulent flows.
This conclusion is questionable since it is based on  simulations
evolving for a relative short time. 

\subsection{Frame moving simulations}

\begin{figure}
\centering
\vskip -0.0cm
\hskip -1.8cm
\psfrag{ylab} {$Q_i$}
\psfrag{xlab}{ $  $}
\includegraphics[width=6.5cm,clip,angle=0]{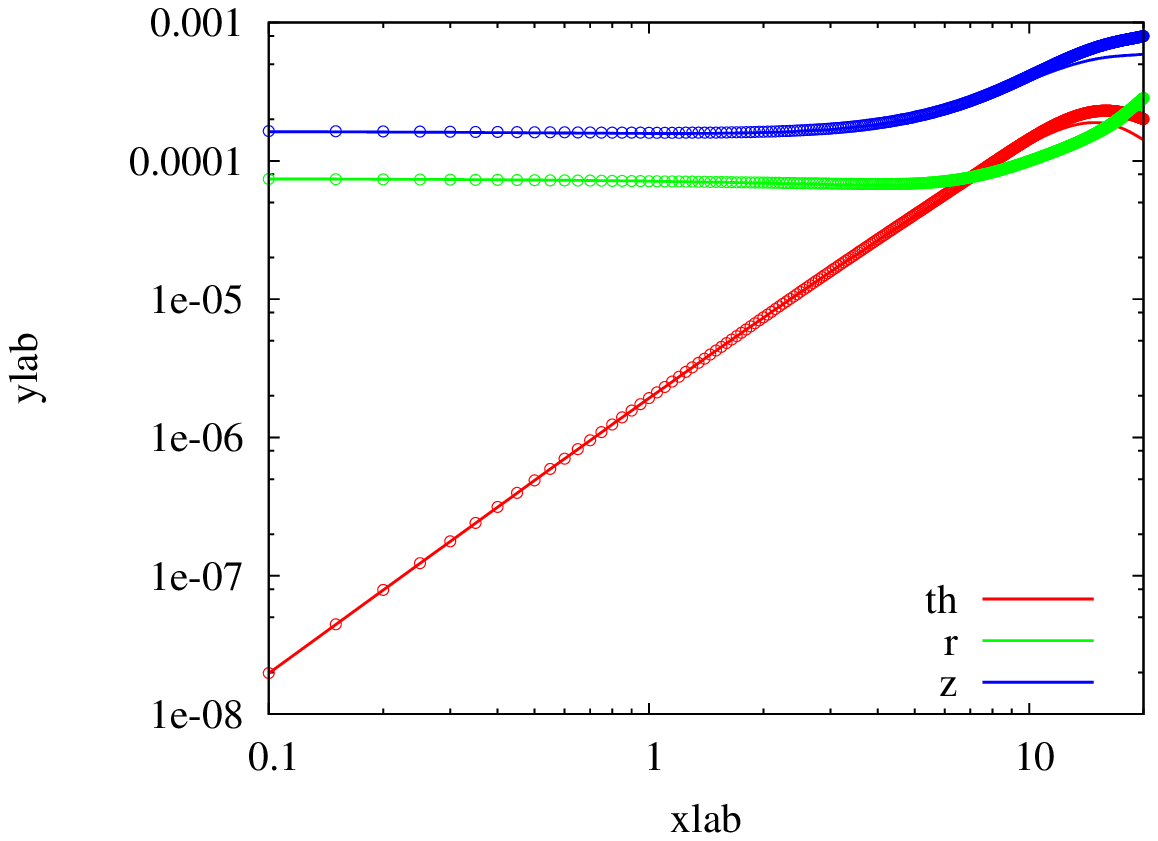}
\hskip 0.0cm
\psfrag{ylab} {$ $}
\psfrag{xlab}{ $  $}
\includegraphics[width=6.5cm,clip,angle=0]{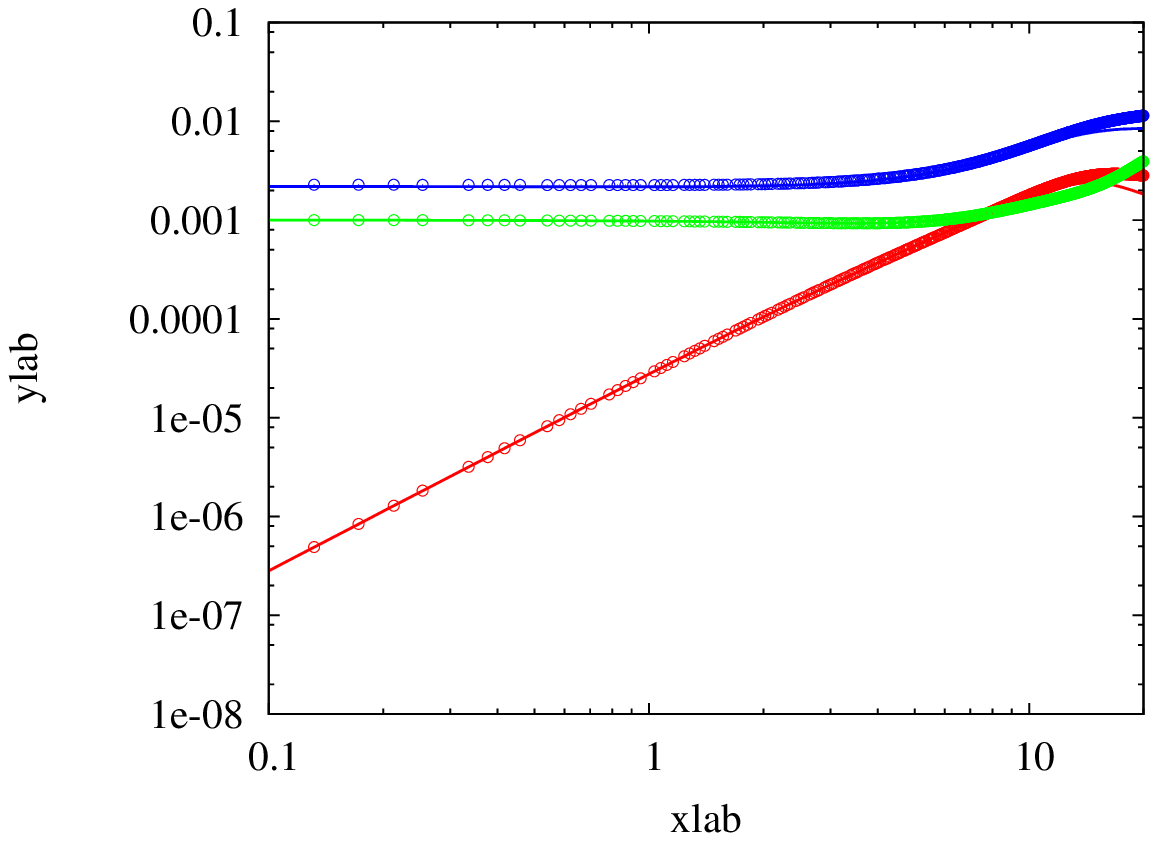}
\vskip -0.2cm
\hskip 3.0cm  a)   \hskip 6.5cm  b)
\vskip 0.0cm
\hskip -1.8cm
\psfrag{ylab} {$\Omega_i$}
\psfrag{xlab}{ $ t$}
\includegraphics[width=6.5cm,clip,angle=0]{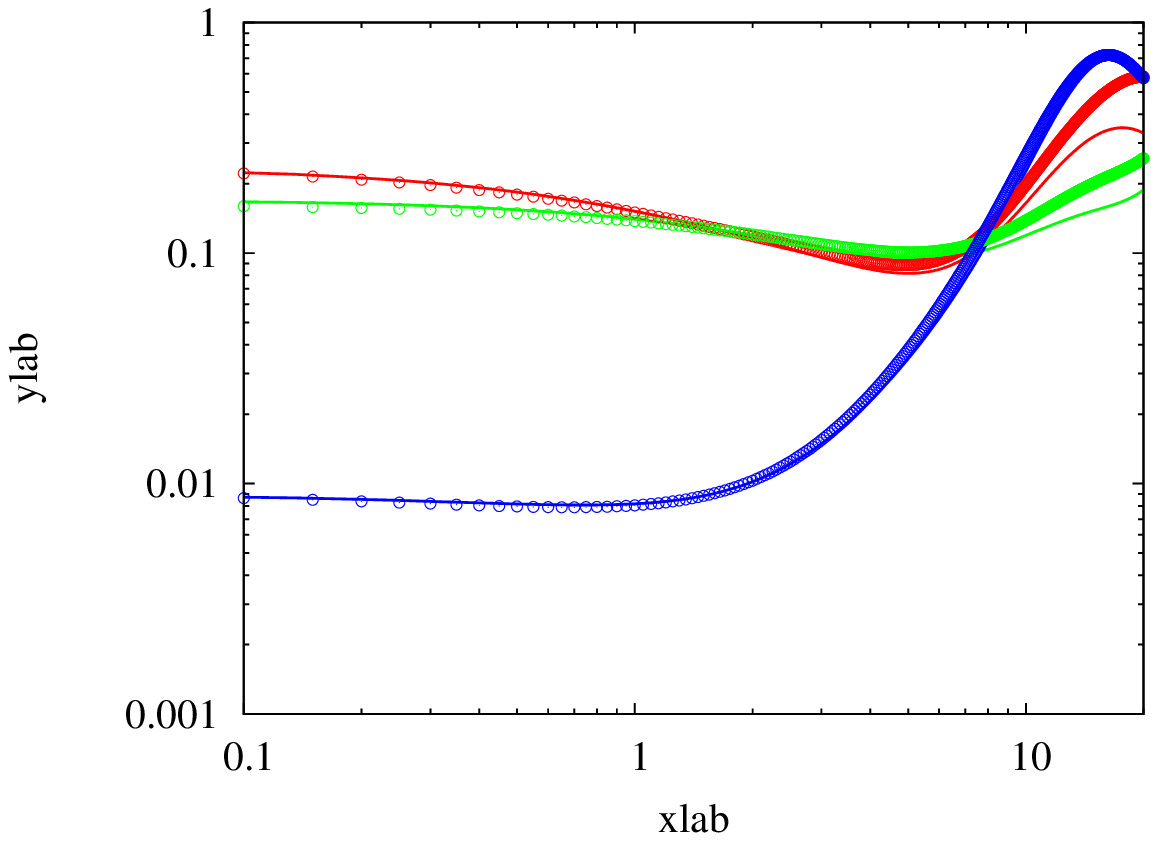}
\hskip 0.0cm
\psfrag{ylab} {$ $}
\psfrag{xlab}{ $ t$}
\includegraphics[width=6.5cm,clip,angle=0]{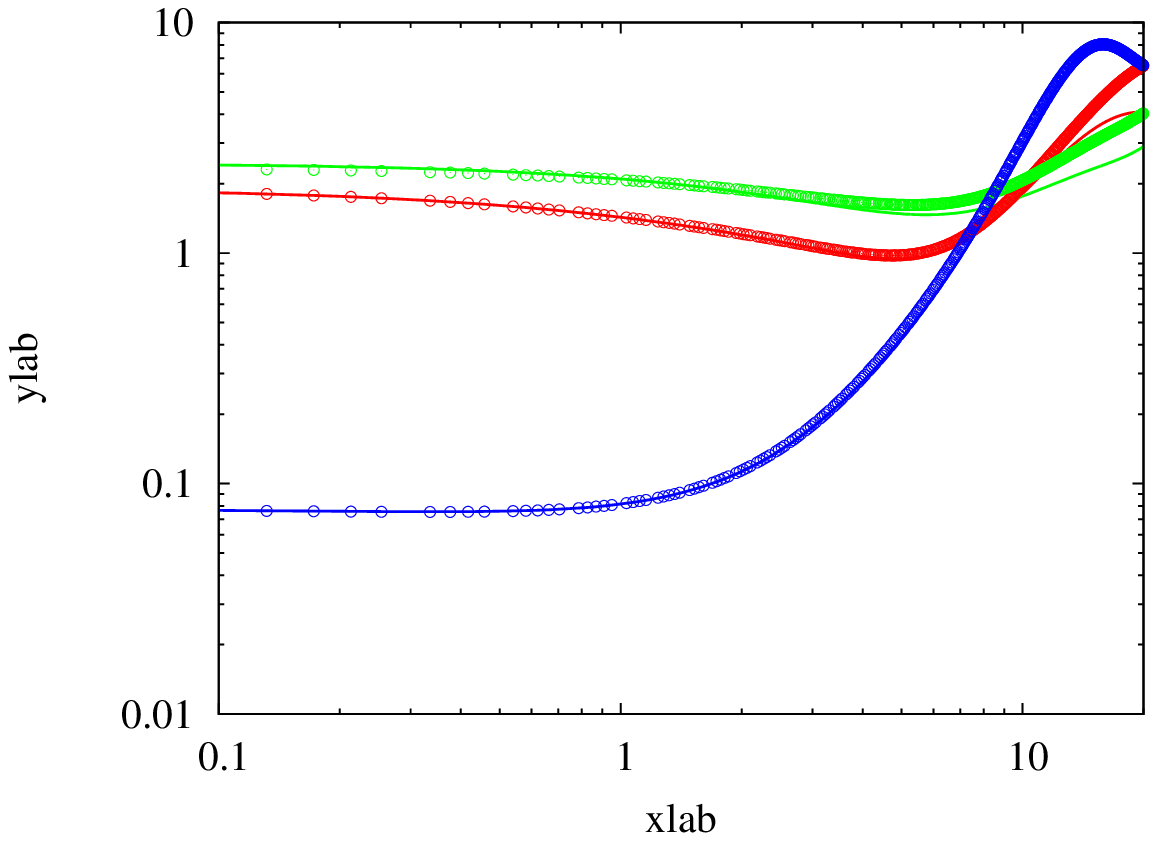}
\vskip -0.2cm
\hskip 3.0cm  c)   \hskip 6.5cm  d)
\caption{  Time evolution for $n_a=4$ of the velocity  $Q_i$ for,
a) $\delta_0=0.01$, b) $\delta_0=0.03$ and vorticity  $\Omega_i$,
c) $\delta_0=0.01$, d) $\delta_0=0.03$; red azimuthal, green radial
and blue axial components, lines fixed open symbols moving frame.
}
\label{fig10}
\end{figure}

To analyse the space evolution of the  Hill vortex
for  long time the computational domain in the streamwise
direction should be long enough do not let the vortex
to leave the computational box. For instance to travel
for $t=100$, that is five time longer than in the previous sections,
a greater computational  effort is necessary. 
A more productive way, by using the same length in $z$,
consists on  solving the Navier-Stokes equations
in a reference frame moving with the theoretical velocity
$U=-2/15$.  The aim of a long evolution is to verify whether the 
secondary motion becomes strong enough not only
to deform the external surface but also to destroy  
the large scale structure, giving rise to small scales 
propagating everywhere, that is to produce a fully turbulent flows. 
The $n_a=4$ disturbances have been simulated for
$\delta_0=0.01$ and $\delta_0=0.03$ amplitudes. 
The former is devoted to investigate whether, for small disturbances,
a stop on the growth of the strength of the secondary
vorticity components may occur. The latter one produces
a faster growth and  symmetry breaking. The two cases 
allow to see whether an universal behaviour can be
established.  The results of space and time developing simulations
should have similar trends.  The eventual small differences can be detected by
looking at the time evolution of the velocity
and vorticity mean square in the whole domain. 
If $\overline{u_i}=<u_i>$ and
$\overline{\omega_i}=<\omega_i>$ 
the mean square are defined as 
$Q_i=\frac{\Sigma_i\Sigma_j\Sigma_k}{N_1N_2N_3} (u_i-\overline{u_i})^2$,
$\Omega_i=\frac{\Sigma_i\Sigma_j\Sigma_k}{N_1N_2N_3} 
(\omega_i-\overline{\omega_i})^2$. Figure \ref{fig10}
indeed shows the rather good agreement of the time
evolution in the different frame of reference. It is worth
to remind that without azimuthal disturbances
the  mean square are zero. In addition the evolution
of $Q_i$ (figure \ref{fig10}a and figure \ref{fig10}b)
and of $\Omega_i$ (figure \ref{fig10}c and figure \ref{fig10}d) 
demonstrate that, in the first $20$ time units, for low ($\delta_0=0.01$)
and high ($\delta_0=0.03$) amplitude, the
evolution of each  component has similar trend. 
However the scale of the coordinates
evinces that for $\delta_0=0.03$ values almost ten times higher
than those for $\delta_0=0.01$, implying an eventual scaling
with the corresponding values at $t=0$ that are proportional to the amplitude of
the disturbance. 

\begin{figure}
\centering
\vskip -0.0cm
\hskip -1.8cm
\psfrag{ylab} {$K/k_0$}
\psfrag{xlab}{ $ t $}
\includegraphics[width=6.5cm,clip,angle=0]{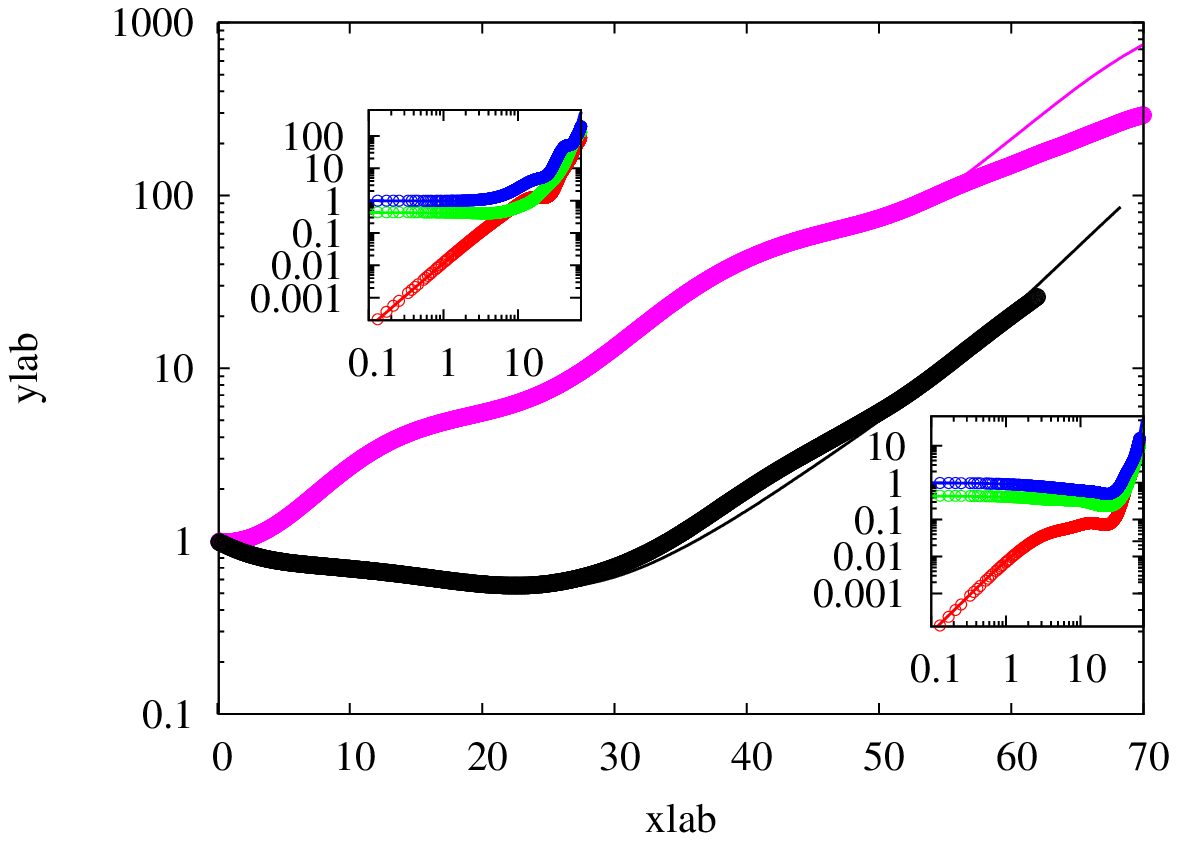}
\hskip 0.5cm
\psfrag{ylab} {$\chi/\chi_0 $}
\psfrag{xlab}{ $ t $}
\includegraphics[width=6.5cm,clip,angle=0]{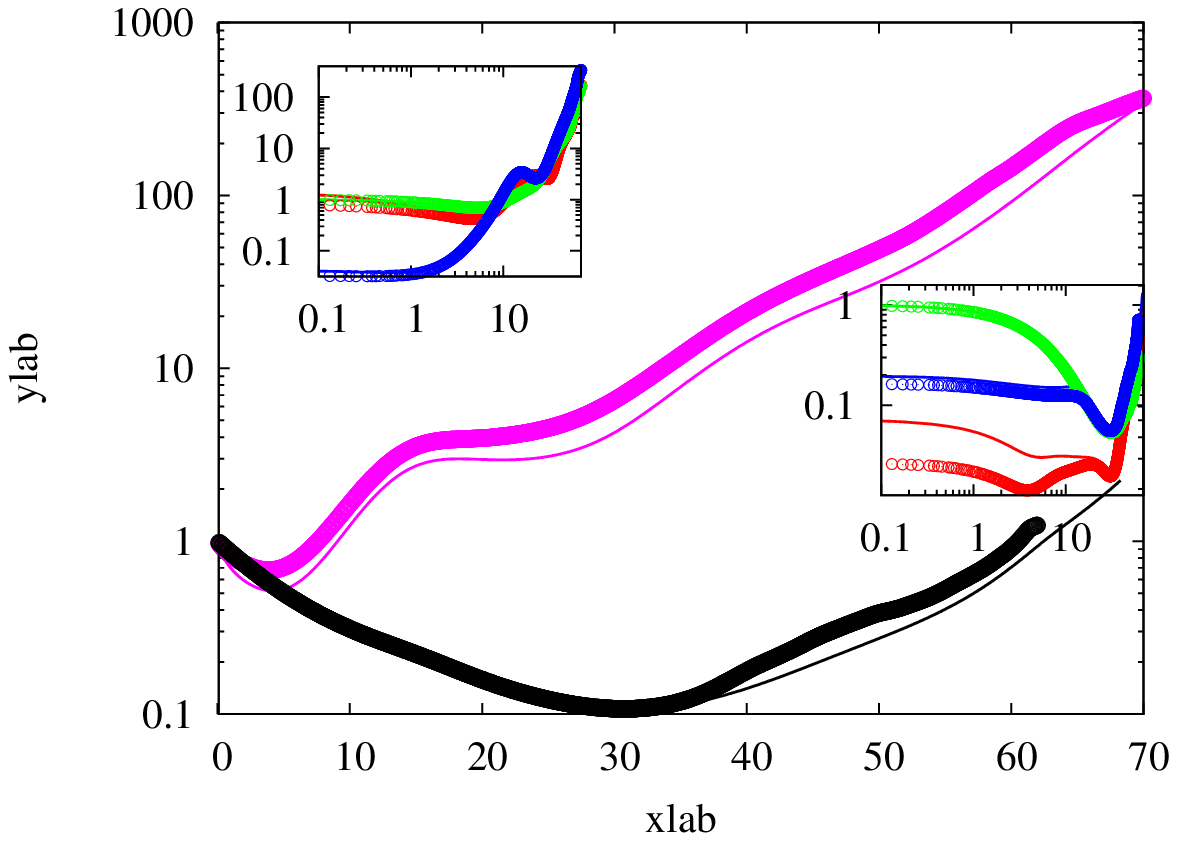}
\vskip 0.0cm
\hskip 3.0cm  a)   \hskip 6.5cm  b)
\vskip 0.00cm
\caption{  Time evolution of a) turbulent kinetic energy
b) enstrophy scaled by the corresponding values at $t=0$
purple $n_a=4$: symbols $ \delta_0=0.03$, lines $\delta_0=0.01$; 
black random: symbols $\delta_0=0.09$,lines $\delta_0=0.01$;
inset in a) mean square velocity components scaled by
$Q_z(t=0)$, red $Q_\theta$, green $Q_r$, blue $Q_z$ left 
$n_a=4$ right random the same colours and symbols for b).
}
\label{fig11}
\end{figure}

The aim of the DNS is to investigate whether the Hill vortex,
while is travelling
for a long distance, maintains a  shape similar to
the initial one  or instead is disrupted.
In addition, it was noticed that, in a short interval
of time, the azimuthal symmetry is preserved, implying that,
the small round-off errors are not amplified by the growth of 
secondary vorticity.  We wish to point out that, on purpose, 
single precision has been used to verify
for how long the symmetry is preserved. To enhance
the destruction of the Hill vortex  
large amplitude disturbances ($\delta_0=0.09$ and $\delta_0=0.03$)
with a random distribution
of azimuthal wave numbers have been introduced at $t=0$.         
The profiles of the total turbulent kinetic energy $K=\Sigma Q_i$
and the enstrophy $\chi=\Sigma \Omega_i$ versus time,
in figure \ref{fig11}a and figure \ref{fig11}b,
give a first impression of the influence  
of the disturbance on a long time evolution.
To investigate the dependence on the amplitude and on the number 
of the number $n_a$ of waves excited, $K$ and $\chi$
have been scaled with their values at $t=0$ ($K_0$ and $\chi_0$). 
The figures show that an independence on the amplitude
of the disturbance does exist, in
fact, in both the figures \ref{fig11} the profiles of the
lines (small amplitude) and of the symbols (large amplitude)
are almost coincident. On the other hand, the trend of the
random wave number disturbances differs from that for
$n_a=4$.  Similarity and differences can be appreciated
also for each mean square component, as it is shown in the insets of
figure \ref{fig11}. When a single wave number is excited
the initial radial and axial velocity mean square components remain
constant for few time units, while the azimuthal
components grows exponentially. Later on all three $Q_i$ components
grow up to $t=60$ (figure \ref{fig11}a) for both amplitudes,
as it is enlightened by the coincidence of the purple line and symbols.
For $t> 60$ the two curves diverge.
Flow visualizations in orthogonal planes to the $z$ axis
allow to understand the reasons of this separation.
For random disturbances the time history is independent
on $\delta_0$. The initial decrease of $K$ can be ascribed
to the transfer of energy among scales due to the different
energy level for each wave number at $t=0$. Figure \ref{fig11}b
shows for $\chi/\chi_0$ a trend similar to that of $K/K_0$, with a
drop in the first time units greater than that for $K/K_0$.
In addition for a single wave number excitation the 
trends of each component of $\chi/\chi_0$ differ from those
with disturbances at random wave number. In the former
case at $t=0$ the amount of $\Omega_z$ is rather small,
through a fast growth it reaches the level
of the others, and all three continue to grow.
The growth for $t>30$  implies an enhancement of the vorticity components
in well localised regions as it is corroborated by visualizations.
The inset relative to the random disturbances has still
a constant evolution in the first few time units with
a greater $\Omega_z$ than that for $n_a=4$. The interesting
result is that during the fast growth for $t>30$ all three
reach values $O(1)$, while for $n_a=4$ a level $O(100)$
is reached.  For $t>55$ $K/K_0$, in the case of $n_a=4$,
show a different trend for $\delta_0=0.01$ (lines) and
for $\delta_0=0.03$ (symbols) due to the symmetry persistence for 
the former, implying vorticity components very intense in 
localised small regions. On the other hand, for the
$\delta_0=0.03$ the symmetry is lost at $t\approx 55$ and the growth
of $K/K_0$ reduces and  becomes closer to that for random
disturbances.

\begin{figure}
\centering
\vskip -0.0cm
\hskip -2.0cm
\includegraphics[width=1.7cm,clip,angle=90]{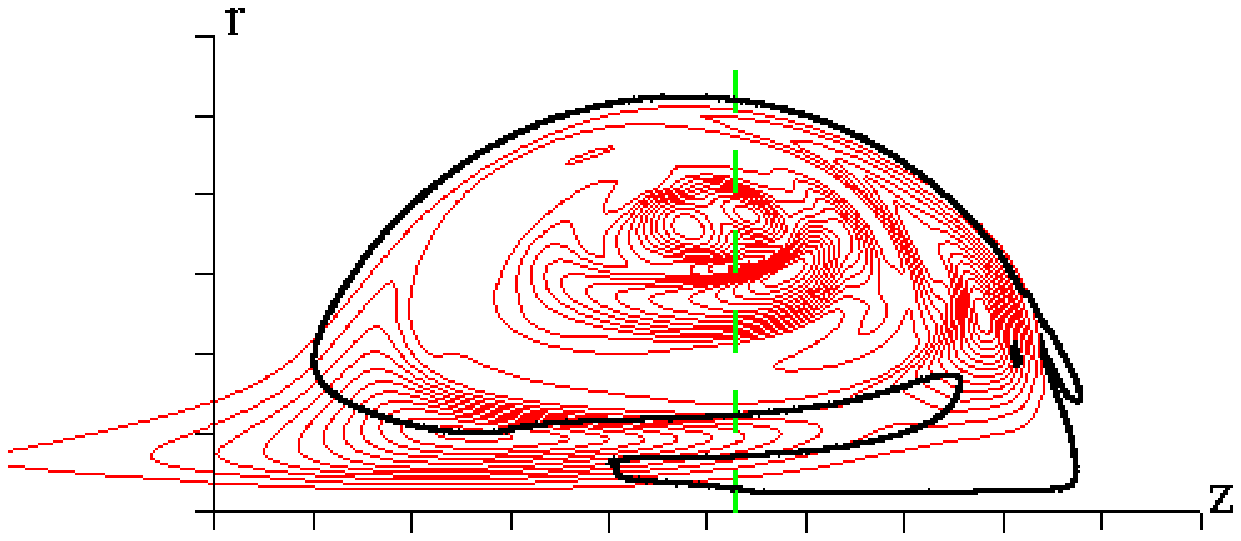}
\hskip -0.5cm
\includegraphics[width=1.7cm,clip,angle=90]{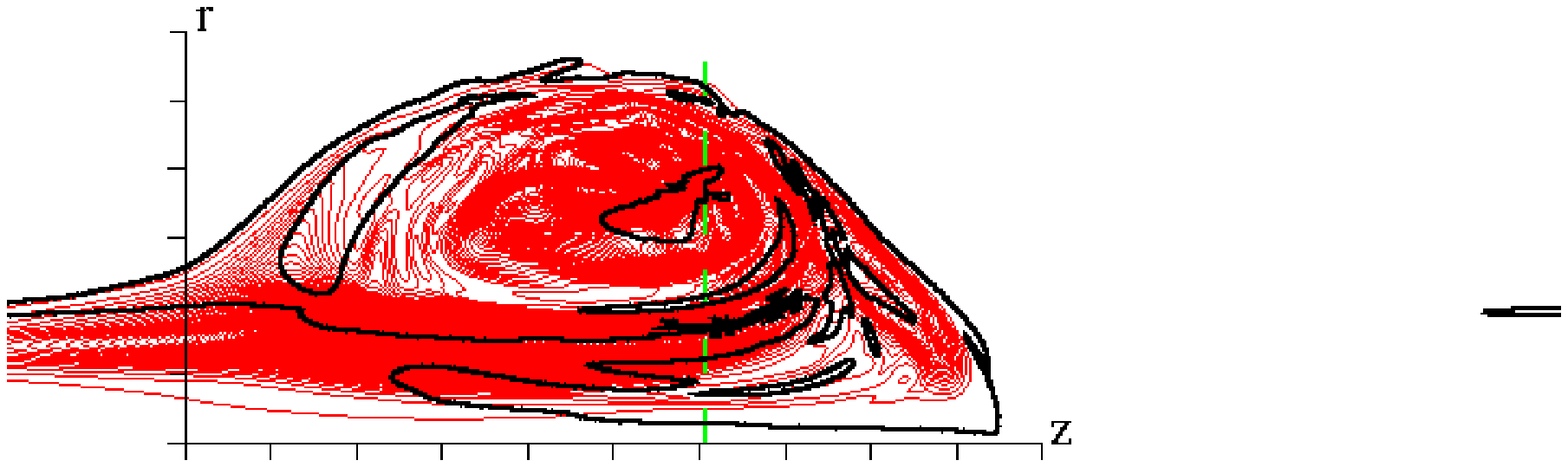}
\hskip -1.5cm
\includegraphics[width=1.7cm,clip,angle=90]{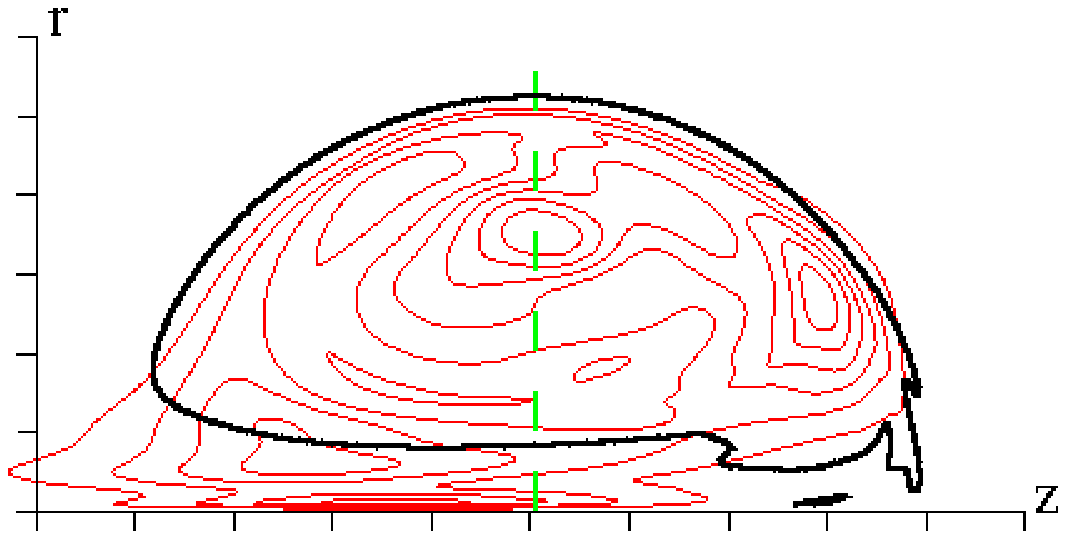}
\hskip -0.5cm
\includegraphics[width=1.7cm,clip,angle=90]{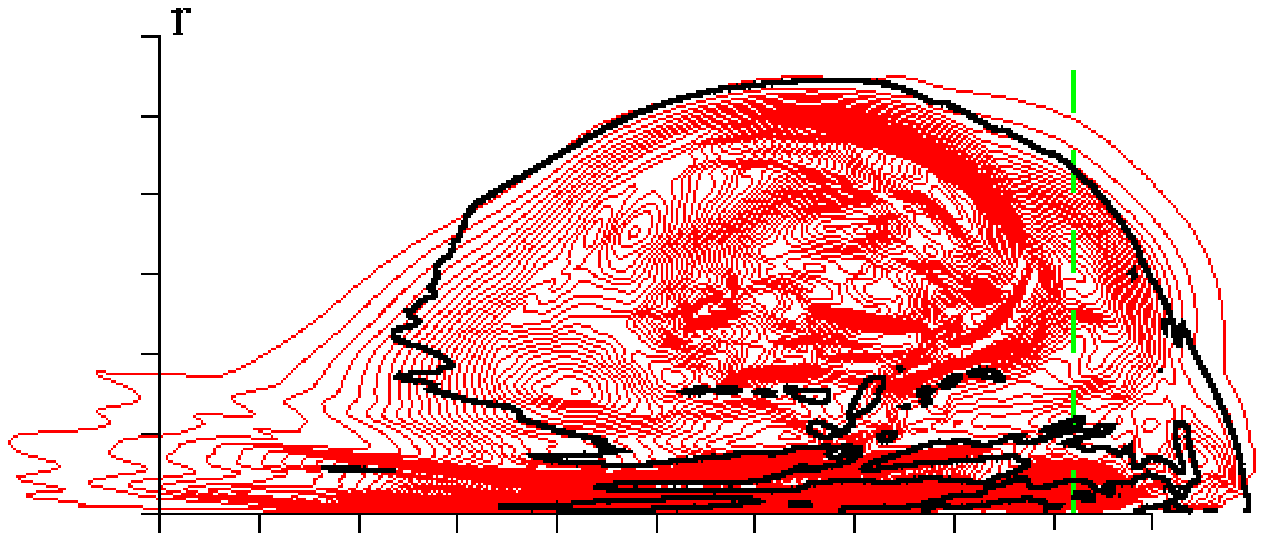}
\vskip 0.25cm
\hskip 1.0cm  a)   \hskip 3.0cm  b)   \hskip 3.0cm c) \hskip 3.0cm d)
\vskip 0.1cm
\vskip -0.0cm
\hskip -2.0cm
\includegraphics[width=3.2cm,clip,angle=90]{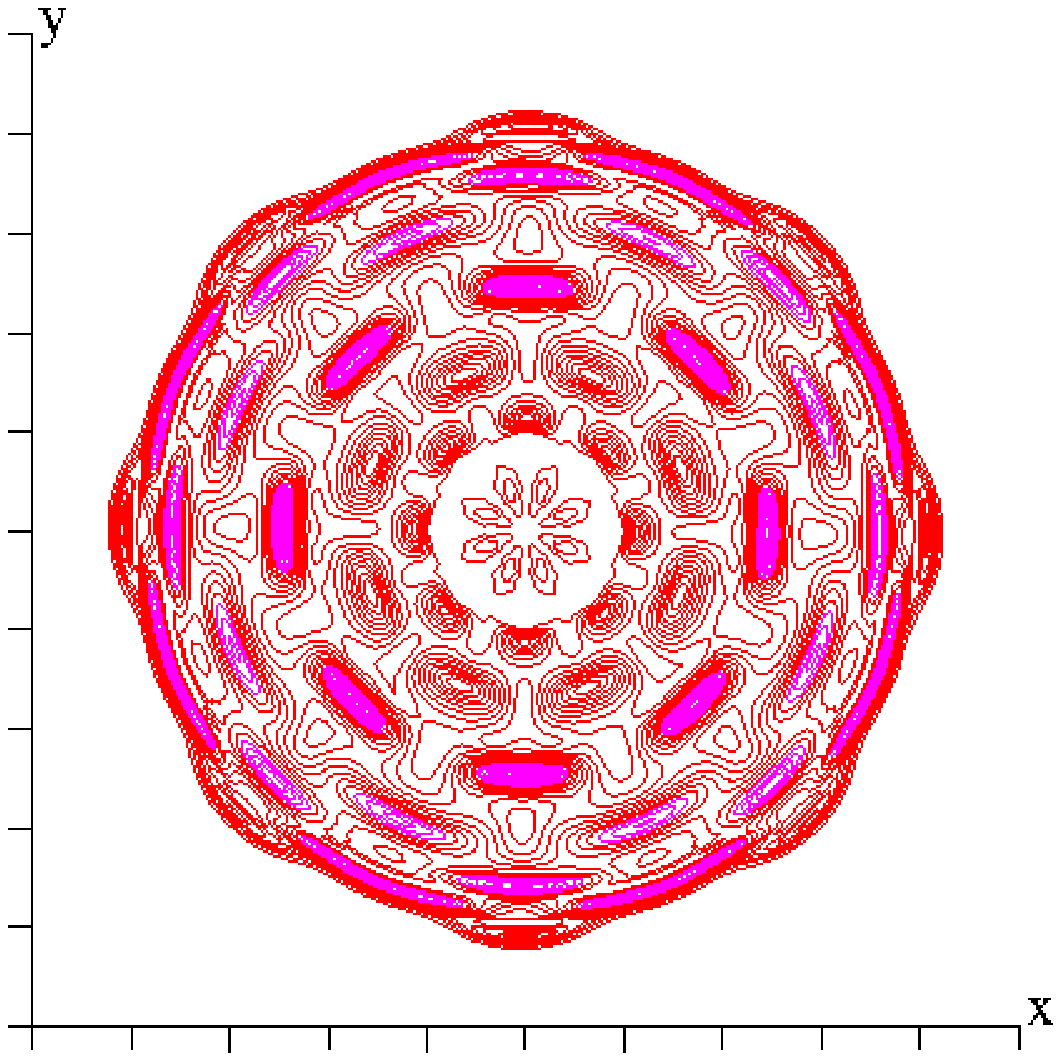}
\hskip 0.25cm
\includegraphics[width=3.2cm,clip,angle=90]{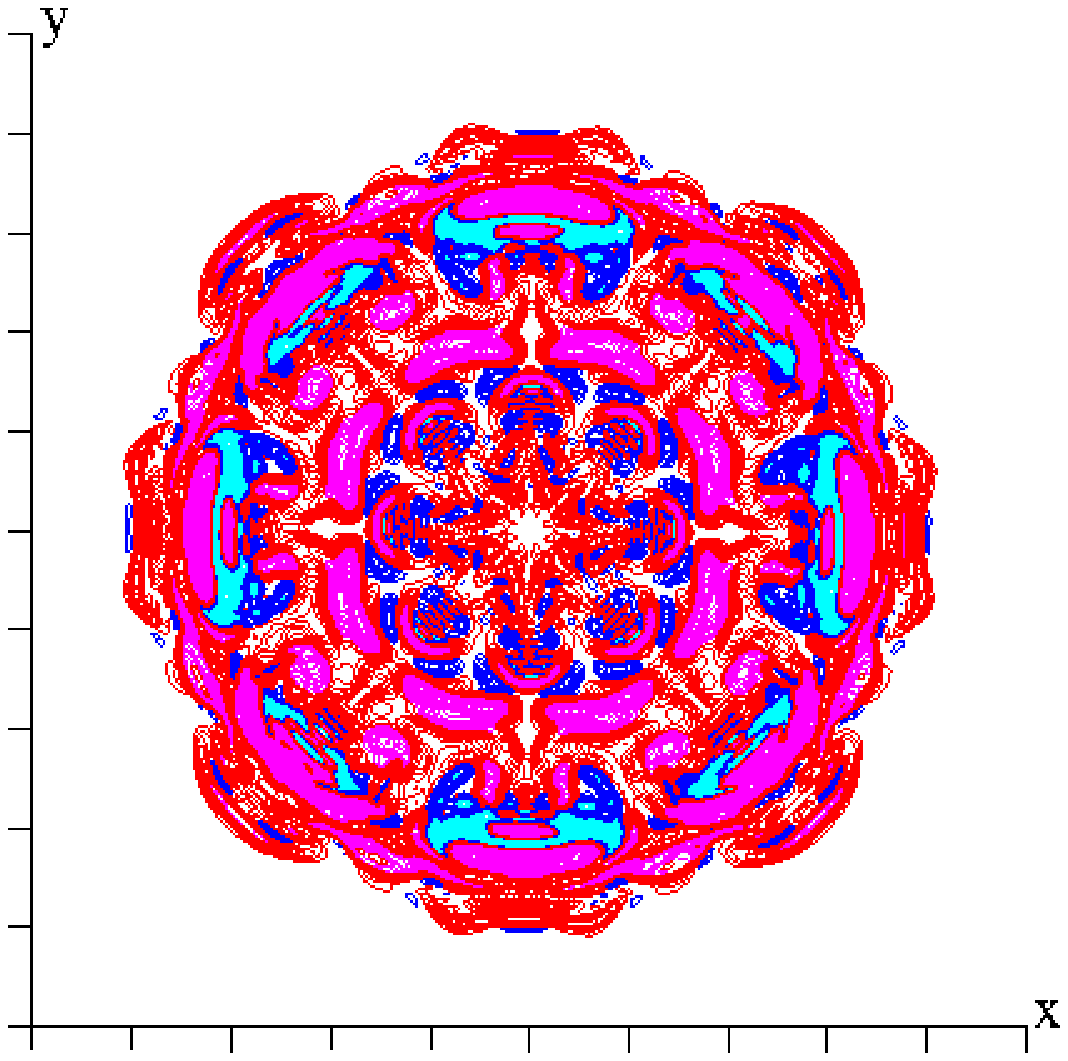}
\hskip 0.25cm
\includegraphics[width=3.2cm,clip,angle=90]{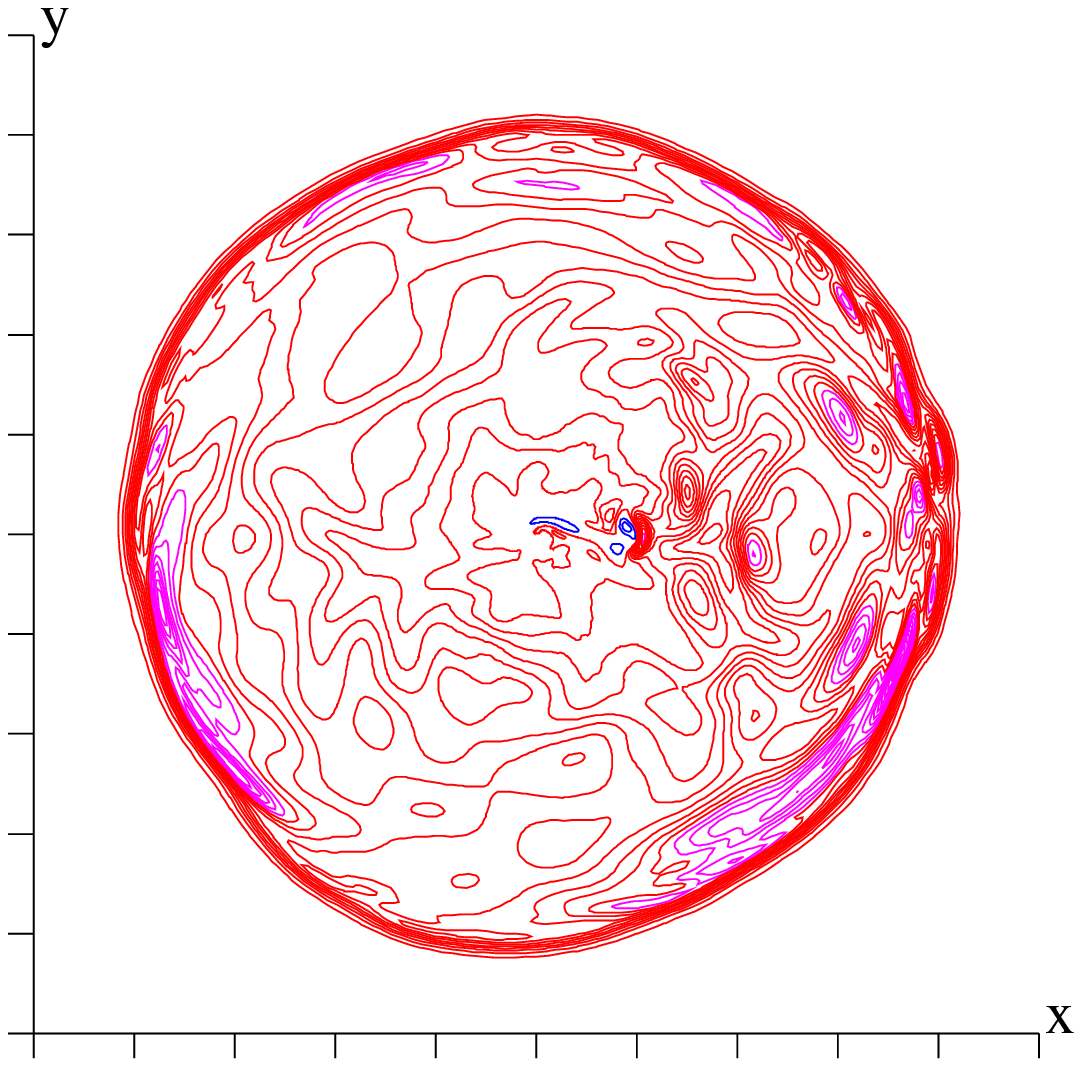}
\hskip 0.25cm
\includegraphics[width=3.2cm,clip,angle=90]{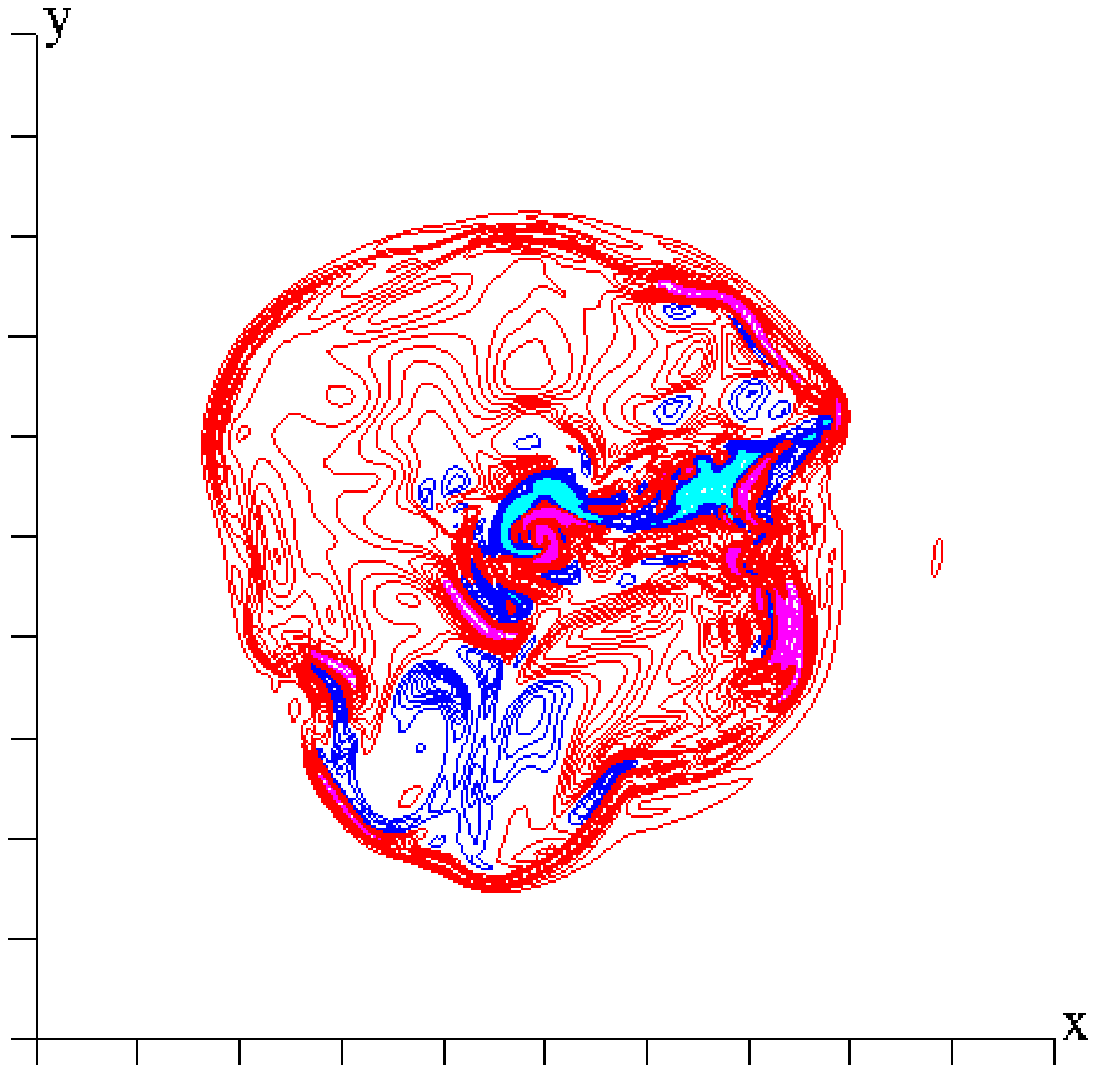}
\vskip 0.25cm
\hskip 1.0cm  e)   \hskip 3.0cm  f)   \hskip 3.0cm g) \hskip 3.0cm h)
\vskip 0.1cm
\vskip -0.0cm
\hskip -2.0cm
\includegraphics[width=3.2cm,clip,angle=90]{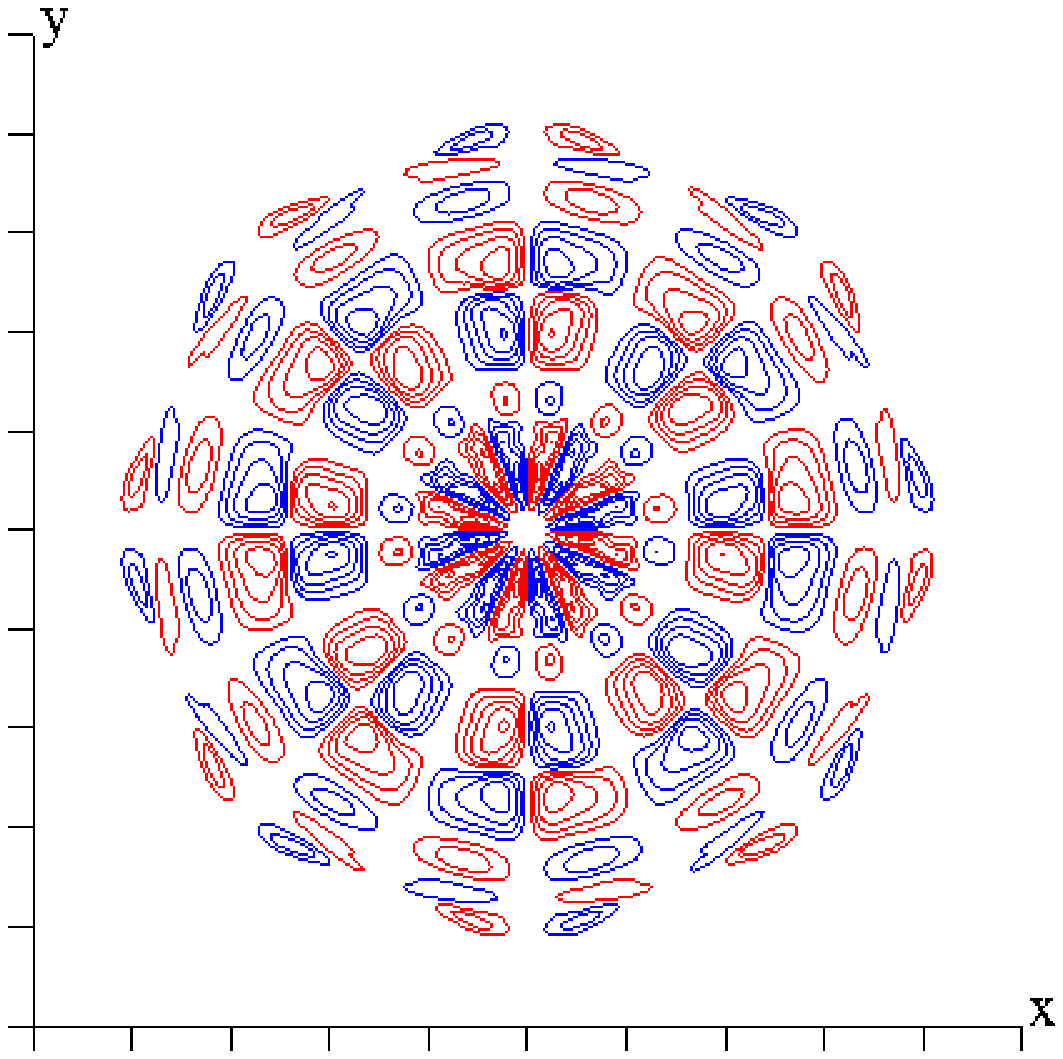}
\hskip 0.25cm
\includegraphics[width=3.2cm,clip,angle=90]{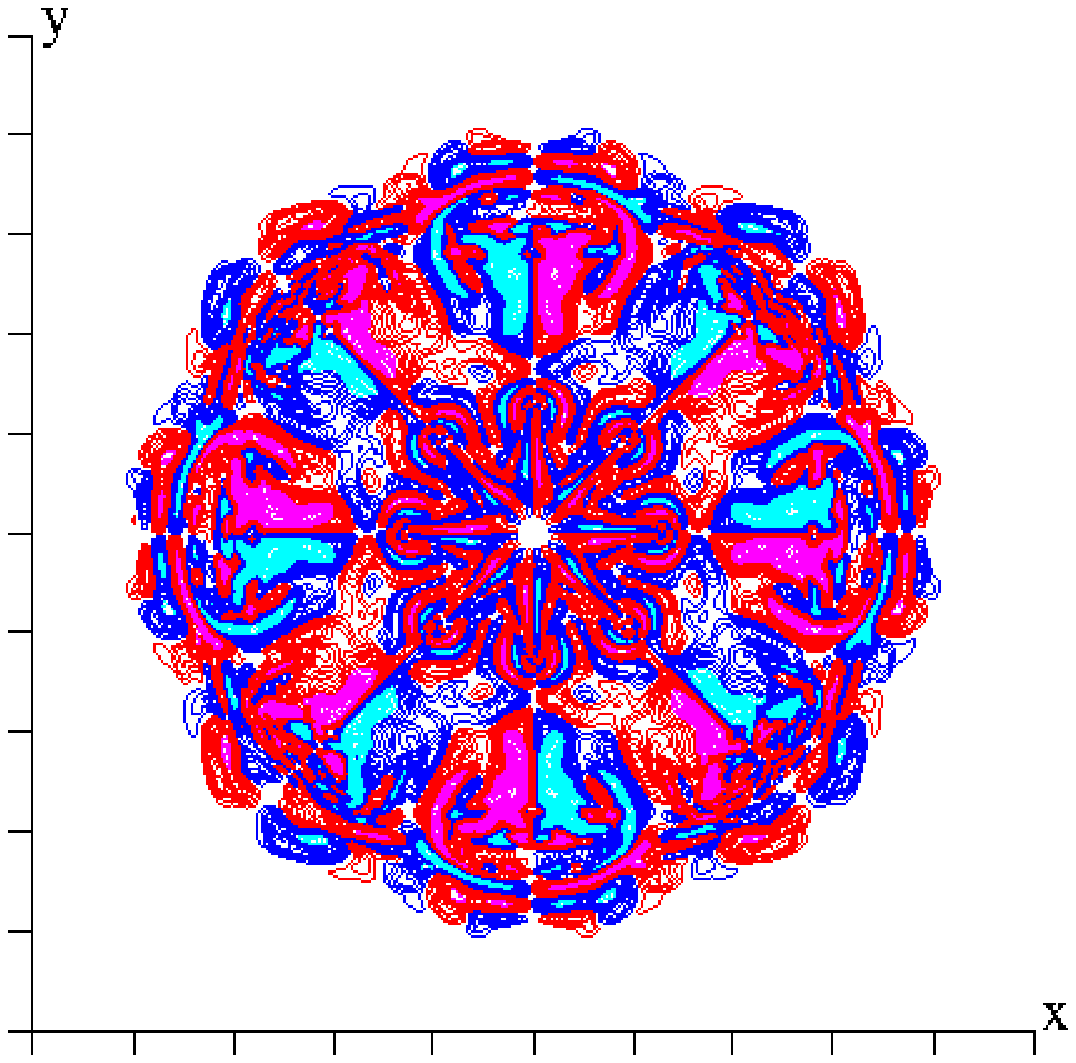}
\hskip 0.25cm
\includegraphics[width=3.2cm,clip,angle=90]{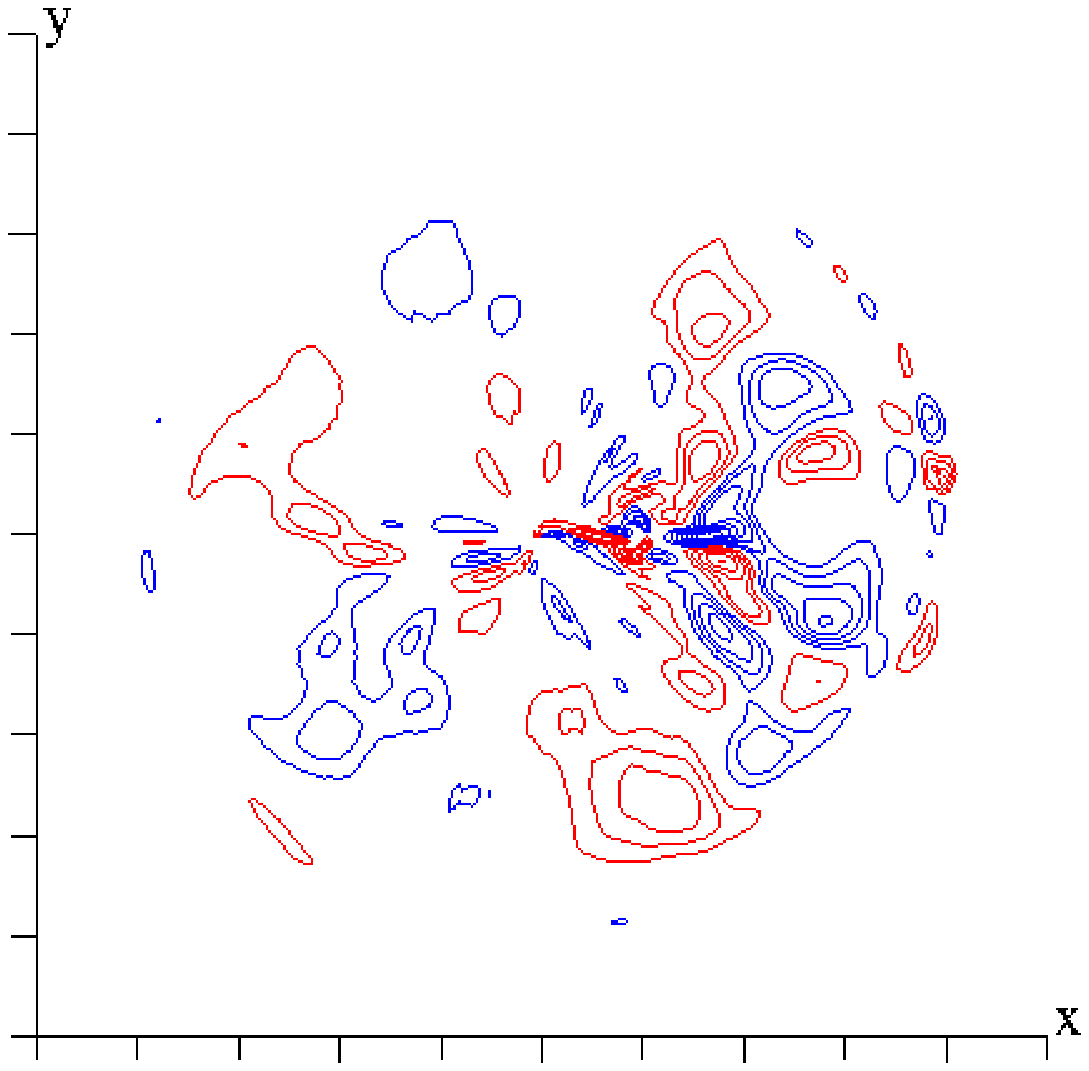}
\hskip 0.25cm
\includegraphics[width=3.2cm,clip,angle=90]{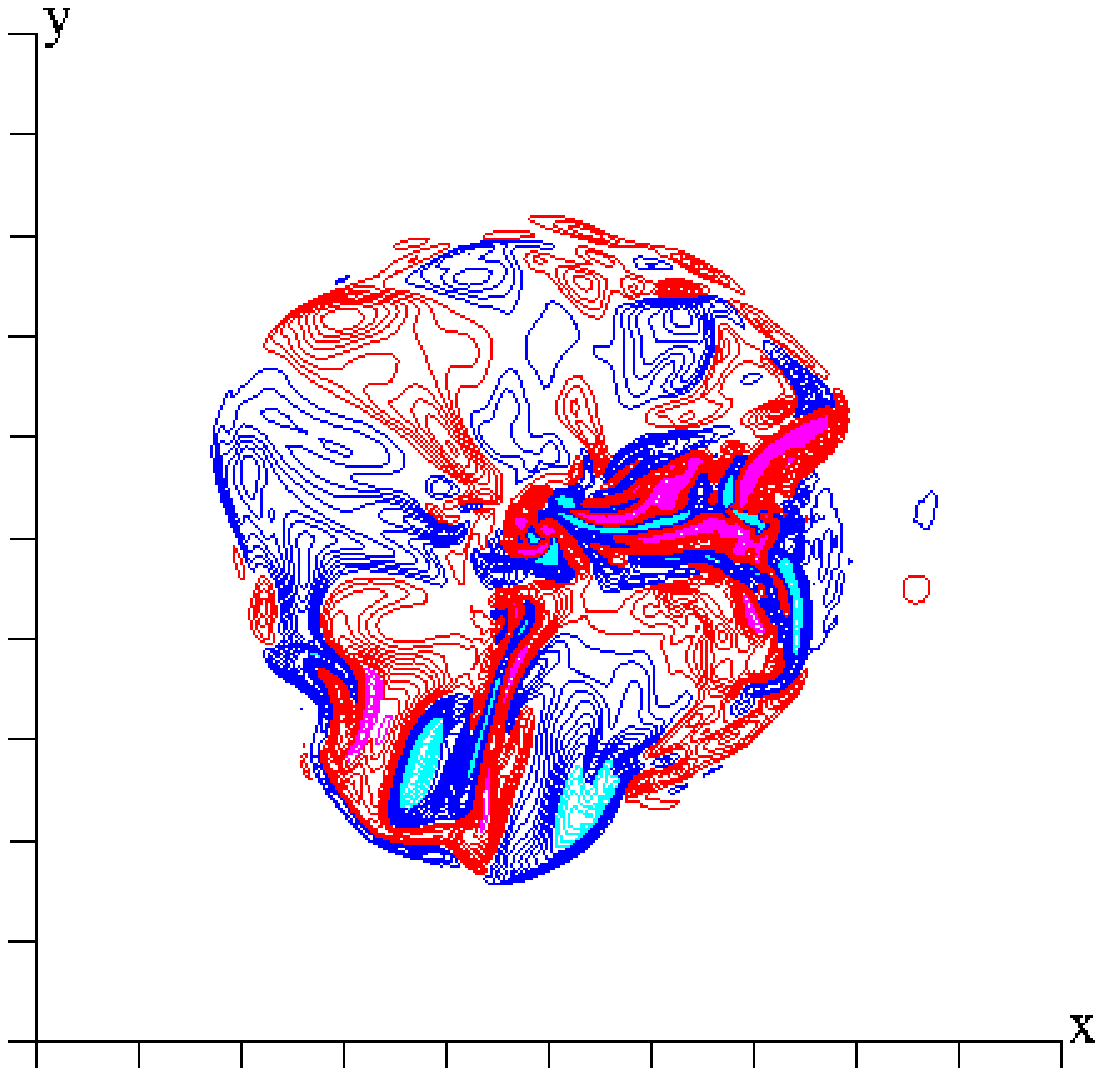}
\vskip 0.25cm
\hskip 1.0cm  i)   \hskip 3.0cm  j)   \hskip 3.0cm k) \hskip 3.0cm l)
\vskip 0.1cm
\caption{ Contours at $t=60$ of $\Sigma <u_i^2>$  (red $\Delta=.00005$)
superimposed to $<\omega_\theta>=0.1$ black, for $-1.75< z-z_c<0.75$
and $0<r<1.2$:  in a) $n_a=4, \delta_0=0.01$,
b) $n_a=4, \delta_0=0.03$ , c)  random  $\delta_0=0.03$,
d)  random  $\delta_0=0.09$;
the green lines correspond to the location where the
$\omega_\theta$ and $\omega_r$ contours ($-1.2<x<1.2$
,$-1.2<y<1.2$) in e,i) as a), f,j) as b),
g,k) as c) and h,l) as d) ;
$\Delta=0.1$ red  up to $\omt<1$, magenta $\omt>1.$,
black $\omt=1$; 
$\Delta=-0.1$ blue for $\omega_\theta <0$,  cyan for $\omega_\theta <-1$. 
}
\label{fig12}
\end{figure}

To understand better the above discussion on the time
behaviour of $K$ and $\chi$
that are linked to the generation of a hierarchy of vortical
structures is worth looking to the vorticity field, directly  
connected to the flow structures.
In the top panels of figure \ref{fig12} the contours
of $<\omega_\theta>=0.1$ (black line) demonstrate
that, the Hill vortex  strongly perturbed with $n_a=4$
does not survive for a long time. In fact
its characteristic shape at $t=60$ is destroyed in the inner 
region. This modification is more visible in figure \ref{fig12}b than in figure \ref{fig12}a.
In correspondence to the
large undulation of $<\omega_\theta>=0.1$,
at a distance $r=0.2$ from the axis 
the fluctuating kinetic energy (the red contours)
is large  (figure \ref{fig12}a) and increases in figure \ref{fig12}b.
The increase for $\delta_0=0.03$ of the velocity fluctuations
produces a large deformation of
$<\omega_\theta>$ in the interior of the vortex
and in a greater measure near the front. 
To understand the flow structures at the interior
of the vortex visualizations of the vorticity components
in $x-y$ planes
at the streamwise location of the maximum of $<\omega_\theta>$
may help.
The magenta patches in figure \ref{fig12}e indicate  
regions with $\omega_\theta > 1$, implying that 
the circular distribution of $\omt$ at $t=0$ and of
the similar one  at $t=20$ (figure \ref{fig5}a)
is destroyed at $t=60$. This excess of
vorticity is generated by the convective
and stretching terms in Eq.(\ref{eqot}). It has been
observed, that also at $t=60$ $C_\theta$ prevails
on $S_\theta$. For $\delta_0=0.01$ there is a reduction of the distance
between the contour lines of $<\omega_\theta>$ in a $r-z$ plane,
from $r=0$ up to the center of the vortex.
The thickness of these layers is close to  that in figure \ref{fig12}e.
On the other hand, from the center to the external surface of the vortex
the separation of the contours is close to
that at $t=0$. For $n_a=4$ and 
$\delta_0=0.03$ even the regions near the surface
present intensification in certain layers, 
and, in particular, near the front. This intensification
can be understood by the $\omt$ visualization in
figure \ref{fig12}f that enlight the formation  
of $\omega_\theta<0$ (blue lines) in certain regions close to
other with $\omega_\theta<-1$ (cyan lines). Close to these
patches there are other location with $\omega_\theta>1$.
This condition with high vorticity gradients is very unstable. 
High gradients of all the  vorticity components are
produced as it is shown by the contours of $\omr$ in
figure \ref{fig12}j. The figure of
$\omz$ (not shown), similar to that of $\omr$ and of $\omt$,
demonstrate that inside the vortex the secondary motion
is comparable to the primary one.
In these two figures it can be appreciated that the
symmetry is no longer good as that for $\delta_0=0.01$. 
Later on the vorticity field is similar to that produced
by random disturbances. The difference is that for $t>60$
and for $n_a=4$ in figure \ref{fig11} the values
of $K$ and $\chi$ are greater than those for random disturbances,
and as a consequence  the  initial structure
of the vortex is no longer    recognisable.
By random disturbances with amplitude $\delta_0=0.03$  the growth
of secondary vorticity is reduced (figure \ref{fig11}) and 
the vorticity field is  not concentrated in
localised structures (figure \ref{fig12}g and figure \ref{fig12}k).
The low  level of $K$ in figure \ref{fig11}a
is corroborated by the reduced number of contours
in figure \ref{fig12}c and that of
$\chi$ in figure \ref{fig11}b by their reduction in 
figure \ref{fig12}g and figure \ref{fig12}k.
Figure \ref{fig12}c shows, through the contour of
$<\omt>=0.1$, an external boundary without a wake and
with a shape  similar to that at $t=0$.
However the convoluted contours of $\omega_\theta$ 
at the interior (figure \ref{fig12}g) are very
different from the circular one at $t=0$. By random
disturbances with $\delta_0=0.09$  the corresponding figures,
figure \ref{fig12}d,figure \ref{fig12}h  and figure \ref{fig12}l
demonstrate that at the interior is rather difficult to 
recognise any structure typical of the Hill vortex.

\section{Conclusions}

Direct numerical simulations of the Navier-Stokes equations have
been performed to investigate whether the results,
obtained by simplified equations, on the instability of
Hill vortices are reproduced by the real conditions with
a small amount of viscosity. Indeed, it has been observed that by
imposing polar disturbances  on the surfaces a tail on the rear of
the vortex forms. At high Reynolds number 
the elongation of the spike agrees with that predicted
by the simplified Euler equations. During the evolution 
the polar disturbances are convected in the wake and
large part of the vortex surface has a shape similar
to the initial one. The two convective terms in the transport 
equation of $\omt$ are mainly localised near the surface,
with that due to $u_r$ 
greater than that due to $u_z$. In the interior of
the vortex the convective terms are zero implying that the 
initial vorticity does not change.
From these axisymmetric simulations
it may be inferred that the Hill vortex behaves as
the Lamb dipole in two-dimensions, that is, both shed an
amount vorticity in a wake and, with exception of a thin
layer near the boundary the vorticity
distribution in the central region do not change.
By adding to the polar azimuthal disturbances on the surfaces 
we are in presence of three-dimensional flows,
that is characterised by the growth of a secondary motion.
Even the three-dimensional disturbances are convected
far from the surfaces and accordingly the structure of the
spike in the rear is modified. The DNS allows to follow
the growth of the secondary motion through visualizations of
the vorticity field at the interior of the vortex.
The disturbances on the surface induce, as it is expected,
a weak secondary vorticity nearby, that does not largely
increase in time. 
On the other hand, a strong secondary vorticity
field is  generated near the axis. These act as travelling
waves growing in amplitude and moving towards the surface.
By increasing the amplitude
of the disturbance the evolution is faster,  but it has
been  observed that for Hill vortices travelling 
for a short distance the vortex is not destroyed,
that is does not create a hierarchy of large and
small scale structures typical of turbulent
flows. The transition from
ordered to disordered structures is due to the
action of the several terms in the vorticity
transport equations. The evaluation of
all the non-linear terms in the transport equation
of the primary azimuthal vorticity has shown
that the convective terms prevail on the vortex stretching
and tilting. The effect of the latter  increases with the amplitude
of the disturbance.

The conclusion drawn by short time simulation have
a relative validity, to be more convincing
should be based on simulations lasting for
a long time. These are inefficient, from computational aspects
in a fixed reference frame. 
The simulations were repeated in a reference frame moving
with the theoretical translational speed of the Hill vortex.
In these circumstances it has been found that the vortex
perturbed by three-dimensional disturbances
at a fixed  and at several wave numbers randomly chosen 
leads to situation that at later time give rise to
the complete destruction of the vortex. At
the very large Reynolds number here
chosen to be as much as possible close to the
Euler equations, the growth of the secondary vorticity 
components is high and the kinetic energy associated with them 
is not dissipated. The velocity components in certain
small regions become very large and the numerical stability 
restrictions on the $\Delta t$, due to the explicit treatment of the
non-linear terms, limit the time of the simulations.
A first estimate of a Taylor microscale Reynolds number
give values order $200$ and to see how far we are
from a turbulent flows spectra in the homogeneous
direction should be evaluated. This will investigated
in the near future at low Reynolds numbers.

\section{Acknowledgments}
This work was inspired by a discussion with Keith Moffat
during his visit in Roma approximately 10 years ago.
The authors wish to thank Sergio Pirozzoli and George Carnevale
for the fruitful discussions. 
We acknowledge that some of the results reported in this paper have been
achieved using the PRACE Research Infrastructure resource GALILEO
based at CINECA, Casalecchio di Reno, Italy.
\bibliographystyle{jfm}
\bibliography{references}

\begin{thebibliography}{23}
\expandafter\ifx\csname natexlab\endcsname\relax\def\natexlab#1{#1}\fi

\bibitem[Cavazza {\em et~al.\/}(1992)Cavazza, van \& Orlandi]{cavazza92}
{\sc Cavazza, P., van, G.J.F.~Heijst \& Orlandi, P.} 1992 The stability of
  vortex dipoles. {\em Proceedings of the 11th Australasian Fluid Mech. Conf.,
  Hobart, Australia\/} p.~67.

\bibitem[Crow(1970)]{CROW}
{\sc Crow, S.C.} 1970 Stability theory for a pair of trailing vortices. {\em
  AIAA J.\/} {\bf 8}, 2172--2179.

\bibitem[van Dyke(1982)]{vandyke_82}
{\sc van Dyke, M.} 1982 {\em An album of fluid motion\/}. The Parabolic Press,
  Satnford Ca.

\bibitem[Fukuyu {\em et~al.\/}(1994)Fukuyu, Ruzi \& Kanai]{fukuyu94}
{\sc Fukuyu, A., Ruzi, T. \& Kanai, A.} 1994 The response of hill’s vortex to
  a small three dimensional disturbance. {\em J. Phys. Soc. Japan\/} {\bf 63,},
  510–527.

\bibitem[Hill(1894)]{hill1894}
{\sc Hill, M.J.M.} 1894 On a spherical vortex. {\em Phil. Trans. R. Soc. Lond.
  A\/} {\bf 185}, 213–245.

\bibitem[Lamb(1932)]{lamb32}
{\sc Lamb, H.} 1932 {\em Hydrodynamics\/}. Cambridge University Press.

\bibitem[Leweke \& Williamson(1998)]{lew_wil98}
{\sc Leweke, T. \& Williamson, C.H.K.} 1998 Cooperative elliptic instability of
  a vortex pair. {\em J. Fluid Mech.\/} {\bf 360}, 85--119.

\bibitem[Melander {\em et~al.\/}(1987)Melander, McWilliams \&
  Zabusky]{melander87}
{\sc Melander, M.~V., McWilliams, J.C. \& Zabusky, N.J.} 1987 Axisymmetrization
  and vorticity gradient intensification of an isolated two-dimensional vortex
  through filamentation. {\em J.Fluid Mech.\/} pp. 137--159.

\bibitem[Moffat \& Moore(1978)]{moff78}
{\sc Moffat, H.K. \& Moore, D.W.} 1978 The response of hill’s spherical
  vortex to a small axisymmetric disturbance. {\em J. Fluid Mech.\/} {\bf 87},
  749--760.

\bibitem[Orlandi(2000)]{orlandi00}
{\sc Orlandi, P.} 2000 {\em Fluid fow phenomena: a numerical toolkit\/}.
  Kluwer.

\bibitem[Orlandi {\em et~al.\/}(2001)Orlandi, Carnevale, Lele \&
  Shariff]{or_ca_le_sh01}
{\sc Orlandi, P., Carnevale, G.~F., Lele, S.K \& Shariff, K.} 2001 Thermal
  perturbation of trailing vortices. {\em Eur. J. Mech. B - Fluids\/} {\bf 20},
  511--524.

\bibitem[Orlandi \& Fatica(1997)]{orlandi97}
{\sc Orlandi, P. \& Fatica, M.} 1997 Direct simulations of a turbulent pipe
  rotating along the axis. {\em J. Fluid Mech.\/} {\bf 343}, 43--72.

\bibitem[Orlandi {\em et~al.\/}(2012)Orlandi, Pirozzoli \&
  Carnevale]{orlandi12}
{\sc Orlandi, P., Pirozzoli, S. \& Carnevale, G.F.} 2012 Vortex events in euler
  and navierstokes simulations with smooth initial conditions. {\em J.\ Fluid\
  Mech.\/} {\bf 690}, 288--320.

\bibitem[Orlandi \& Verzicco(1993)]{orl_verz93}
{\sc Orlandi, P. \& Verzicco, R.} 1993 Vortex ring impinging on a wall:
  axisymmetric and three-dimensional simulations. {\em J. Fluid Mech\/} {\bf
  256}, 615--645.

\bibitem[Pozrikidis(1986)]{pozrikidis86}
{\sc Pozrikidis, C.} 1986 The nonlinear instability of hill’s vortex. {\em J.
  Fluid Mech.\/} {\bf 168}, 337–367.

\bibitem[Protas \& Elcrat(2016)]{protas16}
{\sc Protas, B. \& Elcrat, A.} 2016 Linear stability of hill’s vortex to
  axisymmetric perturbations. {\em J. Fluid Mech\/} {\bf 799}, 579–60.

\bibitem[Ren \& Lu(2015)]{ren15}
{\sc Ren, Heng \& Lu, Xi-Yun} 2015 Dynamics and instability of a vortex ring
  impinging on a wall. {\em Commun. Comput. Phys.\/} {\bf 18}, 1122--1146.

\bibitem[Rozi(1999)]{rozi99}
{\sc Rozi, T.} 1999 Evolution of the surface of hill’s vortex subjected to a
  small three-dimensional disturbance for the cases of m = 0, 2, 3 and 4. {\em
  J. Phys. Soc. Japan\/} {\bf 68}, 2940.

\bibitem[Shariff {\em et~al.\/}(1994)Shariff, Verzicco \&
  Orlandi]{sh_ver_orl94}
{\sc Shariff, K., Verzicco, R. \& Orlandi, P.} 1994 A numerical study of
  three-dimensional vortex ring instabilities: viscous corrections and early
  non-linear stage. {\em J. Fluid Mech.\/} {\bf 279}, 351--374.

\bibitem[Stanaway {\em et~al.\/}(1988)Stanaway, Shariff \& Hussain]{stanaway88}
{\sc Stanaway, S., Shariff, K. \& Hussain, F.} 1988 Head-on collision of
  viscous vortex rings. {\em Proceedings of CTR summer school 1988\/} .

\bibitem[Verzicco \& Orlandi(1996)]{verzicco96}
{\sc Verzicco, R. \& Orlandi, P.} 1996 A finite difference scheme for direct
  simulation in cylindrical coordinates. {\em J. Comp. Phys.\/} {\bf 123},
  402--414.

\bibitem[Widnall {\em et~al.\/}(1974)Widnall, Bliss \& Tsai]{widnal74}
{\sc Widnall, S.E., Bliss, D.B. \& Tsai, C.-Y.} 1974 The instability of short
  waves on a vortex ring. {\em J. Fluid Mech.\/} {\bf 66}, 35--47.

\bibitem[Wray(1987)]{wray87}
{\sc Wray, A.A} 1987 Very low storage time-advancement schemes. {\em Internal
  Report, NASA Ames Research Center, Moffett Field, California\/} .

\end{thebibliography}
\end{document}